\documentclass[11pt,a4paper]{article}
\usepackage[utf8]{inputenc}\usepackage[T1]{fontenc}\usepackage{amsmath,amssymb}\usepackage{booktabs}\usepackage{graphicx}
\usepackage{hyperref}\usepackage[margin=2.4cm]{geometry}\usepackage{microtype}\usepackage{xcolor}\usepackage[section]{placeins}
\title{Early-Buyer Groups and Fast-Wallet Teams on a Bonding-Curve Launchpad: A Prospective Measurement Study of pump.fun on Solana}
\author{Fail Gafarov\\ Institute of Computational Mathematics and Information Technologies \\
Kazan Federal University \\
Kazan, Russia \\
fail.gafarov@kpfu.ru }

\begin{document}\maketitle
\begin{abstract}
We study coordinated groups of wallets on a bonding-curve launchpad (pump.fun on Solana), using a complete record of 200\,443 token launches, 15\,854\,121 curve trades and 51\,025\,958 SOL transfers over 88.3 hours. A prospective, multi-channel discovery procedure (co-buying, shared funding, joint signing, transfers, consolidation) yields 23\,710 wallet groups. Without ground-truth labels we validate them by out-of-sample link prediction, independent non-trading links and a comparison with naive co-buy clustering at equal numbers of flagged tokens: raw clustering collapses into one giant component, yet it ranks tokens at least as well as the groups, so the value of the structure is descriptive. Group activity is highly concentrated and splits into serial creators and long-lived rings of early buyers. Known members buy earlier than other buyers of the same token and extract more cash than an activity-matched placebo set, while late buyers' net cash flow is negative; tokens with members show a pump-like price path, but the contrast vanishes when controls are restricted to comparable tokens, so we report it as an association. A follower entering at the earliest group signal loses money out of sample in all 32 configurations and does no better than random entry. A few pools of tens to over a hundred fast wallets are associated with a two- to four-fold higher chance of reaching 20 SOL, also within creators, with a smaller excess at 60 SOL and none for graduation; on a final hold-out the team rules keep a precision of about one half to three quarters, yet many signals arrive when the curve already holds 20 SOL, one team supplies most of them, and they add no information beyond early curve state. We also document evaluation errors that turn the same data into apparent profit.
\end{abstract}
\section{Introduction}
Launchpads that issue tokens on a bonding curve have turned token creation into a mass-market activity: a token can be created, priced and traded within seconds, with no listing step and no initial liquidity provider. On pump.fun, the dominant launchpad on Solana \cite{mancino2025}, most launches never trade meaningfully (in our window 53\% of complete launches never hold more than 1 SOL in their curve), while a small fraction run up quickly and a smaller one migrates to a conventional automated market maker. Public discussion attributes much of the price formation to coordinated actors -- groups of wallets that create tokens, buy within the first seconds, and sell into the demand they trigger -- but a systematic, population-level measurement of such groups is missing; concurrent studies characterise manipulation strategies and bundle traces \cite{szwajcok2026,hu2026melt} or label fraud \cite{chen2026rug,li2026rug}, and none attaches a validated group structure to the price path and to the profit of an outside trader. This paper provides one, and reports where the structure does and does not carry information.

We ask six questions. (Q1) Can coordinated wallet groups be recovered from public on-chain data without ground-truth labels, and how should such a recovery be validated? (Q2) How concentrated is group activity, and what types of groups exist? (Q3) How do groups fund, launch and withdraw (the \emph{campaign} cycle)? (Q4) Do tokens in which known group members participate follow a distinct price path, and who gains and who loses? (Q5) Does knowing about a group let an outside trader profit once realistic execution costs are paid? (Q6) Are there large teams of fast wallets that are associated with tokens reaching a high curve level, does the discovery find them whole, and is their success selection or effect?

\textbf{Primary and exploratory questions.} Q1 and Q5 were fixed before the analysis (the validation design and the follower simulation); the answer to Q5 is a negative result and does not depend on tuning. Q2--Q4 are descriptive. Q6 grew during the work: the team layer, its refinements and the case studies were developed on the same window, so they are exploratory, and Section 8.10 therefore adds a hold-out protocol in which the rules are frozen before the last block of the window is touched.

\textbf{Contributions.} (i) A prospective (time-consistent), multi-channel discovery procedure with a validation design that does not need labels (Section 4.3), and a comparison with naive co-buy clustering at equal numbers of flagged tokens (Section 5.1). (ii) A population-level description of groups and their funding--launch--consolidation campaigns over 200\,443 launches. (iii) Permutation-based and covariate-adjusted comparisons of timing, price path and cash flow, with an explicit account of what the adjustment does and does not remove (Sections 6.4 and 9). (iv) A negative result with out-of-sample, group-clustered intervals: the signal is detectable but not exploitable by a follower. (v) A catalogue of evaluation pitfalls, including a measured demonstration of how ignoring the trader's own position and a label leak produce profits that vanish under correct accounting. (vi) A team layer that repairs the fragmentation of small groups, case studies of four large fast-wallet teams, and a final hold-out evaluation with timing, concentration and incremental-value checks (Sections 8.1--8.10). The checks lead to a more modest reading of the team signal than a first pass suggested; we say so where it matters.

\textbf{Scope.} The data cover a single window of 88.3 hours. We therefore make claims about this window and discuss external validity separately (Section 10). "Group" always means a set of wallets that behave in a coordinated way according to explicit, observable rules; we do not claim common ownership, and we do not label any group as manipulative.

\textbf{Terminology.} \emph{Manipulation} has a legal and an economic meaning that this paper cannot establish: it would require intent and the wrongful effect on price, which on-chain data do not show. We therefore speak of \emph{coordination} (wallets acting together by observable rules), of a \emph{pump-like price path} (a fast rise followed by a fast fall after a reference time) and of \emph{cash extraction} (net SOL taken out of a curve by a set of wallets). We call a scheme pump-and-dump-like only when all three appear together, and we state the test applied each time.

\section{Background and related work}
Our question sits between four literatures: the economics and measurement of market manipulation, the security literature on scam tokens and sniper bots, work on pump.fun and memecoins that appeared in the last two years, and the network-science methods used to recover coordinated groups from public ledgers. We summarise each and say what separates our design from it; Section 10 returns to the points where earlier results and ours agree or differ.

\textbf{Bonding-curve launchpads and automated market makers.} A token on pump.fun trades against a constant-product curve with virtual reserves; buyers and sellers interact with the curve, not with each other, so a sale never requires a counterparty but its proceeds are bounded by the real SOL held in the curve. When the curve has accumulated a fixed amount of SOL the token \emph{graduates} and liquidity moves to an AMM pool. Constant-product markets have been analysed formally, including conditions under which their prices must closely track a reference market price \cite{angeris2019} and their use as price oracles \cite{angeris2020}, and the design space of AMM-based exchanges has been surveyed \cite{xu2023}. A launchpad curve differs in two ways that matter here: it exists from the first second of a token's life with no liquidity provider, and the SOL it holds is created by the buyers that arrive, so early profit is a transfer from later buyers to earlier ones rather than a return to liquidity provision. Our execution model (Section 4.5) uses the exact post-event reserves recorded by the protocol.

\textbf{Market manipulation in theory and in stock markets.} Pump-and-dump manipulation is old: models show how a trader can profit by moving a price with trades alone \cite{allen1992}, or by exploiting the credibility of privileged information \cite{benabou1992}, and the economic question of what separates manipulation from legitimate informed trading \cite{kyle1985} has been framed as one of harm to both price accuracy and liquidity \cite{kyle2008}. Evidence from enforcement actions in equity markets shows that manipulators are typically insiders or other plausibly informed parties and that prices rise during the manipulation and fall afterwards \cite{aggarwal2006}. A bonding-curve token differs from a listed stock in that its price is set by a deterministic rule that any participant can compute, so an informed party can know the effect of its own trade exactly.

\textbf{Pump-and-dump schemes in cryptocurrency markets.} Early measurement studied coordinated pumps announced in messaging channels and executed on centralised exchanges \cite{xu2019,li2019,kamps2018,hamrick2021}. Later work characterised the communities that organise them \cite{lamorgia2023,nizzoli2020,mirtaheri2021}, built detectors that flag a pump as it unfolds \cite{lamorgia2020,nghiem2021} or predict the target coin before a scheduled pump \cite{hu2023}, traced the organisers of pumps through information-diffusion graphs \cite{fu2025}, and followed coins after the pump \cite{clough2023}. Economic analyses ask why people join: overconfidence and gambling preferences can explain participation despite negative expected returns \cite{dhawan2023}. Outside pump channels, price manipulation has been documented on a single large exchange \cite{gandal2018} and through Tether purchases timed after market downturns \cite{griffin2020}. On decentralised exchanges, Telegram-coordinated campaigns have recently been found to rely on honeypot tokens that victims cannot sell rather than on price manipulation \cite{cernera2026}. Our setting differs in three ways: there is no announcement channel, so coordination has to be inferred from on-chain behaviour; the venue is a deterministic curve rather than an order book; and the unit of observation is the entire population of launches in a window rather than a list of announced events.

\textbf{Scam tokens, rug pulls and sniper bots.} Scam tokens on a decentralised exchange have been detected and characterised on Uniswap \cite{xia2021}; rug pulls on Uniswap have been predicted with machine-learning models before they occur \cite{mazorra2022} and detected by fusing contract code with token-flow graphs \cite{wu2025rphunter}, and sniper bots that buy new tokens in the first seconds after listing have been analysed across Ethereum and the BNB Smart Chain \cite{cernera2023,cernera2023www,cernera2025}. Fraud schemes outside token markets have been profiled on Bitcoin and Ethereum \cite{vasek2015,bartoletti2020}, and attacks on decentralised finance have been systematised \cite{zhou2023sok}. On Solana, a dataset of suspicious liquidity pools is annotated with inactivity indicators \cite{alhaidari2025}, a measurement study distils three behavioural patterns of rug pulls on Solana, one of them pump-and-dump \cite{chen2026rug}, and a gradient-boosting model predicts fraudulent memecoins from the first five minutes of trading \cite{li2026rug}. These studies attach fraud labels to tokens; we attach structure to wallets and ask what the structure does to the price path, and who profits.

\textbf{Pump.fun and memecoins.} Descriptive studies show how much of Solana's token creation and DEX activity pump.fun accounts for and how few of its tokens move on to a major exchange \cite{mancino2025}, and a cross-chain analysis of meme coins finds artificial growth strategies such as wash trading and pool-based price inflation among high-return tokens \cite{mongardini2025}. Marino et al. model the probability that a launch graduates from the SOL locked in its curve and behavioural covariates \cite{marino2026}. The study closest to ours in subject is a large measurement of pump.fun manipulation that identifies five strategy classes (wash trading, creator-address obfuscation, coordinated selling, copycat coins and social-media manipulation) and a market for manipulation tooling \cite{szwajcok2026}; a Solana dataset adds bundle traces that link accounts controlled by the same entity and reports that coordinated accounts hold on average more than a third of a token's supply \cite{hu2026melt}. Our concealment analysis (Section 8.8) complements both from the side of cash flows. A recent preprint series detects persistent cohorts of early buyers and audits the labels and the coverage of pump.fun collection pipelines \cite{kamat2026,kamat2026b,kamat2026c}; the cohort paper's author has withdrawn its numerical results pending a corrected analysis (version 4, October 2026), so we use it only as a pointer to the research question, and we take the audits as a reason to report completeness flags for every token (Section 3). To our knowledge no study combines multi-channel group discovery that includes funding campaigns, a label-free validation design, permutation-based dynamics, and follower profitability under an exact-reserve execution model.

\textbf{Wash trading, front-running and extractable value.} Wash trading has been quantified on decentralised exchanges \cite{victor2021}, centralised venues \cite{cong2023} and NFT markets \cite{wachter2022}; all three identify coordinated activity from trade data, as we do. Front-running and miner or validator extractable value show how ordering within a block can transfer value \cite{daian2020,eskandari2020,zhou2021,qin2022}, and countermeasures have been classified \cite{heimbach2022}. Our time resolution (seconds) does not capture submission order, tips or bundle structure, although the position of a transaction in its block is recorded.

\textbf{Address clustering and Sybil detection.} Heuristic clustering of addresses goes back to payment studies on Bitcoin \cite{meiklejohn2013,reid2011,ron2013,androulaki2013,harrigan2016,moser2013}, and has been adapted to Ethereum \cite{victor2020,beres2021} and Zcash \cite{kappos2018}; graph neural networks have been applied to financial forensics on labelled Bitcoin transactions \cite{weber2019}. Multiple identities controlled by one actor are the Sybil problem \cite{douceur2002}, studied in airdrops through hunter profits and subgraph features \cite{luo2025,liu2025}. Those methods rely on ownership heuristics or on labels. We use behavioural rather than structural-ownership evidence and keep the semantics explicit: a group is a hypothesis about coordination, validated without labels.

\textbf{Community detection, coordination and link prediction.} We cluster with modularity-based methods \cite{newman2004,newman2006,blondel2008,traag2019} and are careful with their known limits: a resolution limit \cite{fortunato2007}, dependence on the random seed, which a consensus over many runs reduces \cite{lancichinetti2012}, and the difficulty of validating a partition without ground truth \cite{fortunato2010}. Methods for finding coordinated networks of accounts on social media use similarity of account behaviour as evidence \cite{pacheco2021,nizzoli2021,cresci2020}; graph-based anomaly detection is surveyed in \cite{akoglu2015}. Pairs that co-buy again are a link-prediction problem \cite{liben2007,adamic2003,zhang2018}, and graph neural networks \cite{kipf2017,hamilton2017,velickovic2018} are the natural learned alternative, which we test in Section 8.9.

\textbf{Evaluating trading signals.} Backtests of a signal selected from many candidates overstate performance \cite{bailey2017,harvey2016}, which is why reality checks and tests of superior predictive ability exist \cite{white2000,hansen2005}, and why false-discovery control is standard in multiple testing \cite{benjamini1995}. Time-ordered data need splits that respect the span of the labels: purging and an embargo \cite{lopezdeprado2018}, and out-of-sample evaluation designs that suit dependent data \cite{bergmeir2012,cerqueira2020}. Our statistical tools follow this line: bootstrap resampling \cite{efron1979,politis1994} at the level at which observations are dependent \cite{cameron2008,cameron2015}, exact permutation p-values that cannot be zero \cite{phipson2010}, Wilson intervals for proportions \cite{wilson1927}, matching and inverse-probability weighting against controls \cite{rosenbaum1983,stuart2010,ho2007}, empirical-Bayes shrinkage for small samples \cite{efron1975}, and precision-recall summaries for rare positive classes \cite{davis2006,saito2015}.

\textbf{Positioning.} The closest studies label manipulation strategies, bundles or fraud at the token level \cite{szwajcok2026,hu2026melt,chen2026rug,li2026rug}. We take the wallet group as the unit, validate it without labels against a baseline compared at equal numbers of flagged tokens, link it to price paths and to cash flows, evaluate an outside follower under an exact-reserve execution model, and report what survives a final hold-out, a signal-timing check and an incremental-value test. Where our results are negative or weaker than a first pass suggested, we say so (Sections 5.1, 6.4 and 8.10).

\section{Data and measurement}
\subsection{The collection module}
The data come from a purpose-built collector that follows the chain block by block rather than through the programs' public interfaces. It subscribes to every confirmed block of Solana over a WebSocket connection, and falls back to remote-procedure-call requests for any slot that arrives incomplete, late or not at all; a slot is never declared empty unless the parent chain of the finalised blocks proves that it was skipped. Each block is decoded once, by one set of decoders, into the events of the two protocols (token creations, curve buys and sells, graduations, AMM trades) and into all native SOL transfers and token transfers of the block, including transfers made inside nested program calls. Writing is idempotent: every row has a natural key (transaction signature and position of the instruction), batches are written in the order of their dependencies, and durable checkpoints let the process resume after a restart without gaps or duplicates. A coverage table records every processed slot with its parent and its number of transactions, so that missing blocks can be detected after the fact instead of assumed away, and the collector keeps counters for its own throughput and backlog. The module is read-only with respect to the chain and independent of any analysis code.

The decoder reads the instruction data and the event records that the protocol emits for every trade. This matters for the execution model of Section 4.5: each curve trade carries the reserves of the curve \emph{after} the trade as reported by the protocol itself, so prices and curve balances are observed, not reconstructed, and the arithmetic of a hypothetical trade can be checked against what the protocol did.

\subsection{Data structure}
\begin{table}[htbp]
\centering\small
\setlength{\tabcolsep}{3pt}
\caption{Structure of the collected data. Every table carries the slot and the position of the event in the block, so that events of different tables can be ordered and joined by transaction.}\label{tab:datastructure}
\begin{tabular}{lp{205pt}r}
\toprule
Table & Content & Rows in the window \\
\midrule
Blocks and transactions & one row per transaction that touches the protocols: signature, slot, position in the block, block time, fee payer, signers, fee and estimated priority fee, programs invoked & 21 183 236 \\
Block coverage & one row per processed slot: parent slot, number of transactions, skipped flag; used to prove that no slot was missed & one per slot \\
Tokens & one row per creation: mint, creator and signer, initial buy, bonding-curve account, metadata, completeness flag & 200 443 \\
Curve trades & one row per buy or sell: trader, SOL and token amounts, protocol-reported reserves after the event (virtual and real), fee and creator fee, slot and position in the block & 15 854 121 \\
Graduations & one row per migration of a token from the curve to the AMM & 5 862 \\
AMM trades & trades of migrated tokens, plus one-minute aggregates & 5 565 871 \\
SOL transfers & native transfers of SOL with slot, position, instruction path (including inner calls), sender, receiver, amount and kind; trade legs inside protocol transactions are marked & 51 025 958 \\
Token transfers & transfers of SPL tokens between owners, used for transfer evidence & one row per transfer \\
\bottomrule
\end{tabular}
\end{table}

The records form a relational database whose tables are linked by the transaction and by the token (Table~\ref{tab:datastructure}). Every row carries the slot, the block time (in seconds) and the position of the transaction in its block, so that curve trades, transfers and creations of one transaction can be joined and ordered. A token is identified by its mint address; its creation row holds the creator, the signer that paid for the transaction, the creator's initial purchase and the account of its bonding curve. From these tables the analysis builds a set of derived tables (one row per token, per wallet and token, per pair of wallets, per group), documented column by column in the replication package. All wallet and mint addresses are replaced by salted hashes in the public release.

\textbf{Quality controls.} The analysed snapshot spans 2026-09-26T19:00:23 to 2026-09-30T11:16:00 UTC (88.3 hours). No change of the source tables was detected between the start and the end of the analysis run and read-only watermarks were fixed at the start; this is evidence of stability, not a transactionally consistent snapshot, and we list it among the limitations. The two halves of the window have similar token statistics (Table~\ref{tab:dataaux}), which we use later for out-of-sample validation. Every token carries a completeness flag, set when its observation is cut (for example only its creation or part of its early trades were seen).

\begin{table}[htbp]
\centering\small
\setlength{\tabcolsep}{3pt}
\caption{Dataset used in the study (single observation window).}\label{tab:dataset}
\begin{tabular}{p{214pt}p{214pt}}
\toprule
Item & Value \\
\midrule
Observation window (UTC) & 2026-09-26T19:00:23 -- 2026-09-30T11:16:00 \\
Duration (hours) & 88.3 \\
Tokens created / with complete creation record & 200,443 / 154,045 \\
Bonding-curve trades & 15,854,121 \\
System-program SOL transfers & 51,025,958 \\
PumpSwap trades / graduations & 5,565,871 / 5,862 \\
Protocol transactions & 21,183,236 \\
Tokens whose curve reached 20 / 60 SOL & 7.7\% / 2.4\% \\
Tokens with at most 3 trades & 26.3\% \\
\bottomrule
\end{tabular}
\end{table}

\textbf{Incomplete tokens.} About a quarter of the launches are incomplete; their median number of recorded trades is one against 20 for complete tokens, and they almost never reach 20 SOL (Table~\ref{tab:dataaux}). Their low apparent success is partly a consequence of the truncated observation, so all outcome statistics use complete tokens only. We examined how the exclusion arises (Table~\ref{tab:incompleteness}). The incomplete flag is almost a function of one creation-time variable: launches in which the creator made an initial buy are never incomplete (the buy is a recorded trade), while about 60\% of the launches without an initial buy are, that is launches in which nothing was recorded after creation; the other creation-time variables (time of day, the creator's earlier tokens) add little. Incompleteness therefore removes launches with no recorded buyer. Two consequences follow. First, base rates are conditional on at least some recorded trading. Second, the contrasts of the paper compare group tokens, which by definition have at least two members and two other buyers in the first minute, with control tokens that have at least four early buyers; neither can be incomplete, so the exclusion cannot act differently on them. What it does affect is the denominator: a lift (the share of signal tokens that reach a level divided by the base rate) is between one and about 1.3 times larger when the base rate is computed over all launches, depending on what is assumed about the incomplete ones (panel B of Table~\ref{tab:incompleteness}), and its ordering across methods is unchanged. We did not model the censoring mechanism beyond this, because the launches excluded are, by their records, launches on which no behaviour was observed. The numbers of tokens, trades and wallets that appear in different parts of the paper differ because different populations are used (complete tokens, second half, tokens with recorded early trades); Table~\ref{tab:reconcile} reconciles them.

\begin{figure}[tbp]
\centering
\includegraphics[width=\linewidth]{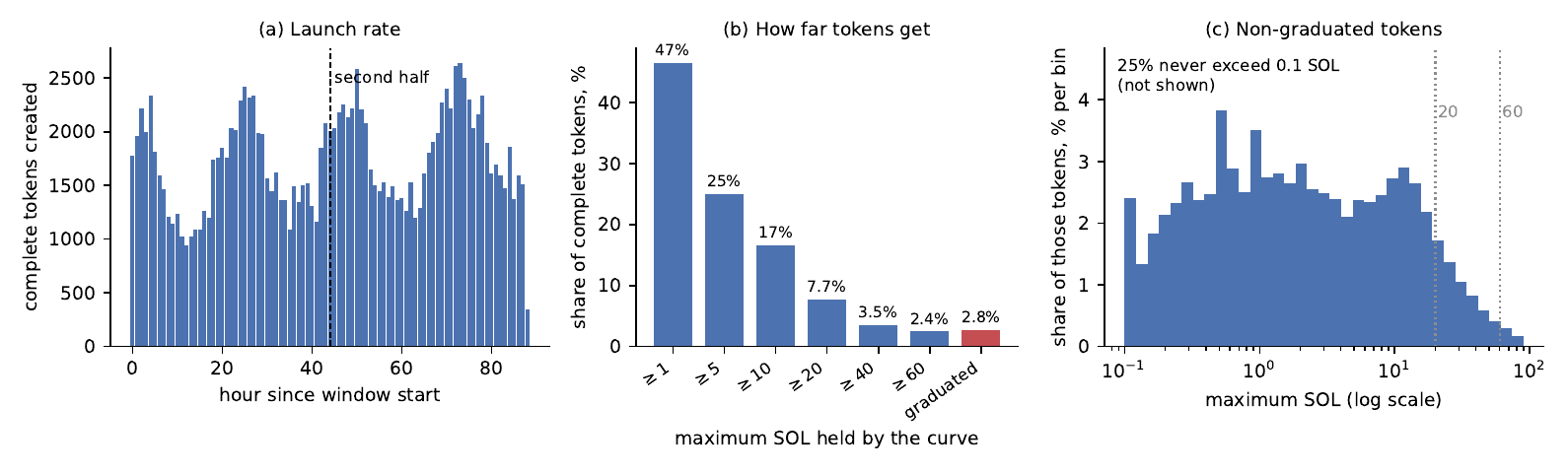}
\caption{The data. (a) Complete tokens created per hour; the dashed line marks the boundary of the two halves of the window. (b) How far tokens get: share of complete tokens whose curve held at least the given SOL at some time, and share that graduated. (c) Distribution of the maximum SOL held by the curve of the complete tokens that did not graduate (log axis; a graduated token's maximum is fixed by the graduation threshold, so it is shown as a separate bar in (b)).}\label{fig:dataset}
\end{figure}

The figure shows that launching is a steady activity with a pronounced daily rhythm, that the large majority of tokens stop far below the levels discussed later (a quarter reach 5 SOL, fewer than one in ten 20 SOL, a few per cent graduate), and that the maxima of non-graduated tokens are spread smoothly over several orders of magnitude. There is no feature of the distribution at 20 SOL: the levels of 20 and 60 SOL used as outcomes are round numbers that separate tokens that are clearly alive from tokens that run far, not thresholds of the protocol, and the results for 60 SOL are reported next to those for 20 SOL.

\textbf{Time resolution and ordering.} Block time has a resolution of one second. The position of a transaction inside its block is recorded, so the order in which two transactions of the same block were executed is observed; what is not observed is the order in which they were submitted, the tips that determine inclusion, and whether transactions were sent as one bundle. The follower simulation of Section 7 therefore works in whole seconds and cannot represent the latency advantage of a trader that is included in the same block, and Sections 8.6--8.7 use the position inside the block as evidence of coordination, not as a measure of latency.

\textbf{Terminology: prospective, not causal.} Every group, edge and signal used in the main analysis is derived by replaying events in time order; a decision at time $t$ uses only information available before $t$. We call this \emph{prospective} (time-consistent) analysis. It rules out look-ahead; it does not identify causal effects. The associations in Section 6 (members enter earlier, tokens with members follow a different price path, members extract cash) are descriptive: they may reflect token popularity, creator behaviour or market conditions, and we control for observables with permutation nulls, matched controls and a sensitivity analysis, but we make no claim of causal identification. A separate retrospective attribution (using the final graph) is reported only for descriptive group sizes and is never used to evaluate trading.

\section{Methods}
\subsection{Evidence channels and group formation}
Pairs of wallets accumulate observable evidence: \textbf{co-buying} (first buys of the same token close in time, both early in the token's life), \textbf{co-selling}, \textbf{common funding} (a shared funder for the same token through a few hops within a day), \textbf{joint signing} of creation or early trades, direct \textbf{SOL transfers}, \textbf{consolidation} (several funded wallets sending to one destination shortly after selling), and token transfers. A pair is accepted when explicit repetition rules hold (for example at least three shared tokens and a minimum overlap of the two wallets' token sets for co-buying, or repeated common funding); the exact thresholds are listed in the replication package and varied in Section 9. Wallets with very many distinct SOL counterparties are treated as \emph{infrastructure} from the moment they reach a threshold: they cannot be the core of a group or a bridge between groups, and may remain members only with independent repeated evidence. Groups grow from a stable core by wallets that are supported by at least two core members; a single bridge never merges two populations.

\subsection{Earliest prospective signal}
For each token we select the group attribution that was available earliest (ties broken by attribution score), recording the group size at that moment and whether the group is linked to the creator. This \emph{earliest signal} is the only quantity used in the follower analysis.

\subsection{Label-free validation design}
Because wallet ownership is unknown, we use four independent checks. (V1) \textbf{Link prediction} \cite{liben2007}: pairs accepted using only first-half evidence are scored on whether they co-buy again in the second half, compared with all other co-buying pairs and stratified by first-half co-buy count. (V2) \textbf{Independent links}: pairs connected by non-trading evidence (shared funding campaign, joint signing, transfers) are compared with activity-matched random pairs. (V3) \textbf{Baselines}: connected components of the raw co-buy graph at several count thresholds, and Louvain communities \cite{blondel2008} and components of the first-half co-buy graph, evaluated as group signals on second-half tokens at the same numbers of flagged tokens as the discovery groups (Section 5.1). (V4) \textbf{Parameter sensitivity}: the stability of the edge set and of signal quality under changes of the hub threshold and of the discovery parameters.

\subsection{Dynamics measures}
\emph{Membership} is defined without reference to the token studied: a wallet is a \textbf{member} at token creation if it belongs to a group of at least two wallets whose membership was known before that creation. A \textbf{group token} has at least two members and two other buyers among the first-minute buyers; a \textbf{control token} has no members among at least four early buyers. Timing is the lag of each wallet's first buy after creation; the null distribution is obtained by permuting member labels within each token. Price-path measures are computed only after a 60-s reference time, so early mechanical effects cannot create them, and are compared with controls stratified by curve SOL at the reference and by time-of-day block. Profit uses each wallet's cash flow (sales minus purchases) within the token; we report cash flow rather than mark-to-market because unsold inventory has no executable value at the last spot price. A placebo set of the same size drawn from the other early buyers gives the comparison.

\subsection{Execution model}
The follower buys 0.05 SOL against the exact post-event reserves at the first curve observation at or after the later of the signal and four seconds after creation, delayed by a latency of 1, 2, 3 or 5 s, and exits by a take-profit, stop or time rule, again after latency. Costs follow two conventions. The \emph{project} convention applies the larger of the event fee and 2\% per side and 1.5\% slippage per side; the \emph{event} convention uses the fee recorded in the event and 0.3\% slippage. The trader's own position is added to the reserve state used for the exit (SOL in, tokens out), because an exit quoted on a path that omits the entry would find less SOL in the pool than the trader deposited. Net return per trade is $r=(P_\text{out}-P_\text{in}-f_\text{in}-f_\text{out}-2c)/P_\text{in}$ with network fee $c$. When latencies are compared, the comparison is made on the same set of signals and the same policy, so that a different set of executable trades does not masquerade as a latency effect.

\subsection{Statistics}
Tokens of one group are dependent; all intervals for group-level quantities are bootstraps that resample groups \cite{efron1979,cameron2008,cameron2015}; for the matched contrasts of Section 6.4 we also resample creators, because one creator launches many tokens. Policy selection uses the first 70\% of signals in time and evaluation the last 30\%. Link prediction compares first and second half. Predictive models use a chronological 55/15/30 split with early stopping on the middle part and selection thresholds fixed on it. We report effect sizes with intervals rather than significance stars. Proportions carry Wilson intervals \cite{wilson1927}. A permutation p-value of zero means that no permuted statistic reached the observed one, that is $p<1/(B+1)$ for $B$ permutations \cite{phipson2010}. Time-ordered splits remove observations whose labels overlap the evaluation period \cite{lopezdeprado2018,bergmeir2012,cerqueira2020}.

\emph{Multiplicity.} Many specifications are examined, so the intervals are descriptive and no multiplicity-corrected significance is claimed \cite{bailey2017,harvey2016,benjamini1995}. We separate the questions fixed before the analysis (validation design, follower simulation, the primary contrasts of timing and cash flow) from the ones that were refined while looking at the same window (the team layer and its refinements, case studies, parameter sweeps); the latter are labelled exploratory, and wherever a rule was selected from many candidates the figure that carries the conclusion is the one measured on data that played no part in the selection (Section 8.10). The ledger of all tests with their status, data and outcome is in Table~\ref{tab:ledgerA (questions 1 to 5)} and Table~\ref{tab:ledgerB (question 6)}.

\emph{Unmeasured confounding.} For associations that matter for the conclusions we report the E-value of the risk ratio \cite{vanderweele2017}: the smallest risk ratio that an unmeasured confounder would need with both the exposure and the outcome to explain the estimate away. An E-value is a statement about the sensitivity of a point estimate to a hypothetical confounder, not evidence that none exists.

\subsection{Team layer and fast-wallet search}
Section 8 adds two constructions. The \emph{team layer} merges groups that share wallets (or have a high member overlap) and is built from evidence qualified before the middle of the window; Louvain communities \cite{blondel2008} of the accepted-edge graph are the comparison. The \emph{fast-wallet search} does not use the groups: wallets that bought within 5 s of token creation in many complete tokens are linked when they co-occur far more often than their activity implies, and clustered with Louvain \cite{blondel2008}. A token carries a team signal if at least three team members are among the buyers of the first minute (first 5 s for the fast-wallet communities). Selection is separated from effect by inverse-probability weighting of the team's tokens against controls without any member of a large team, 1:5 nearest-neighbour matching with a caliper \cite{rosenbaum1983,stuart2010,ho2007}, and a within-creator contrast, all using covariates measured strictly before the team's entry age; intervals are bootstraps over tokens or creators.

\subsection{Hold-out protocol, equal budgets and sensitivity analyses}
\emph{Three blocks.} For the team layer, the window is split in time into a block A (the first half, used to build teams, estimate reputations and select scouts), a block B (the next part, used to tune the few parameters of the rules) and a block C (the end of the window), which is evaluated once, with the rules and parameters frozen, using teams rebuilt from all evidence before C. The parameters of the first manuscript version are evaluated on C as well, to show how much of the earlier result was due to tuning.

\emph{Signal timing.} A team signal is available when the third member has bought. The curve may already hold 20 SOL by then, in which case the "outcome" happened before or at the signal. We therefore report the share of signals that fire after the curve has reached 20 SOL, the share of the others that reach it later, and the base rate of the same kind of tokens (conditional on being below 20 SOL at the same age). A strictly \emph{prospective} outcome is used throughout the comparisons of methods: among tokens still below 20 SOL at the end of the first minute, whether the curve reaches 20 SOL later.

\emph{Equal budgets.} Methods that flag different numbers of tokens are compared at equal numbers K of flagged tokens (precision of the K highest scored tokens, with ties at the cut-off broken at random in expectation) and by average precision over all tokens \cite{davis2006,saito2015}. Each method scores a token by the number of members of one of its groups among the first-minute buyers, and the team layer by the shrunk reputation of its group.

\emph{Concentration.} Because one team can dominate the signals, the rules are re-evaluated after dropping each team in turn (leave-one-team-out) and without the largest team.

\emph{Incremental value.} To test whether group features carry information beyond how busy the first minute was, a regularised logistic model of the prospective outcome on first-minute features that contain no group information is compared with the same model with group features added, on the final block, with hour-block bootstraps.

\emph{Sensitivity of the matched contrasts.} The price-path contrast of Section 6.4 is re-estimated with weights built only from covariates fixed at creation (creator's initial buy, creator history, time of day) or measured before two seconds after creation, with and without trimming to the region where treated and control propensities overlap, with effective sample sizes, intervals that resample creators, and E-values.

\section{Results I: validity and structure of groups}
\subsection{Validation without labels}
\emph{Naive co-buy clustering fails as a partition.} Connecting wallets that co-bought at least three times and taking connected components puts 94\% of the connected wallets into one component, and raising the threshold to twenty co-buys still leaves 82\% in the largest one (Table~\ref{tab:baseline}). The discovery rules instead produce a large number of groups whose largest member count is modest, so the core-and-support design does real work in keeping distinct populations apart.

\begin{figure}[tbp]
\centering
\includegraphics[width=\linewidth]{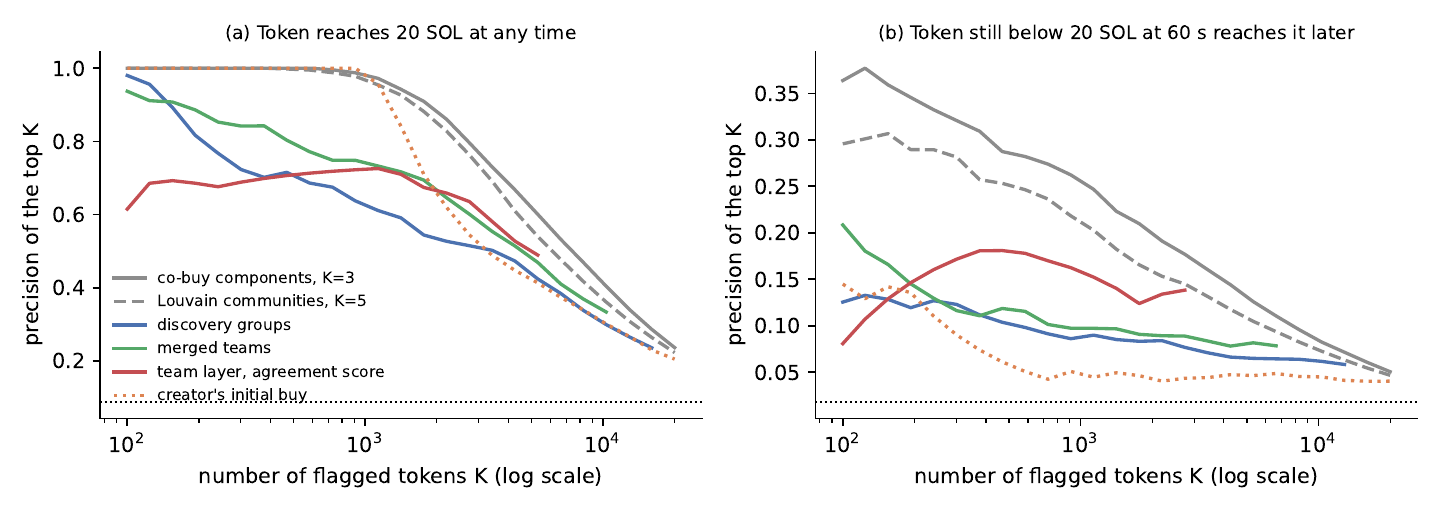}
\caption{Precision of the K highest-scored tokens against the number of flagged tokens K (log scale), second half of the window. Left: the token reaches 20 SOL at any time. Right: prospective outcome, among tokens still below 20 SOL at 60 s. Dotted horizontal line: base rate. The creator's initial buy is nearly perfect on the left because a large initial buy puts the curve at 20 SOL at once; on the right it carries little information.}\label{fig:budget}
\end{figure}

\emph{As a signal, naive clustering is not worse.} A partition is not the same as a signal, so we also asked how well each grouping predicts the outcome of second-half tokens when it is built from first-half data only. A token is flagged by a grouping when at least two members of one group are among its first-minute buyers, and the methods are compared at \emph{equal numbers of flagged tokens} (Section 4.8; Table~\ref{tab:equalbudget}, Figure~\ref{fig:budget}). Two outcomes are used: the token reaches 20 SOL at any time, and the strictly prospective one, in which only tokens that are still below 20 SOL after the first minute count and the question is whether the curve reaches 20 SOL later. The base rate of the second outcome is below two per cent.

\begin{table}[htbp]
\centering\footnotesize
\setlength{\tabcolsep}{3pt}
\caption{Methods compared at equal numbers of flagged tokens on the second half of the window. Each method scores every token by the number of members of one group among the buyers of the first minute (team layer: shrunk reputation). AP: average precision over all tokens; P@K: precision of the K highest-scored tokens. Left block: outcome 'reaches 20 SOL at any time' (the creator's initial buy is mechanically related to it). Right block: prospective outcome, among tokens still below 20 SOL at 60 s, 'reaches 20 SOL later'. The last row gives the base rates.}\label{tab:equalbudget}
\begin{tabular}{p{128pt}rrrrrrrrr}
\toprule
Method & \shortstack{Flagged \\[0pt] tokens} & AP & P@500 & P@2000 & \shortstack{Flagged \\[0pt] tokens} & AP & P@250 & P@500 & P@2000 \\
\midrule
co-buy components, K=3 & 45\,532 & 0.585 & 100.0\,\% & 88.2\,\% & 41\,551 & 0.162 & 32.4\,\% & 29.3\,\% & 19.9\,\% \\
Louvain communities, K=5 & 41\,729 & 0.522 & 99.8\,\% & 84.8\,\% & 37\,786 & 0.119 & 28.9\,\% & 25.1\,\% & 15.8\,\% \\
discovery groups & 16\,921 & 0.317 & 71.0\,\% & 53.4\,\% & 13\,831 & 0.055 & 12.8\,\% & 10.2\,\% & 8.4\,\% \\
merged teams & 10\,694 & 0.370 & 78.5\,\% & 66.4\,\% & 7\,766 & 0.057 & 12.8\,\% & 11.9\,\% & 8.9\,\% \\
team layer, agreement score & 5\,411 & 0.328 & 70.9\,\% & 67.2\,\% & 3\,208 & 0.065 & 16.2\,\% & 18.5\,\% & 12.8\,\% \\
reference: creator's initial buy & 69\,223 & 0.423 & 100.0\,\% & 64.8\,\% & 64\,282 & 0.044 & 10.7\,\% & 5.8\,\% & 4.3\,\% \\
base rate &  & 0.088 & 8.8\,\% & 8.8\,\% &  & 0.018 & 1.8\,\% & 1.8\,\% & 1.8\,\% \\
\bottomrule
\end{tabular}
\end{table}

The result is not the one a first look at single rules suggested. For both outcomes, the components and Louvain communities of the naive co-buy graph rank tokens better than the discovery groups at every budget we examined, and the margin is large for the prospective outcome (precision of the 500 highest-scored tokens about three in ten against one in ten). The reason is visible in what the score measures: the number of early buyers that belong to a large, densely co-buying population is itself a measure of how crowded and active the first minute is, which is what predicts a further rise. Among the grouping methods that carry a group structure, the team layer with its reputation score is the best at small budgets for the prospective outcome and clearly beats the creator's initial buy, but it does not beat the naive scores. We therefore no longer claim that the discovery rules improve the \emph{predictive quality} of the signal. What they provide is the structure: a partition into groups of limited size, with evidence behind every edge, that can be examined, tested for persistence and used to ask who does what (Sections 5.2--8). Whether that structure carries information beyond the crowd size is tested directly in Section 8.10, with a negative answer for the prospective outcome.

\begin{figure}[tbp]
\centering
\includegraphics[width=\linewidth]{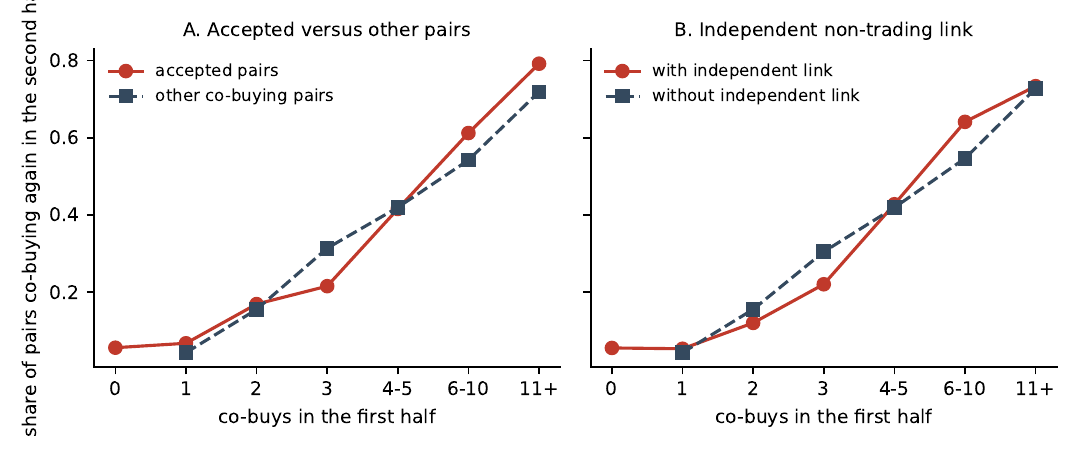}
\caption{Persistence of co-buying pairs on the second half of the window, by the number of co-buys in the first half. Left: pairs accepted with first-half evidence versus other co-buying pairs. Right: pairs with an independent non-trading link versus pairs without. Values in Table~\ref{tab:persistence}.}\label{fig:persistence}
\end{figure}

\emph{Accepted edges persist, but not beyond what co-buy counts predict.} Pairs accepted using first-half evidence co-buy again in the second half far more often than other co-buying pairs, but this contrast is confounded by the number of first-half co-buys: stratified on it (Table~\ref{tab:persistence}, Figure~\ref{fig:persistence}), the persistence ratio of accepted to other pairs is close to one for pairs with several co-buys and larger only for pairs with a single co-buy, and the AUC of predicting second-half co-buying is nearly the same for the co-buy count alone (0.70) as when accepted edges are ranked first (0.70). The additional channels and the overlap filter therefore do not improve this predictive test.

\emph{Independent links.} Pairs connected by non-trading evidence (shared funding campaign, joint signing or direct transfer) co-buy in the second half about 27\% of the time against 0.6\% for activity-matched random pairs, so co-buying coincides with evidence of shared control that is not itself trading. At equal co-buy counts, however, the non-trading link adds no persistence (Table~\ref{tab:persistence}): the persistence is carried by the co-buy history that such pairs also have.

\emph{Do the channels carry independent information?} We tested the conditional information of each evidence channel by regressing second-half co-buying on the first-half co-buy count and one indicator per acceptance reason (Table~\ref{tab:channelinfo}). Acceptance by the discovery rules adds nothing at the same co-buy count (odds ratio 0.94, interval including one), as the stratified comparison above suggested. Individual channels differ: entry-and-exit coordination, co-buy with consolidation and transfer-and-trade evidence raise the odds, joint authorization and the plain recurrent co-buy reason lower them, and adding the wallets' activity changes several signs. The honest reading is that the channels are partly a measure of how busy and how repetitive a pair is, and that persistence of behaviour, shared infrastructure and coordinated control are not separated by these data; a group is therefore a hypothesis about coordination, and the evidence for single-wallet groups (43\% of all groups) is only that a creator launches repeatedly. Results are reported separately for groups of at least two wallets wherever it matters.

\emph{The hub threshold is not a lever.} Changing the infrastructure threshold over two orders of magnitude (or removing it) leaves the accepted edge set largely unchanged and the class of the earliest signal unchanged for nearly all tokens (Table~\ref{tab:hubs}); removing it altogether made the funding search intractable on a 12-hour slice, so hubs matter for tractability, not for the groups found.

\subsection{Concentration and types}
\begin{figure}[tbp]
\centering
\includegraphics[width=\linewidth]{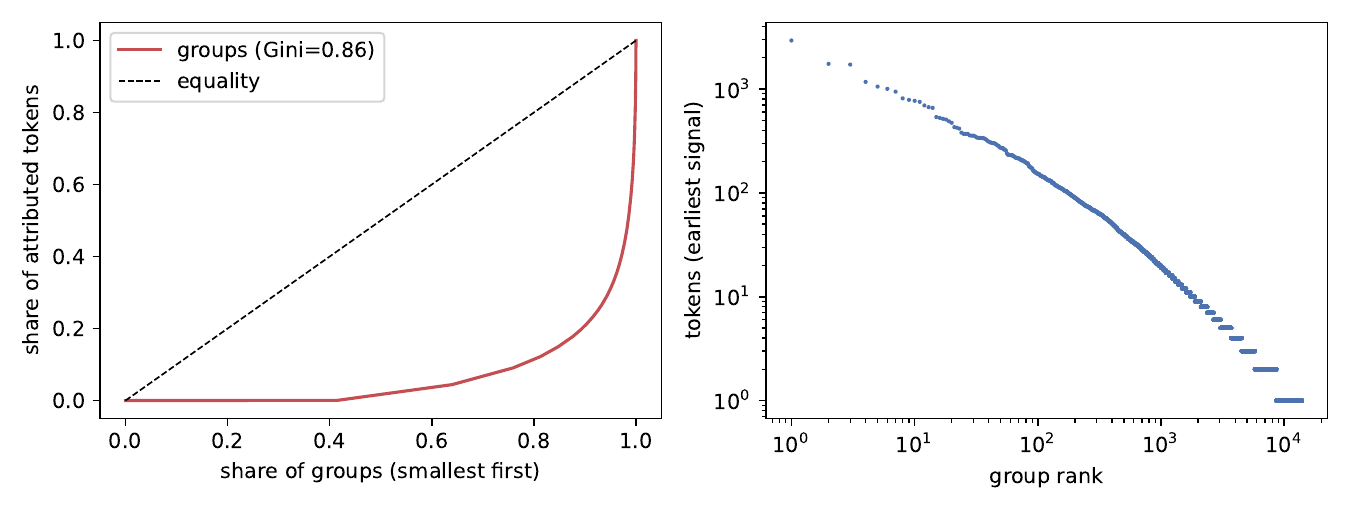}
\caption{Left: Lorenz curve of earliest-signal tokens over groups. Right: rank--size plot (log--log) of tokens per group.}\label{fig:concentration}
\end{figure}

13\,940 of the 23\,710 groups are the earliest signal for at least one token, and the activity is highly concentrated: the Gini coefficient of tokens per group is 0.86 against 0.19 for the same number of tokens spread uniformly over the same groups, and the top 1\% of groups account for 43\% of the attributed tokens (Figure~\ref{fig:concentration}). 43\% of the groups consist of a single wallet.

\begin{figure}[tbp]
\centering
\includegraphics[width=\linewidth]{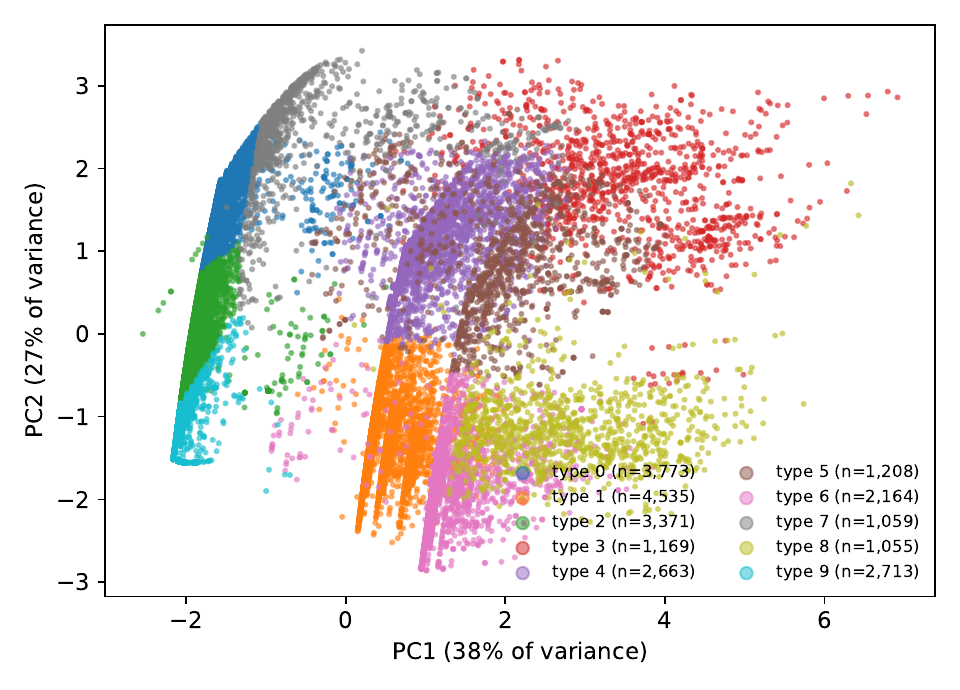}
\caption{Groups projected on the first two principal components of the standardised typology features, coloured by cluster.}\label{fig:typology}
\end{figure}

\emph{Types.} Clustering group-level features (size, token count, creator-linked and buyer-only shares, rate and lifetime, density, share of multi-channel edges) with k-means chooses its number of clusters by silhouette \cite{rousseeuw1987} and is stable under subsampling (mean adjusted Rand index \cite{hubert1985} 0.97). Table~\ref{tab:typology} shows four families: (a) \textbf{serial creators}, single wallets linked to the creator of their tokens, which supply most earliest signals and rarely see their tokens reach 20 SOL; (b) \textbf{long-lived rings of early buyers} of a few to some tens of wallets that stay active for days and whose tokens reach 20 SOL much more often; (c) \textbf{short-lived buyer pairs and triples}, active for minutes to hours, with a high reach rate on few tokens; and (d) a \textbf{mixed} cluster of small groups partly linked to creators. Cluster membership was not used to select any tested rule.

\subsection{Campaigns: funding, launch, consolidation}
\begin{table}[htbp]
\centering\small
\setlength{\tabcolsep}{3pt}
\caption{Funding--launch--consolidation campaigns and their continuity across launches.}\label{tab:campaigns}
\begin{tabular}{lr}
\toprule
Statistic & Value \\
\midrule
Campaigns (funder -$>$ $\ge$3 workers for a token) & 100,643 \\
Campaigns with an observed consolidation (fan-in) & 40,427 (40.2\%) \\
Median / 99th pct. funded workers per campaign & 4 / 44 \\
Campaign-to-campaign links & 1,301,618 \\
  supported cycle continuity ($\ge$2 reused workers) & 38.2\% \\
  candidate full rotation / shared service & 61.8\% \\
Random campaign pairs sharing any worker & 0.9985\% \\
Linked pairs sharing any worker & 39.7\% \\
Worker wallets seen in exactly one campaign & 8.4\% \\
Lineages (connected supported links): count / largest & 234 / 3,091 \\
\bottomrule
\end{tabular}
\end{table}

\begin{figure}[tbp]
\centering
\includegraphics[width=\linewidth]{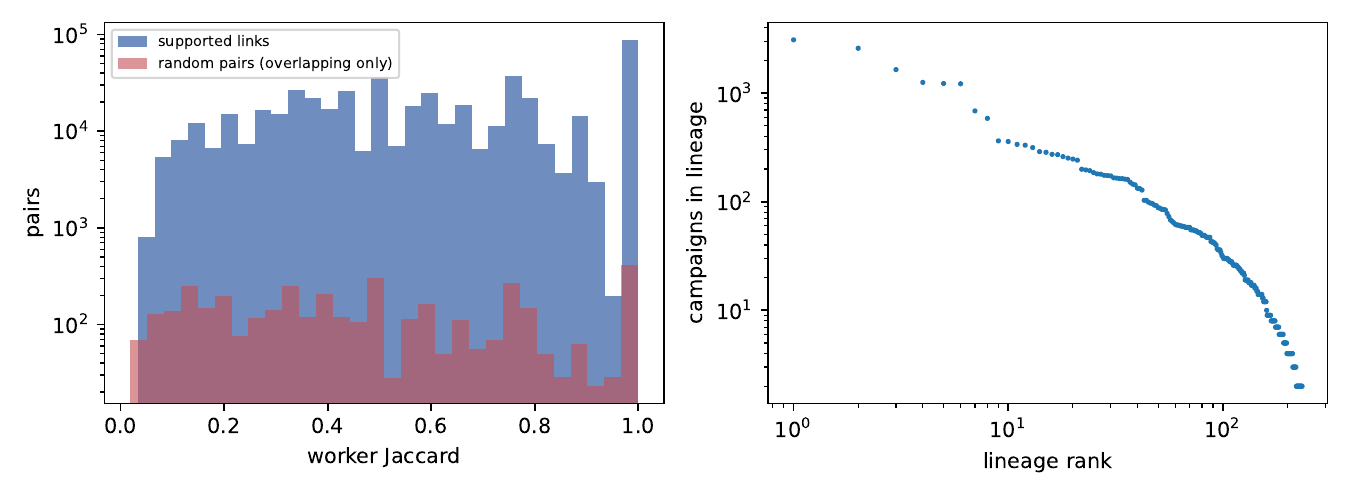}
\caption{Left: worker-wallet Jaccard overlap between consecutive campaigns for supported links and for random campaign pairs that overlap. Right: rank--size of campaign lineages (connected supported links).}\label{fig:campaigns}
\end{figure}

A campaign is a funder that supplies at least three wallets which then buy the same token. We find about 100\,643 campaigns; roughly 40\% show a fan-in to a common destination within the consolidation window. They reuse wallets heavily: a relatively small pool of worker wallets serves all campaigns, only 8\% of them appear in a single campaign and a substantial minority in five or more, whereas two random campaigns share a worker only 1.0\% of the time. (Links classified as \emph{supported continuity} require at least two reused workers by construction, so their overlap is not an independent test; the informative statistic is the small worker pool.) Supported links chain campaigns into lineages of up to 3\,091 campaigns, consistent with farms that recycle a pool of funded wallets across launches.

\subsection{Group size, creator link and timeliness}
\begin{figure}[tbp]
\centering
\includegraphics[width=\linewidth]{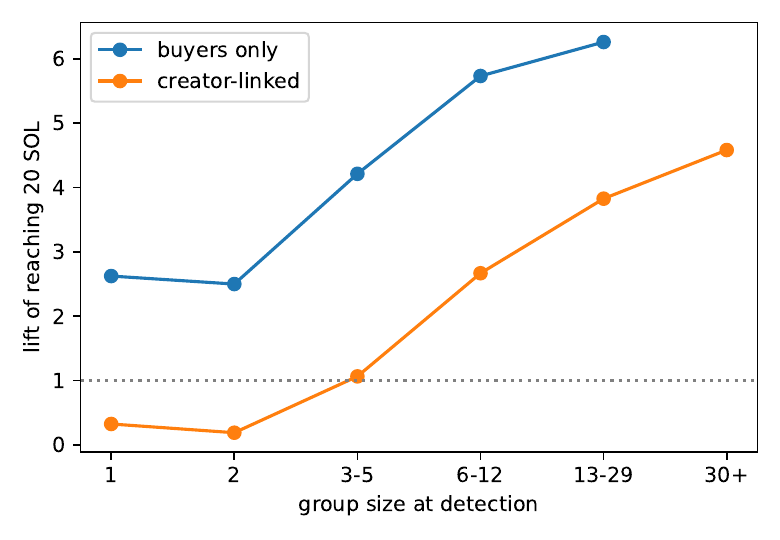}
\caption{Lift of reaching 20 SOL relative to all complete tokens, by group size at detection, separately for buyers-only and creator-linked earliest signals.}\label{fig:dose}
\end{figure}

The probability that a token reaches 20 SOL increases with the size of the group present at detection and is much higher for buyers-only signals than for creator-linked ones (Table~\ref{tab:doseday}, Figure~\ref{fig:dose}): large buyers-only groups lead to a 20-SOL curve for about 44\% of tokens against 7.7\% overall, whereas creator-linked single wallets, which supply most signals, are far \emph{below} the base rate. A logistic regression with group-clustered errors that also controls for the log initial buy and time of day gives an odds ratio per log-wallet clearly above one for group size and clearly below one for a creator link (Table~\ref{tab:logit}).

\emph{Single-wallet creator groups dominate the signals.} Most of the earliest signals are creator-linked and come from a group of a single wallet, which is also the creator's wallet. These are serial token creators rather than coordinated groups in the sense of the questions above; only 26\% of the signals come from groups of at least two wallets and 14\% from groups of at least five. All claims about coordination therefore rest on the multi-wallet subset; we report the dynamics results below separately for groups of at least 2, 3 and 5 wallets (Section 6.4) and the follower analysis for buyers-only signals as well as for all signals.

\begin{table}[htbp]
\centering\small
\setlength{\tabcolsep}{3pt}
\caption{Coverage and quality of the earliest prospective signal as a function of the allowed detection age.}\label{tab:timeliness}
\begin{tabular}{rrrrrrr}
\toprule
\shortstack{Allowed \\[0pt] age (s)} & \shortstack{Detected \\[0pt] tokens} & \shortstack{Share of \\[0pt] complete} & Reach 20 & Lift 20 & \shortstack{Buyers-only \\[0pt] signals} & \shortstack{Reach 20, \\[0pt] buyers-only} \\
\midrule
4 & 117\,831 & 76.5\,\% & 7.3\,\% & 0.95 & 14\,348 & 31.7\,\% \\
10 & 120\,332 & 78.1\,\% & 7.6\,\% & 0.98 & 16\,786 & 30.2\,\% \\
30 & 122\,490 & 79.5\,\% & 7.8\,\% & 1.01 & 18\,817 & 29.0\,\% \\
60 & 123\,752 & 80.3\,\% & 7.8\,\% & 1.02 & 19\,865 & 28.4\,\% \\
\bottomrule
\end{tabular}
\end{table}

\emph{Timeliness.} A prospective signal exists for a share of tokens within 4 s and a larger share within 60 s (Table~\ref{tab:timeliness}). Because most signals are serial creators, the lift of reaching 20 SOL among all detected tokens is close to one, i.e. no better than the base rate, while buyers-only signals keep a high reach rate.

\begin{figure}[tbp]
\centering
\includegraphics[width=\linewidth]{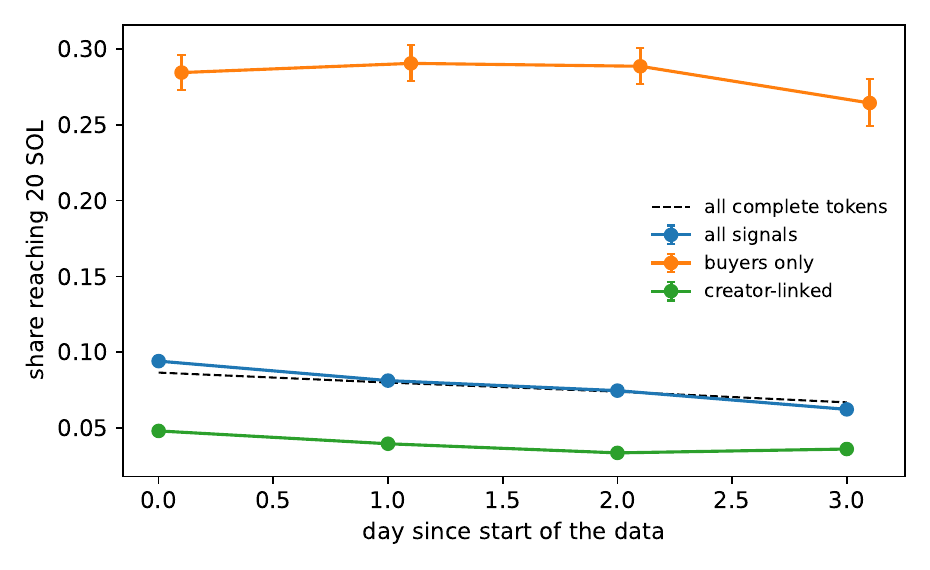}
\caption{Share of earliest-signal tokens that reach 20 SOL by day of creation, with Wilson intervals, for all signals, buyers-only and creator-linked signals; dashed line: all complete tokens of the same day.}\label{fig:daystab}
\end{figure}

\emph{Stability over the window.} The window is short, so we check whether the main contrast depends on a particular day. The lift of buyers-only signals is of the same order on each of the four observed days and in each chronological third (Table~\ref{tab:doseday}, Figure~\ref{fig:daystab}); creator-linked signals stay below the base rate on every day. The pattern is therefore stable within the window, but this is not a test across regimes: all four days belong to one market state. About 51\% of the signals come from groups first known on an earlier day than the token.

\section{Results II: price paths and cash flows around groups}
The comparisons of this section are associations. A token that attracts known group members differs from other tokens before the members arrive, and the members' own buying changes the early state of the token, so we describe what the data show, say what the adjustment removes, and state in Section 6.4 how much of the price-path result survives when the comparison is restricted to tokens that have comparable controls.

\subsection{Entry timing}
Among the tokens with at least two known members and two other buyers in the first minute, members' first buys come about 11 s after creation on average, earlier than those of the other early buyers (15 s) by about 4.3 s (token-bootstrap interval 4.2--4.4). Permuting member labels within each token gives a mean difference near zero; the member is earlier in 72\% of the tokens, and 28\% of the tokens show a one-sided permutation $p<0.05$ (5\% expected under the null). Membership was fixed before the token existed, so the effect is not produced by the way groups are detected.

\subsection{Price-path signature}
After the 60-s reference, tokens with known members are more likely to remain active, reach a higher post-reference peak and fall faster afterwards than controls matched on curve SOL and time of day (Table~\ref{tab:h5}). The probability of a peak of at least twice the reference price is 9\% for group tokens against 3\% for the matched controls, and the price one minute after the peak retains a smaller share of the peak. This is the descriptive signature of a fast rise followed by a faster fall; Section 6.4 asks how much of it remains once the controls are required to resemble the group tokens in their first minute.

\subsection{Who gains and who loses}
\begin{figure}[tbp]
\centering
\includegraphics[width=\linewidth]{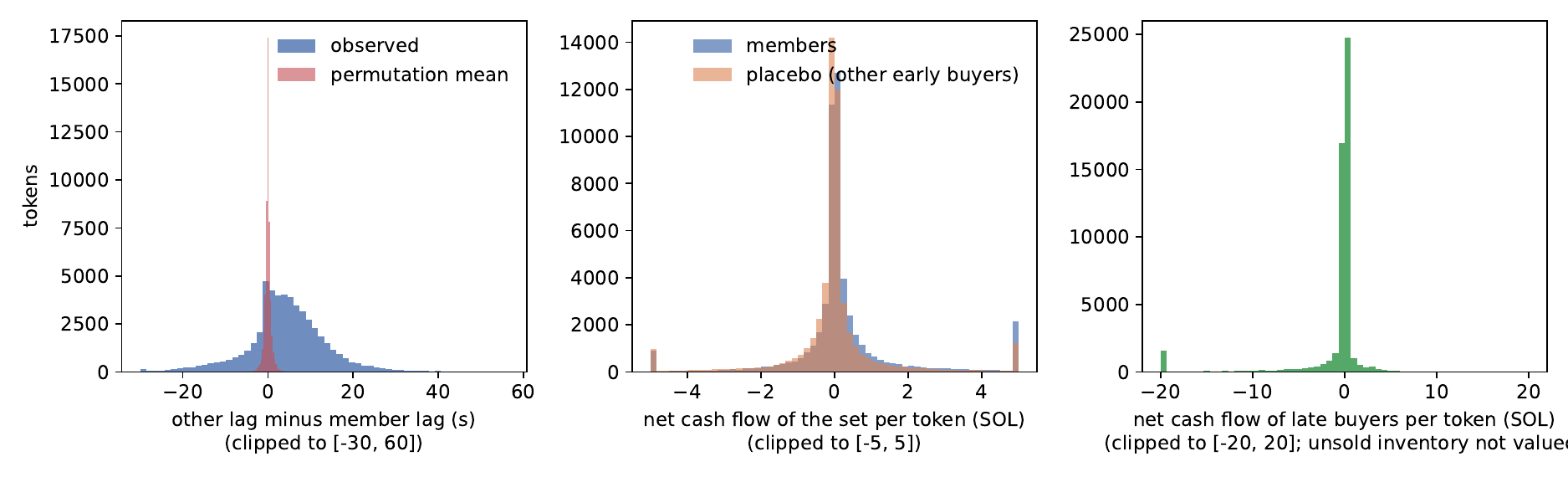}
\caption{Left: within-token difference in first-buy lag between other early buyers and members (observed) and its permutation mean. Middle: token-level cash result of the member set and of an equally sized placebo set of other early buyers (clipped to $\pm5$ SOL). Right: net cash result of late buyers (after the first minute).}\label{fig:dynamics}
\end{figure}

For group tokens the members' net cash flow (sales minus purchases) is positive on average (0.21 SOL per token), whereas an equally sized placebo set of other early buyers of the same tokens ends slightly below zero (-0.07 SOL); the paired difference is 0.28 SOL per token (interval 0.22--0.33). This simple comparison overstates the advantage: against a placebo set matched on first-buy time and buy volume (Section 6.4) it falls to 0.17 SOL per token, and this stricter figure is the one to quote. In roughly 30\% of the tokens members extract cash while the late buyers' cash flow is negative (Figure~\ref{fig:dynamics}). These are cash flows: late buyers often still hold tokens that we do not value, so the late-buyer figure is not a realised loss. The member--placebo comparison is within the same token and on the same basis, so it is unaffected by inventory valuation and by the choice of control tokens. Marking inventory at the last spot price inflates all groups (it values large unsold positions at a price they could not realise) and is reported only in the data files.

\subsection{Covariate-balanced comparison and profit measures}
The comparison of Section 6.2 stratifies controls only on curve SOL and time of day. As a stronger control we estimate, with a logistic model, the probability that a token has known members from eleven covariates measured before the 60-s reference time (curve SOL, number of events, buyers and sellers, share of the largest buyer, volatility of the early price, size of the first buy, the creator's prospectively computed history, and time-of-day terms) and weight the controls by the odds of this probability (inverse-probability weighting for the effect on the treated \cite{rosenbaum1983,stuart2010}). We repeat the analysis for groups of at least 2, 3 and 5 wallets.

\begin{figure}[tbp]
\centering
\includegraphics[width=\linewidth]{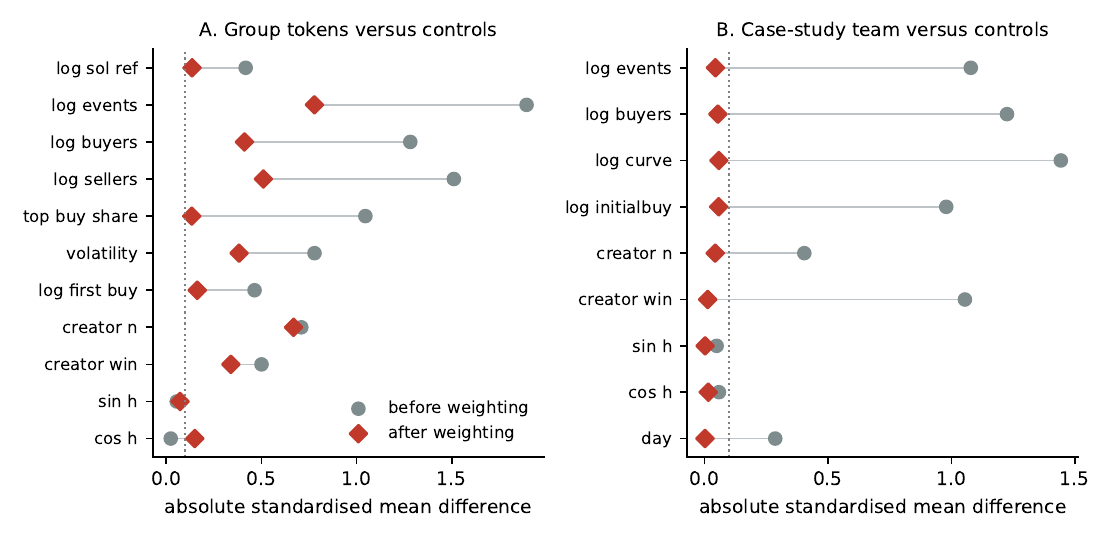}
\caption{Covariate balance before and after inverse-probability weighting (absolute standardised mean difference; dotted line: 0.1). Left: tokens with known group members against controls. Right: tokens of the case-study team against controls without any member of a large team.}\label{fig:balance}
\end{figure}

Balance improves but is far from achieved (Figure~\ref{fig:balance}): the largest absolute standardised difference falls from about 1.9 to 0.8 (a difference of 0.1 is the usual tolerance), with the number of events, the number of sellers and the creator's history remaining the most imbalanced. Tokens that attract known members are busier before the reference time than almost any control, so the overlap of the covariate distributions is only partial.

\begin{table}[htbp]
\centering\small
\setlength{\tabcolsep}{3pt}
\caption{Price-path measures after the 60-s reference time for tokens with known group members (groups of at least two wallets; larger groups in Table~\ref{tab:h5matchedall}) versus inverse-probability-weighted controls; difference with 95\% bootstrap interval over tokens (propensity model held fixed).}\label{tab:h5matched}
\begin{tabular}{p{153pt}rrr}
\toprule
Measure & Group tokens & Weighted controls & Difference [95\% CI] \\
\midrule
active after reference (share) & 85.9\,\% & 71.0\,\% & +14.9 [+12.5, +17.6] pp \\
peak multiple $\ge$ 2 (share) & 8.8\,\% & 2.8\,\% & +6.0 [+4.6, +7.3] pp \\
peak multiple $\ge$ 5 (share) & 1.7\,\% & 0.4\,\% & +1.4 [+0.9, +1.8] pp \\
peak multiple (price / reference price) & 1.46 & 1.11 & +0.35 [+0.25, +0.50] \\
time to peak (s) & 48.7 & 49.7 & $-$1.0 [$-$6.9, +4.7] \\
price 60 s after peak / peak & 0.82 & 0.92 & $-$0.10 [$-$0.11, $-$0.09] \\
price at 10 min / reference price & 0.99 & 0.96 & +0.03 [$-$0.05, +0.15] \\
\bottomrule
\end{tabular}
\end{table}

\begin{figure}[tbp]
\centering
\includegraphics[width=\linewidth]{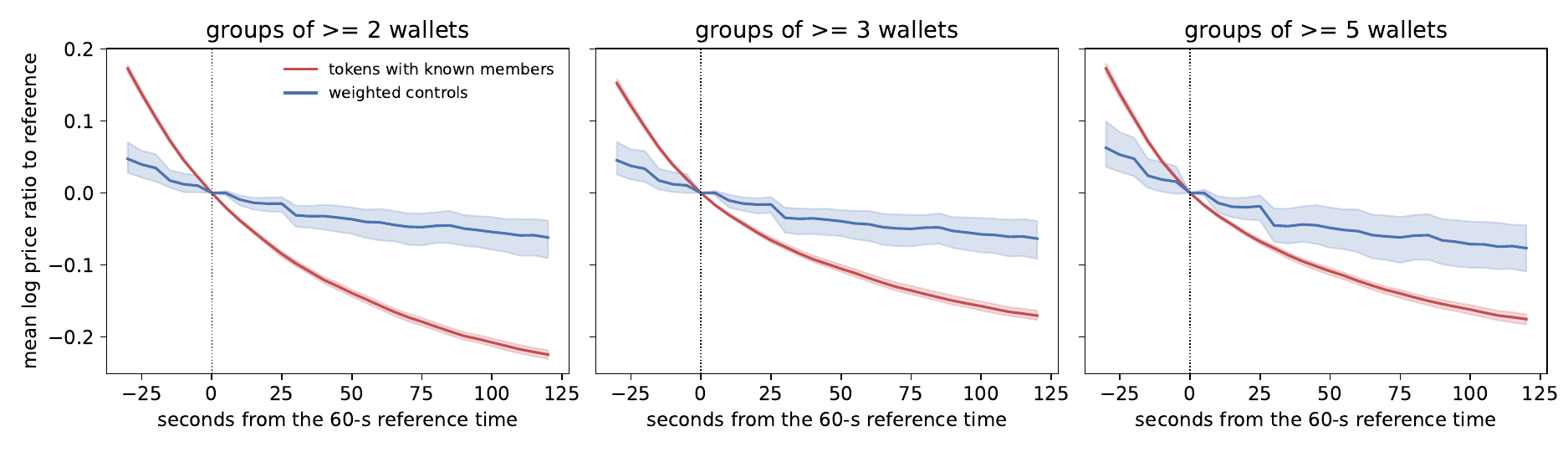}
\caption{Mean log price ratio to the reference price around the 60-s reference time for tokens with known group members (red) and inverse-probability-weighted controls (blue), for groups of at least 2, 3 and 5 wallets; shaded: 95\% bootstrap intervals over tokens.}\label{fig:eventstudy}
\end{figure}

With these weights the signature persists in the full comparison (Table~\ref{tab:h5matched}, Figure~\ref{fig:eventstudy}; larger groups in Table~\ref{tab:h5matchedall}): the share of tokens with a peak of at least twice the reference price is 8.8\% against 2.8\% for weighted controls, the price one minute after the peak retains 82\% of the peak against 92\%, and the price after ten minutes is not distinguishable from the weighted controls. The time to the peak no longer differs.

\emph{How much of it survives a stricter comparison?} Three features of this comparison need testing. Part of the covariates (events, buyers and curve SOL in the first minute) are themselves influenced by the members' buying, so adjusting for them may remove part of an effect and not only selection. The weights are extreme for many group tokens because comparable controls are rare. And the intervals resample tokens, whereas one creator launches many. Table~\ref{tab:matchedsens} therefore repeats the contrast with weights built only from what is fixed at creation, with weights that add the state of the curve before two seconds, with and without trimming to the region where the two groups overlap, and with intervals that resample creators.

\begin{table}[htbp]
\centering\scriptsize
\setlength{\tabcolsep}{3pt}
\caption{Sensitivity of the matched price-path contrast (tokens with at least two known group members against controls without any member). Peak: probability that the price reaches at least twice the reference price after the first minute; retention: price one minute after the peak relative to the peak. Covariates fixed at creation: creator's initial buy, creator history, time of day; the third set adds the state of the curve strictly before 2 s. Trimmed: only tokens with propensity inside the support of the controls. Intervals resample creators; intervals that resample tokens differ by less than one percentage point. The E-value is the minimum risk ratio that an unmeasured confounder would need with both membership and outcome to explain the point estimate away.}\label{tab:matchedsens}
\begin{tabular}{p{77pt}rrrrrrr}
\toprule
Covariates in the weights & \shortstack{Group \\[0pt] tokens} & \shortstack{Effective \\[0pt] controls} & \shortstack{Peak $\ge$ 2x: \\[0pt] group} & \shortstack{weighted \\[0pt] control} & \shortstack{Difference, pp \\[0pt] [CI over creators]} & E-value & \shortstack{Retention after peak, \\[0pt] diff. pp [CI]} \\
\midrule
original eleven covariates & 51\,912 & 1\,106 & 8.8\,\% & 2.8\,\% & +6.0 [+4.6, +7.4] & 5.7 & $-$9.8 [$-$11.1, $-$8.2] \\
original eleven covariates, trimmed & 18\,521 & 1\,466 & 2.7\,\% & 2.1\,\% & +0.7 [$-$0.3, +1.6] & 2.0 & $-$0.4 [$-$1.2, +0.7] \\
creation covariates only & 51\,912 & 3\,013 & 8.8\,\% & 1.3\,\% & +7.6 [+6.8, +8.2] & 13.6 & $-$13.4 [$-$13.8, $-$12.9] \\
creation covariates only, trimmed & 40\,865 & 3\,591 & 9.5\,\% & 1.2\,\% & +8.3 [+7.7, +8.9] & 15.8 & $-$14.7 [$-$15.1, $-$14.3] \\
creation covariates plus curve state before 2 s & 51\,912 & 2\,486 & 8.8\,\% & 1.2\,\% & +7.7 [+6.9, +8.3] & 14.6 & $-$13.8 [$-$14.3, $-$13.3] \\
creation covariates plus curve state before 2 s, trimmed & 37\,516 & 2\,789 & 9.2\,\% & 1.2\,\% & +8.0 [+7.4, +8.6] & 15.4 & $-$14.5 [$-$14.9, $-$14.0] \\
\bottomrule
\end{tabular}
\end{table}

The answer has two parts. When only creation-time covariates (the creator's first buy and history, time of day) are used, the contrast is large and stable: the probability of a doubling is several times higher for group tokens, the fall after the peak is steeper, the creator-level intervals are as narrow as the token-level ones, and the E-value is large, so an unmeasured confounder would have to be associated with both membership and the doubling by risk ratios above ten to remove it. But adding the state of the first minute and restricting the comparison to the region where group tokens and controls overlap removes the difference: in that region the probability of a doubling is similar and the interval includes zero, and so does the difference in retention after the peak. The honest reading is that tokens with known members have pump-like price paths, that this holds when they are compared with controls that look alike at creation, but that the data cannot show that the members add anything beyond the early activity that makes the token busy in the first place: where the busy tokens have no comparable controls, the contrast rests on extrapolation by the weights. The estimates of Table~\ref{tab:h5matched} should therefore be read as conditional associations of the whole group-token population, not as effects of the members' presence.

\emph{Profit measures.} Cash flow understates what a holder keeps and overstates what a seller realises, so we report three bases for the members, for a placebo set of the same size matched on first-buy time and buy volume, and for late buyers (Table~\ref{tab:profit}): (i) cash flow (sales minus purchases); (ii) the realised result, sales minus the average cost of the tokens sold; and (iii) a liquidation basis, cash flow plus the value of the remaining inventory if each wallet alone sold it into the curve at the last observed reserves, net of the event fee and capped by the SOL in the curve. The third basis is an optimistic bound because other holders would sell at the same time.

\begin{table}[htbp]
\centering\footnotesize
\setlength{\tabcolsep}{3pt}
\caption{Mean per-token result (SOL) of known group members, of a placebo set of the same size matched on first-buy time and buy volume, and of late buyers, under three valuation bases; difference members minus placebo with 95\% bootstrap interval over tokens. Liquidation basis values remaining inventory as if each wallet alone sold it into the curve at the last reserves (an optimistic bound).}\label{tab:profit}
\begin{tabular}{lrrrrrr}
\toprule
\multicolumn{7}{l}{\emph{Tokens with at least 2 known members}} \\
\midrule
Measure & Tokens & Members & Placebo & Difference [95\% CI] & Late buyers & \shortstack{Positive, \% \\[0pt] members / placebo} \\
\midrule
cash flow & 51\,912 & 0.21 & 0.04 & +0.17 [+0.12, +0.23] & $-$2.34 & 57\,/\,44 \\
realised result & 51\,912 & 1.00 & 0.62 & +0.38 [+0.33, +0.43] & 0.18 & 65\,/\,50 \\
liquidation basis & 51\,912 & 0.23 & 0.07 & +0.16 [+0.10, +0.22] & $-$2.16 & 57\,/\,45 \\
\bottomrule
\end{tabular}
\\[0.8ex]
\begin{tabular}{lrrrrrr}
\toprule
\multicolumn{7}{l}{\emph{Tokens with at least 3 known members}} \\
\midrule
Measure & Tokens & Members & Placebo & Difference [95\% CI] & Late buyers & \shortstack{Positive, \% \\[0pt] members / placebo} \\
\midrule
cash flow & 42\,551 & 0.22 & 0.04 & +0.18 [+0.13, +0.25] & $-$2.83 & 55\,/\,46 \\
realised result & 42\,551 & 0.91 & 0.61 & +0.30 [+0.26, +0.35] & 0.20 & 57\,/\,53 \\
liquidation basis & 42\,551 & 0.24 & 0.07 & +0.18 [+0.12, +0.24] & $-$2.62 & 55\,/\,46 \\
\bottomrule
\end{tabular}
\\[0.8ex]
\begin{tabular}{lrrrrrr}
\toprule
\multicolumn{7}{l}{\emph{Tokens with at least 5 known members}} \\
\midrule
Measure & Tokens & Members & Placebo & Difference [95\% CI] & Late buyers & \shortstack{Positive, \% \\[0pt] members / placebo} \\
\midrule
cash flow & 33\,816 & 0.33 & 0.02 & +0.30 [+0.24, +0.37] & $-$3.39 & 56\,/\,43 \\
realised result & 33\,816 & 0.95 & 0.51 & +0.44 [+0.40, +0.49] & 0.21 & 59\,/\,49 \\
liquidation basis & 33\,816 & 0.35 & 0.05 & +0.30 [+0.23, +0.36] & $-$3.15 & 56\,/\,44 \\
\bottomrule
\end{tabular}
\end{table}

On all three bases members exceed the placebo set of the same tokens by a small positive amount per token, and larger groups show larger gaps; the late buyers' cash flow is negative on average, also on the liquidation basis, i.e. they do not recover their cash even if each liquidated the remaining inventory alone. Because the comparison is between wallets of the same token it does not depend on the choice of control tokens, so it is the more reliable of the two kinds of evidence in this section. The per-token amounts are small compared with the amounts bought by the late buyers collectively, so most late losses go to parties other than the identified members (other early buyers, the creator, and fees).

\subsection{Graduation}
Graduation is rare (2.8\% of complete tokens; Table~\ref{tab:grad}). It is highest for buyers-only groups (5.6\% for groups of 3--12 wallets and 7.1\% for 13 or more), low for creator-linked groups, and, importantly, as high for tokens with no group signal at all (6.1\%) as for medium-sized buyers-only groups. Presence of a group is therefore not a marker of success compared with organically trading tokens; it is the type of group that matters.

\section{Results III: detectability is not exploitability}
\subsection{A follower at the earliest signal}\label{sec:followers}
\begin{table}[htbp]
\centering\scriptsize
\setlength{\tabcolsep}{3pt}
\caption{Out-of-sample (last 30\% of signals in time) mean net return per trade of entering at the earliest group signal; policy selected on the first 70\%. Interval: bootstrap over groups.}\label{tab:followers}
\begin{tabular}{llrrrr}
\toprule
Signals & Cost & \shortstack{Test \\[0pt] signals} & \shortstack{Net return per trade \\[0pt] latency 1 s [95\% CI]} & \shortstack{Net return per trade \\[0pt] latency 3 s [95\% CI]} & \shortstack{Winning \\[0pt] trades} \\
\midrule
all signals & project & 31\,836 & $-$8.9 [$-$9.5, $-$8.4] pp & $-$10.5 [$-$11.4, $-$9.6] pp & 24\,\% \\
all signals & event & 31\,836 & $-$4.5 [$-$5.1, $-$4.0] pp & $-$6.3 [$-$7.2, $-$5.4] pp & 25\,\% \\
buyers only & project & 4\,867 & $-$7.8 [$-$8.8, $-$6.8] pp & $-$8.0 [$-$9.1, $-$6.8] pp & 27\,\% \\
buyers only & event & 4\,867 & $-$4.0 [$-$5.1, $-$3.0] pp & $-$4.2 [$-$5.3, $-$3.0] pp & 29\,\% \\
buyers only, $\ge$6 members & project & 2\,097 & $-$8.5 [$-$10.1, $-$6.9] pp & $-$8.3 [$-$10.3, $-$6.8] pp & 33\,\% \\
buyers only, $\ge$6 members & event & 2\,097 & $-$4.7 [$-$6.4, $-$3.0] pp & $-$4.6 [$-$6.6, $-$3.0] pp & 35\,\% \\
buyers only, $\ge$13 members & project & 1\,299 & $-$8.6 [$-$10.7, $-$6.9] pp & $-$8.3 [$-$10.6, $-$6.5] pp & 34\,\% \\
buyers only, $\ge$13 members & event & 1\,299 & $-$4.9 [$-$7.0, $-$3.1] pp & $-$4.6 [$-$6.9, $-$2.7] pp & 35\,\% \\
\bottomrule
\end{tabular}
\\[0.4ex]{\footnotesize Policy of all rows: take profit at 1.3 times the entry price, stop at 0.7 times, maximum holding 60 s.}
\end{table}

We simulate a follower that enters at the earliest prospective signal under both cost conventions and four latencies, with the exit rule chosen on the first 70\% of signals. In all 32 configurations the out-of-sample mean net return per trade is negative, between -7\% and -4\% under event-level costs and between -11\% and -8\% under the conservative convention, and the group-bootstrap interval lies entirely below zero in 32 of them (Table~\ref{tab:followers} shows latencies of 1 and 3 s). Restricting to large buyers-only groups raises the win rate but not the mean.

\emph{No better than random entry.} Entering random tokens in the test period with the same policies yields similar losses (project costs: -10\% to -8\%; event costs: -6\% to -4\%). For the same policy and a latency of 1 s the buyers-only follower earns -7.8\% against -7.9\% for random entry under project costs, and -4.0\% against -3.6\% under event costs. By the time a group is detectable its buying is already in the price.

\emph{Latency, on the same trades.} Because the exit policy is re-selected at each latency and trades without a resolvable exit drop out separately, the rows of Table~\ref{tab:followers} at different latencies refer to slightly different trades. We therefore fixed the policy chosen at 1 s and compared the latencies on exactly the same test trades (more than 99\% of the signals resolve at every latency). Every additional second costs about one percentage point of return per trade for all signals (a latency of 5 s instead of 1 s lowers the mean by about 2.6 points under project costs and 2.9 under event costs, intervals clear of zero), and less for buyers-only signals (about one point at 5 s, indistinguishable from zero at 2 and 3 s). The mean return is negative at every latency, including 1 s, so the loss is not an artefact of a slow follower; a follower with latency below the one-second resolution is not represented (Section 3).

\subsection{Validation of the execution model and cost decomposition}
The simulator is exact arithmetic on recorded reserves, so we check it against what happened. Applying the constant-product formula to the post-state of the preceding event of the same token reproduces the token amount of nearly all observed buys (99.6\%) and the SOL amount of nearly all observed sells (98.1\%) to within $10^{-4}$ relative error, with a median relative error far below that (Table~\ref{tab:quotecheck}). The remaining discrepancies are larger for sells; we did not investigate their cause further. What the check cannot validate is behaviour that depends on the trader's own transaction: confirmation delay, priority fees, bundle inclusion and the response of other traders to the trader's order. We did not send real transactions; latency is modelled as a delay in seconds, and real execution may be better or worse.

\begin{table}[htbp]
\centering\small
\setlength{\tabcolsep}{3pt}
\caption{Decomposition of execution costs for a fixed momentum rule (enter after the curve reaches 10 SOL within 120 s of creation, take profit at 1.5x, stop at -20 percent, maximum hold 60 s; 15,000 randomly sampled tokens). Costs are added one at a time; mean return on the position size with 95 percent bootstrap interval over positions. Results hold under these assumptions only and are not a statement about real execution.}\label{tab:costdecomp}
\begin{tabular}{p{171pt}rrrr}
\toprule
Scenario & Positions & Mean return [95\% CI] & Median & Win share \\
\midrule
no fees, no slippage, no latency, 0.05 SOL & 2\,014 & +0.054 [+0.036, +0.073] & $-$0.224 & 41\,\% \\
+ observed protocol fee (floor 0) & 2\,014 & +0.054 [+0.035, +0.070] & $-$0.225 & 41\,\% \\
+ fee floor 2 percent & 2\,014 & +0.037 [+0.019, +0.053] & $-$0.238 & 41\,\% \\
+ 1.5 percent slippage per side & 2\,014 & +0.006 [$-$0.009, +0.022] & $-$0.261 & 40\,\% \\
+ latency 1 s (default) & 1\,975 & $-$0.099 [$-$0.114, $-$0.083] & $-$0.263 & 29\,\% \\
latency 3 s & 1\,973 & $-$0.115 [$-$0.134, $-$0.099] & $-$0.225 & 27\,\% \\
latency 5 s & 1\,973 & $-$0.117 [$-$0.138, $-$0.100] & $-$0.192 & 25\,\% \\
size 0.01 SOL (default costs) & 1\,975 & $-$0.101 [$-$0.117, $-$0.085] & $-$0.265 & 29\,\% \\
size 0.2 SOL (default costs) & 1\,975 & $-$0.099 [$-$0.117, $-$0.085] & $-$0.263 & 29\,\% \\
size 0.5 SOL (default costs) & 1\,975 & $-$0.099 [$-$0.118, $-$0.081] & $-$0.262 & 29\,\% \\
\bottomrule
\end{tabular}
\end{table}

To see which component removes the gross edge, a fixed momentum rule (entry when the curve reaches 10 SOL within 120 s, take profit at 1.5x, stop at -20\%, 60-s maximum hold) is simulated on 15,000 random tokens with costs added one at a time (Table~\ref{tab:costdecomp}). Without any cost the rule earns 5.4\% per trade; the 2\% fee floor takes it to 3.7\%, slippage to 0.6\%, and a latency of one second to -9.9\%. Position sizes between 0.01 and 0.5 SOL change the result by less than one percentage point because these sizes are small relative to the curve. The latency step is the largest, as expected in a market whose price moves within the first seconds. Our conclusion is accordingly stated conditionally: \emph{under these cost and latency assumptions no strategy we tested earns a positive out-of-sample return}; we do not claim that no strategy can profit, in particular with colocated infrastructure or bundle inclusion that our time resolution cannot represent.

\subsection{Learned rankings of entry moments}
We built 369\,036 decision points (tokens at ages from 4 to 240 s) with order-flow, price-path and prospectively computed wallet- and creator-reputation features, and trained gradient-boosted regressors \cite{friedman2001,ke2017} of the net return of a 20- or 60-s hold on the first 55\% of decisions in time. On the last 30\%, the top-ranked decisions earn negative returns under both cost conventions (between -9\% and -4\% under project costs), the best validation cell falls from 2.3\% to -2.9\% out of sample, and no cell has an interval above zero (5 of 24 intervals include zero; Table~\ref{tab:ml}). Adding prospectively computed wallet and creator reputation changes the out-of-sample mean of the top-ranked decisions by a few percentage points without turning any interval positive: reputation carries some information, but it is already priced.

\begin{table}[htbp]
\centering\footnotesize
\setlength{\tabcolsep}{3pt}
\caption{Entry-ranking model with a token-level purged chronological split (train/validation/test = 55/15/30 percent of tokens by creation time; 600-s embargo before each boundary). Mean net return per decision (60-s hold, project cost model) of the top 5 percent of test decisions by predicted return, for feature subsets and for test decisions without any proven wallet; intervals by bootstrap over tokens. Baseline: mean return of all decisions in the subset.}\label{tab:mlv2}
\begin{tabular}{llrrr}
\toprule
Features & Test subset & Decisions & Test mean [95\% CI] & Baseline \\
\midrule
flow & all test decisions & 4\,690 & $-$0.058 [$-$0.071, $-$0.044] & $-$0.133 \\
flow & no proven wallet & 44 & $-$0.091 [$-$0.165, +0.027] & $-$0.094 \\
reputation only & all test decisions & 6\,670 & $-$0.070 [$-$0.079, $-$0.061] & $-$0.133 \\
reputation only & no proven wallet & 1\,431 & $-$0.071 [$-$0.073, $-$0.069] & $-$0.094 \\
flow+reputation & all test decisions & 4\,424 & $-$0.053 [$-$0.073, $-$0.034] & $-$0.133 \\
flow+reputation & no proven wallet & 70 & $-$0.134 [$-$0.207, $-$0.060] & $-$0.094 \\
flow without hour and age & all test decisions & 4\,602 & $-$0.057 [$-$0.068, $-$0.044] & $-$0.133 \\
flow without hour and age & no proven wallet & 37 & $-$0.063 [$-$0.160, +0.072] & $-$0.094 \\
\bottomrule
\end{tabular}
\end{table}

\emph{Stricter protocol.} In the first protocol the split is chronological by decision, so decisions of a token that straddles a boundary can fall into different parts. We therefore repeated the experiment splitting by token: all decisions of a token belong to one part, the parts are ordered by token creation time (55/15/30\%), and tokens created shortly before a boundary are removed from the earlier part so that no training outcome window overlaps the test period (Table~\ref{tab:mlv2}). The top 5\% of test decisions by predicted return still lose money (interval -7\% to -3\%), better than the average decision but below zero. For test decisions without any proven wallet in the early buyers the reputation features carry no information; the ranking there is based on the order flow only and is not better than the unconditional mean. Feature importance (gain share; Figure~\ref{fig:mlimportance}) is led by the mean reputation of early buyers, the curve SOL and the short-term price change; hyperparameters are listed in Table~\ref{tab:mlhp}.

\subsection{How the same data can show a profit}
\begin{table}[htbp]
\centering\small
\setlength{\tabcolsep}{3pt}
\caption{The project's time-consistent portfolio evaluator (1 SOL initial capital, 5\% of capital per trade) under three execution assumptions. Unresolved: share of training exits without an executable quote.}\label{tab:official}
\begin{tabular}{p{115pt}rrrrrrr}
\toprule
Execution model & \shortstack{Training \\[0pt] rows} & \shortstack{Unresolved \\[0pt] share} & Positions & Closed & \shortstack{Capital \\[0pt] at cost} & \shortstack{Mean \\[0pt] return} & \shortstack{Max. \\[0pt] drawdown} \\
\midrule
observed quotes (default) & 755\,468 & 88\,\% & 43 & 33 & 1.198 & 8.2\,\% & 9.7\,\% \\
reserve model, no state-age limit & 500\,676 & 60\,\% & 40 & 30 & 1.121 & 4.5\,\% & 8.7\,\% \\
reserve model + own position & 500\,676 & 3\,\% & 332 & 327 & 0.504 & -4.9\,\% & 54.9\,\% \\
\bottomrule
\end{tabular}
\end{table}

Table~\ref{tab:official} runs a time-consistent portfolio evaluator on the same signals with 1 SOL and 5\% of capital per trade. With the default rule that an exit needs a later observed quote, 88\% of training exits are unresolved and the portfolio shows a small profit on 43 trades. A reserve-based exit without the trader's own position still leaves 60\% unresolved, because the exit state omits the SOL the trader deposited. Adding the own position leaves only 2.8\% of training exits unresolved, increases the number of trades from 40 to 332, and turns the result to a loss (capital 0.50 SOL from 1 SOL). A second pitfall arose in our own exploration: a label column and a feature column shared a name, producing a spurious +300\% mean return with a 100\% win rate in a top-ranked slice; automated checks that features and labels are disjoint and uncorrelated are part of the replication package.

\section{Results IV: large teams of fast wallets}
The groups of Sections 5--7 are mostly small: the median group has two wallets and few of the earliest signals come from groups of five or more. The coordinated actors that motivate the study are different: pools of dozens or hundreds of wallets that buy in the first seconds of many unrelated launches. This section asks whether the pipeline finds them, how they behave, and whether their success is selection or effect. Because this part grew while we examined the same window, Sections 8.1--8.9 are exploratory; Section 8.10 evaluates the resulting rules once on a block of the window that did not take part in their selection, checks when the signals arrive and asks whether they carry information beyond the size of the early crowd.

\begin{figure}[tbp]
\centering
\includegraphics[width=\linewidth]{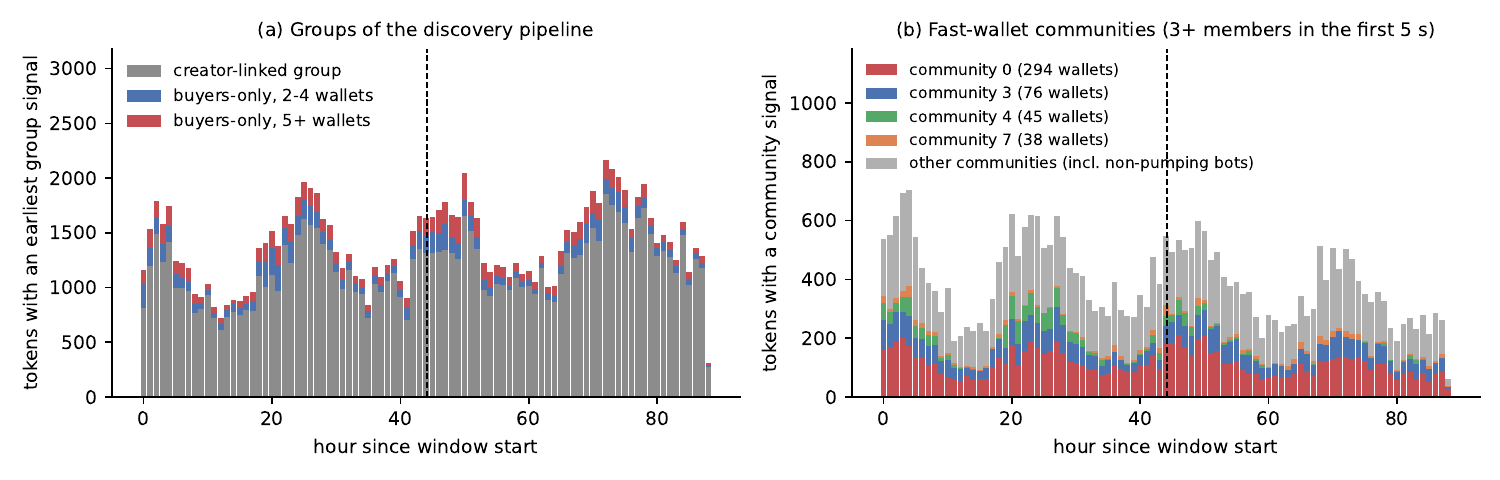}
\caption{Activity of groups and teams over the window. (a) Tokens per hour whose earliest group signal comes from the discovery pipeline, by type of signal. (b) Tokens per hour in which at least three wallets of one fast-wallet community are among the buyers of the first five seconds, for the four communities studied below and all others pooled (the others include bots that do not pump). Dashed line: boundary of the two halves.}\label{fig:activity}
\end{figure}

The figure shows the scale of the phenomenon: creator-linked groups appear in a large fraction of all launches at every hour, while the buyers-only groups and the fast-wallet communities account for a few hundred tokens per hour, with the same daily rhythm as the launch rate and with no sign of a burst or of a disappearance in the window.

\subsection{Fragmentation of groups and a team layer}
In the group builder a group is seeded by the first accepted edge, so its core never exceeds two wallets; a wallet joins only if it has accepted edges to both core wallets, and groups are never merged (one bridge must not union two populations). A team that rotates wallets, or simply uses many wallets that never co-buy repeatedly with the same two, is therefore split into many overlapping small groups: the groups hold tens of thousands of distinct wallets, the median group has two members and about 30\% of the wallets belong to more than one group. Merging in the other direction is not harmless either: the connected components of the accepted-edge graph contain one component of about 19,000 wallets.

We therefore evaluate a \emph{team layer} above the groups, built only from evidence qualified before the middle of the window and evaluated on tokens created after it (signal: at least three members among the first-minute buyers). Merging groups that share wallets is the best of the variants tried (Table~\ref{tab:teamlayer}): for teams of at least 100 wallets it raises the number of signal tokens several-fold at a reach-20 share of 58\% (lift 6.6), and the same holds for teams of at least 30 wallets. Louvain communities of the accepted-edge graph cover more tokens but dilute the signal and produce teams of more than 500 wallets. Adding funder links changes little because few workers are funded repeatedly by one funder. The merge threshold on shared wallets matters, the overlap threshold does not, and the signal is stable over the remaining days. The thresholds were tuned on the evaluation half, so the absolute lifts are optimistic; Section 8.10 gives the figures for an untouched block.

\begin{table}[htbp]
\centering\small
\setlength{\tabcolsep}{3pt}
\caption{Team layer prototype, evaluated on tokens created after the middle of the window with teams built from earlier evidence: signal = at least three members among first-minute buyers. Teams = number of teams of at least the given size; wallets and largest team in wallets; Reach 20 = share of signal tokens reaching 20 SOL; lifts against all complete second-half tokens (base rate of reaching 20 SOL: 0.088).}\label{tab:teamlayer}
\begin{tabular}{lrrrrrrrr}
\toprule
Method & \shortstack{Min. \\[0pt] size} & Teams & Wallets & Largest & \shortstack{Signal \\[0pt] tokens} & Reach 20 & Lift 20 & Lift 60 \\
\midrule
Louvain communities & 30 & 122 & 14\,687 & 565 & 10\,914 & 34.8\,\% & 3.95 & 2.62 \\
current groups & 30 & 161 & 7\,588 & 103 & 979 & 46.7\,\% & 5.30 & 3.26 \\
groups merged by overlap & 30 & 63 & 3\,560 & 174 & 4\,643 & 54.7\,\% & 6.21 & 4.45 \\
Louvain communities & 100 & 43 & 10\,595 & 565 & 8\,179 & 44.2\,\% & 5.01 & 3.41 \\
current groups & 100 & 1 & 103 & 103 & 92 & 90.2\,\% & 10.24 & 13.80 \\
groups merged by overlap & 100 & 8 & 1\,038 & 174 & 3\,937 & 58.3\,\% & 6.61 & 4.99 \\
\bottomrule
\end{tabular}
\end{table}

\subsection{An independent search for fast-wallet teams}
To avoid relying on the group builder we searched directly for pools of fast buyers. About 1\,553 wallets bought within 5 s of creation in many complete tokens; pairs that co-occur far more often than their activity implies were linked and clustered with Louvain, giving 26 communities of at least five wallets, the largest with several hundred (Table~\ref{tab:sniper}). Fast buying alone does not mean a successful team: one community of 133 wallets operates on only 217 creators and rarely sees its tokens reach 20 SOL. The communities that combine many creators, a high tempo and a high reach-20 share are the candidates studied below. Only part of their wallets lies in a single merged pipeline team, so the pipeline sees them only in part.

\begin{table}[htbp]
\centering\small
\setlength{\tabcolsep}{3pt}
\caption{Communities of wallets that repeatedly buy within 5 s of token creation (Louvain on co-occurrence, edges with lift of at least 3): wallets, tokens with at least three members among the first buyers, distinct creators, tokens per hour, reach-20 and reach-60 shares, share of the community's wallets that belong to groups of the discovery pipeline and share in its largest merged team.}\label{tab:sniper}
\begin{tabular}{rrrrrrrrr}
\toprule
Comm. & Wallets & Tokens & Creators & \shortstack{Tokens \\[0pt] per hour} & Reach 20 & Reach 60 & \shortstack{In pipeline \\[0pt] teams} & \shortstack{Largest pipe- \\[0pt] line team} \\
\midrule
0 & 294 & 10\,099 & 5\,580 & 113.1 & 46.7\,\% & 10.8\,\% & 51\,\% & 11\,\% \\
3 & 76 & 5\,217 & 2\,973 & 58.5 & 39.1\,\% & 8.1\,\% & 83\,\% & 18\,\% \\
2 & 85 & 4\,405 & 2\,454 & 49.4 & 21.8\,\% & 3.6\,\% & 71\,\% & 15\,\% \\
1 & 133 & 3\,445 & 217 & 38.7 & 0.8\,\% & 0.0\,\% & 86\,\% & 2\,\% \\
5 & 44 & 2\,119 & 1\,048 & 23.7 & 16.6\,\% & 2.2\,\% & 86\,\% & 9\,\% \\
4 & 45 & 1\,656 & 1\,145 & 19.4 & 57.8\,\% & 14.1\,\% & 82\,\% & 31\,\% \\
6 & 41 & 989 & 122 & 11.1 & 33.9\,\% & 5.7\,\% & 98\,\% & 15\,\% \\
22 & 7 & 946 & 929 & 10.6 & 6.3\,\% & 1.6\,\% & 86\,\% & 57\,\% \\
9 & 29 & 914 & 9 & 10.8 & 34.5\,\% & 5.4\,\% & 90\,\% & 14\,\% \\
7 & 38 & 886 & 567 & 9.9 & 67.5\,\% & 20.2\,\% & 95\,\% & 58\,\% \\
8 & 35 & 785 & 373 & 8.9 & 45.0\,\% & 9.6\,\% & 74\,\% & 14\,\% \\
13 & 18 & 656 & 432 & 7.4 & 19.1\,\% & 4.0\,\% & 83\,\% & 44\,\% \\
\bottomrule
\end{tabular}
\end{table}

\subsection{The largest team: 126 wallets}
The merged team with the most signals consists of 126 wallets and supplies most of the signal tokens of all merged teams of at least 30 wallets. Its tokens were created by 4\,496 different creators; one of the team's wallets was the creator in a negligible share of its tokens. It acts at a rate of about 95 tokens per hour in every hour of the window, about 5\% of all launches. The first member buys a median 2 s after creation and the third after 4 s, with a median of 6 members among the first-minute buyers. The pool is stable: all wallets are active and the median wallet takes part in 316 tokens.

Its tokens reach 20 SOL in 53\% of cases against 9.0\% for all complete tokens, reach 60 SOL in 13\% against 2.8\% and graduate in 8\% against 3.0\%, in every 12-hour block of the window. The members buy and sell several thousand SOL in total, with a cash flow of 1,629 SOL (0.19 SOL per token), positive in 44\% of tokens. The result is concentrated: the ten best wallets account for 77\% of the positive total.

\emph{Selection or effect.} The team's tokens differ from other tokens before the team enters: they already hold more SOL in the curve, their creators bought more at creation and have a better record (Figure~\ref{fig:balance}). We compare them with tokens without any member of a large team among the first-minute buyers, using covariates measured strictly before the team's entry age and three estimators (Table~\ref{tab:casecontrol}). The raw gap in the reach-20 share (53\% against 3.3\%) falls to a difference of 32\% percentage points with inverse-probability weights (interval 31\%--34\%), a similar value with nearest-neighbour matching and 32\% within the 1\,342 creators who have both team and other tokens. Selection on observables therefore explains roughly a third of the gap, and the rest is not explained by anything we can observe before the entry. As a measure of how strong a hidden confounder would have to be, the risk ratio of the weighted estimate (about 2.5) has an E-value of about 4.5, and that of the within-creator contrast (about 3.5) an E-value of about 6.5: only a confounder associated with team presence and with reaching 20 SOL by risk ratios above four could remove the weighted association. The excess is concentrated at the level of 20 SOL: the reach-60 share is higher by 4.5\% percentage points (a positive and clearly non-zero difference, but a small one in absolute terms; E-value about 2.5) and the graduation share by 0.6\% percentage points, which is within noise for the weighted estimate. It is not mechanical either: the team's net capital in the curve at the peak is a small share of the peak, and 51\% of its tokens would still reach 20 SOL if the team's net inflow were removed (ignoring other buyers' reactions). The pattern fits a strategy of pushing a token that already looks alive to about 20 SOL and leaving before it goes further; unobserved information (off-chain attention, creator bundles) that the team could use to choose tokens cannot be excluded, and Section 8.10 shows that a large part of the signals arrive when the curve is already at 20 SOL.

\begin{table}[htbp]
\centering\small
\setlength{\tabcolsep}{3pt}
\caption{Case-study team: share of tokens reaching 20 SOL, reaching 60 SOL and graduating versus controls with no member of any large team among the first-minute buyers. IPW: inverse-probability weighting on pre-entry covariates; nearest neighbour: 1:5 matching with caliper; within creator: creators that have both team and other tokens. Differences with 95\% bootstrap intervals.}\label{tab:casecontrol}
\begin{tabular}{llrrrr}
\toprule
Method & Outcome & Team & Control & Difference, pp [95\% CI] & \shortstack{Matched \\[0pt] pairs} \\
\midrule
IPW (ATT) & reach20 & 53.0\,\% & 20.9\,\% & +32.2 [+30.6, +33.6] &  \\
1:5 nearest neighbour & reach20 & 52.8\,\% & 21.1\,\% & +31.7 [+30.5, +32.7] & 8\,337 \\
IPW (ATT) & reach60 & 12.6\,\% & 8.1\,\% & +4.5 [+3.5, +5.4] &  \\
1:5 nearest neighbour & reach60 & 12.5\,\% & 7.6\,\% & +4.9 [+4.3, +5.7] & 8\,337 \\
IPW (ATT) & grad & 7.9\,\% & 7.3\,\% & +0.6 [$-$0.0, +1.3] &  \\
1:5 nearest neighbour & grad & 7.9\,\% & 7.0\,\% & +0.9 [+0.3, +1.5] & 8\,337 \\
within creator & reach20 & 44.7\,\% & 12.6\,\% & +32.1 [+29.5, +34.7] & 1\,342 \\
within creator & reach60 & 8.7\,\% & 2.6\,\% & +6.0 [+4.5, +7.6] & 1\,342 \\
within creator & grad & 5.1\,\% & 1.9\,\% & +3.2 [+2.0, +4.4] & 1\,342 \\
\bottomrule
\end{tabular}
\end{table}

\subsection{Three further teams}
Three of the communities found in the independent search show the same scheme with a strong, selection-robust excess (Table~\ref{tab:cards}, Table~\ref{tab:candcontrols}). Community 7 (38 wallets) reaches 20 SOL in 67\% of its tokens and 60 SOL in 20\%; its cash flow per token is the highest of the four teams, but the ten best wallets account for most of the positive total. Community 4 (45 wallets) reaches 20 SOL in 58\% of its tokens with a small cash flow, and its wallets are the least persistent. Community 3 (76 wallets) is the broadest: 39\% of its tokens reach 20 SOL and a majority of its wallets have a positive cash flow, unlike the other teams, where most of the wallets lose. After weighting, the reach-20 excess is 41\% percentage points for community 4 (31\% within creators), 50\% for community 7 (26\%) and 22\% for community 3 (11\%). Balance is good for community 4 and acceptable for the other two; for most other candidates it is poor, and their estimates should not be read. Several communities show no excess at all, so buying in the first seconds is not sufficient for being a pump team.

\subsection{Rotation, stability and exploitability}
Among the wallets that the pipeline can see, rotation is low: for merged teams of at least 100 wallets the median share of a block's active wallets that were not active before is small and the median retention of the previous block's wallets is high. This is a lower bound: a wallet becomes a member only after repeated evidence, so fresh wallets cannot appear by construction. As a proxy for unseen rotation we compared the first-minute buyers of team-signal tokens that are not members: 15\% of them are fresh wallets (first trade in the data within an hour before the token) against 21\% for other tokens, so there is no sign of a hidden pool of fresh wallets around the team. Wallet rotation over horizons longer than the window cannot be tested with these data.

Success is persistent in level but only weakly in rank: all large teams exceed the base rate, and among the teams with enough tokens in both halves the Spearman correlation of the reach-20 shares of the two halves is 0.40 (p = 0.05).

None of this helps a follower. A trader who enters after the third member's buy and exits by a take-profit of 30\%, a stop of 30\% or after 60 s earns on average -5\% per trade on the 126-wallet team (interval -6\% to -4\%) and a loss on each of the three other teams. The third member buys a few seconds after creation (median); by the time a follower can act, the price has moved.

\subsection{Refining the team layer: reputation, agreement and scouts}
The merge of Section 8.1 repairs the fragmentation but keeps every merged team, whether it lifts tokens or not. Three further steps raise the precision of the signal at the cost of coverage; all are fitted on the first half of the window and, in this section, evaluated on the second, which is also the half on which their few parameters were chosen.

\emph{In-block adjacency as a second evidence type.} Pairs of fast wallets whose buys sit in the same block at neighbouring positions are independent of co-buying. Such links are stable: of the pairs with at least three adjacent events in the first half, 74\% are adjacent again in the second half, against 0.3\% for random pairs among the same wallets. For single pairs the signal is weak (AUC 0.52 for telling wallets of one community from wallets of different communities against 0.73 for co-buying on the same pairs), so adjacency is used as a graph edge, not as a pair score. Clustering the adjacency graph with Louvain at a high resolution gives fine clusters of at least five wallets; a coarse clustering mixes pumping teams with generic bots, while sub-teams differ in the reach-20 share from a few per cent to more than 90 per cent.

\emph{Reputation.} A team is kept if it has enough first-half signal tokens and their reach-20 share exceeds a threshold. Without a success threshold the merged groups mark 7\,105 second-half tokens with a reach-20 share of 36\%; keeping teams with a first-half share of at least 0.4 leaves 3\,973 tokens at 53\% (lift 6.0); for the fine adjacency clusters a threshold of 0.6 leaves 1\,933 tokens at 67\% (Table~\ref{tab:repoot}).

\emph{Agreement.} A token is flagged by both layers when it has at least three members of a kept merged team and of a kept adjacency cluster. The union of the two selections adds nothing, but their intersection is the cleanest signal: 1\,369 tokens with a reach-20 share of 73\% (lift 8.3). Against controls without any member of a candidate team among the first-minute buyers, matched on covariates measured before the teams' entry, the excess in the reach-20 share is 45\% percentage points (interval 40\%--50\%) and 44\% within creators; for the reach-60 share the two estimators disagree, so we do not interpret it.

\emph{Scouts.} Inside a team the identity of the first member to buy matters. Wallets that were the first team member to buy in at least ten first-half tokens are ranked by the reach-20 share of those tokens, and those in the top quartile are scouts. In the second half, tokens in which a scout buys first reach 20 SOL more often than tokens in which another member is first in seven of nine teams, by between a few and several tens of percentage points, with a permutation p-value below $1/301$ (300 permutations of the first-mover labels within hours; the smallest attainable value \cite{phipson2010}); two teams show no effect (Table~\ref{tab:h7roles}). A few wallets play this role: the ten per cent of wallets with most first-mover tokens account for between a third and three quarters of the first-mover tokens of a team.

\emph{Combining the steps.} Stacking the filters raises the precision monotonically and lowers the coverage (Table~\ref{tab:rulesq05}): merged groups without any filter mark 9\,004 tokens at 35\%, reputation 4\,067 at 53\%, agreement of the two layers 1\,474 at 74\%, and agreement with a scout buying first 891 at 77\%. Agreement is the most efficient step; the scout adds a few points of precision at a large loss of coverage. There is no free gain.

\begin{table}[htbp]
\centering\small
\setlength{\tabcolsep}{3pt}
\caption{Signal rules on second-half tokens with teams, reputations and scouts estimated on the first half (scouts: top half by first-mover reach-20 share).}\label{tab:rulesq05}
\begin{tabular}{p{119pt}rrrrr}
\toprule
Rule & Signal tokens & \shortstack{Share of second-half \\[0pt] tokens} & Reach 20 & Lift 20 & Reach 60 \\
\midrule
R0 merged groups, no filter & 9\,004 & 13.0\,\% & 34.7\,\% & 3.93 & 7.4\,\% \\
R1 reputation & 4\,067 & 5.9\,\% & 53.4\,\% & 6.06 & 12.4\,\% \\
R2 reputation + scout first & 1\,570 & 2.3\,\% & 63.1\,\% & 7.15 & 14.0\,\% \\
R3 adjacency reputation + scout first & 464 & 0.7\,\% & 75.2\,\% & 8.53 & 22.2\,\% \\
R4 agreement & 1\,474 & 2.1\,\% & 73.6\,\% & 8.35 & 17.5\,\% \\
R5 agreement + scout first & 891 & 1.3\,\% & 77.4\,\% & 8.79 & 19.4\,\% \\
R6 union of scout rules & 1\,812 & 2.6\,\% & 64.4\,\% & 7.31 & 15.1\,\% \\
\bottomrule
\end{tabular}
\end{table}

\emph{Stability and a soft reputation.} Coverage of the rules that use the adjacency clusters is not stable: a single Louvain run \cite{blondel2008} gives a wide range of flagged tokens for the agreement rule depending on the seed, at a stable precision. A consensus of 20 runs \cite{lancichinetti2012} makes seven of eight seeds agree but leaves one seed in which a large borderline cluster fails the reputation threshold. We therefore replace the hard threshold by a soft reputation: the success share of a team shrunk towards the pooled share \cite{efron1975} with a prior of 20 pseudo-tokens, and the agreement score is the lower of the two layers' scores. Selecting the top 1000 tokens by this score gives a reach-20 share between 71\% and 75\% across all eight seeds, at a fixed coverage (Table~\ref{tab:softprior}). The strength of the prior was chosen on a validation block that does not include the second half: precision is flat between 5 and 50 pseudo-tokens and lower without shrinkage, and we keep the value 20 fixed in advance because the differences are within noise. The layer is published as a documented module with unit tests, and on the real data its rules reproduce the numbers of the stand-alone analysis.

These refinements are tuned on one window of 88 hours: thresholds, the resolution of the clustering and the scout quantile were chosen with the same data, although every team, reputation and scout was fitted on the first half and evaluated on the second. Absolute precisions in this section are therefore optimistic; Section 8.10 shows how much. None of the rules is usable by a follower: each needs the first minute of trading.

\subsection{Creator, bundle and snipers}
The communities of Sections 8.3 and 8.4 do not act independently of the token's creator. The creator makes an initial buy at creation in 95\% of the complete tokens, and this buy is a large share of all SOL bought in the first 10 s. Its size is the strongest single predictor in the data (Table~\ref{tab:chaincreator}): the share of tokens reaching 20 SOL rises steeply with the creator's initial buy, and so does the share of tokens with sniper communities c0 or c3 among the first buyers (at least three members of one of them in the first 10 s). On a chronological split (first 70\% of tokens for fitting) the initial buy alone predicts the success of the token with an AUC of 0.82 and the entry of the snipers with 0.80; the presence of the snipers alone gives 0.70, and both together 0.88. The snipers therefore carry information beyond the size of the initial buy: within strata of the initial buy their presence multiplies the odds of reaching 20 SOL by 8.0 (Mantel--Haenszel \cite{mantel1959}; Table~\ref{tab:chaincreator}). This agrees with the matched comparisons above, which already conditioned on the initial buy, and it shows why they had to.

\emph{Order of entry.} The sniper communities are not the first actors. The community c2 buys in the same slot as the first trade of the token in 75\% of its tokens, a bundle at creation, and in 19\% of those tokens the creator's own wallet is a member of c2. The snipers c0 and c3 enter a median of a few slots after the first trade. Where c0 and c2 are both present, c0 leads in only 17\% of the joint tokens (Table~\ref{tab:leadlag}). Between the snipers themselves, c0 leads c3 in 62\% of the joint tokens and the first buy of c3 comes after the third buy of c0 in 51\% of them, against 35\% for matched random wallets, while c0 does not react to c3; tokens led by c0 reach 20 SOL in 60\% of cases against 47\% when c3 leads.

\emph{Who drives success.} The configurations differ sharply: tokens with none of the three communities reach 20 SOL in 4.9\% of cases, with c2 alone in 9.5\%, with c0 or c3 alone in 41.6\%, and with c2 together with c0 or c3 in 37.7\%. The bundle at creation is therefore not what lifts the token; the snipers are. The bundle does, however, accompany the snipers: within strata of the initial buy and of the creator's earlier tokens, the snipers are present in 46\% of the tokens with c2 against 18\% without it. Two readings are compatible with this: the snipers react to a visible bundle, or the creators that bundle also choose tokens that attract snipers. Our data cannot separate them. A creator who buys again within the first 10 s is a negative sign: snipers are present less often in those tokens and the reach-20 share is lower.

All of this happens within the first seconds after creation (the bundle in the first slot, the snipers a few slots later), which is why a follower who sees the signal cannot profit from it (Section 7), and why the comparisons of this section must be read conditional on the creator's initial buy.

\subsection{Concealment and cash-out}
Coordinated actors have reasons to hide their traces: to keep copy-traders and analysts from following them, to separate the creator from the buyers, and to move the proceeds away from the wallets that traded. We look at four layers where such hiding would leave a mark: the link between the creator and the team, the funding of the wallets, the cash-out of the proceeds, and the life cycle of the wallets. The evidence is the table of native SOL transfers for the wallets of the studied teams (the 126-wallet merged team and the sniper communities c0--c4 and c7). Two cautions apply throughout. First, most native transfers of a team's wallets are parts of trading transactions (payments to the curve, fees, tips inside the transaction) and are excluded, so that the remaining transfers are those outside the transactions of the trading protocol. Second, the exchange flag of the transfer table is empty in this snapshot, so exchange-like counterparties (hubs) are recognised by their number of distinct senders in the whole window: a destination with at least 200 distinct senders is called a hub, one with 20--199 mid, one with fewer than 20 specific. The numbers behind this section are in Table~\ref{tab:teamprofile}, Table~\ref{tab:chain2} and Table~\ref{tab:pairs}.

\emph{The creator is not hidden behind the team.} The creators of the tokens of a team very rarely exchange SOL with the team's wallets: 0.03\% of the creators of the 126-wallet team's tokens, against 0.04\% for creators of comparable tokens without any studied team; the figures for the sniper communities c4, c7, c3 and c0 are of the same order. The one exception is community c1 (133 wallets, 217 creators), where 6\% of the creators are linked: a creator farm rather than a pumping team. We find no hidden link between creators and the pumping teams.

\emph{Funding is mostly invisible.} In the observed window at most about half of the team wallets receive any native transfer (35\% of the wallets of the 126-wallet team, 21\% of c7), and fewer than one in six receive it before their first early buy; the creator farm c1 is the exception (92\% funded). The wallets were therefore funded before the window starts or by a mechanism that does not appear as a native transfer. Where funding is visible it is concentrated or scattered depending on the team: the 126-wallet team has 108 external funders, whereas five funders supply most of the SOL of community c7. That the funding of most wallets cannot be seen is consistent with concealment, and equally with ordinary funding before the window; our data cannot separate the two.

\emph{Cash-out goes to dedicated low-degree addresses (126-wallet team).} The 126-wallet team sends about 2,664 SOL to external addresses, 6 times what it receives from outside, and 97\% of it goes to specific addresses with fewer than 20 distinct senders in the window; only 1\% goes to hubs. The five largest destinations take 63\% of it. The destinations look like purpose-made receiving addresses, not like exchange deposits. The other communities differ: c1 sends about half of the SOL to hubs (49\%); c4 sends to 24\,418 different addresses, the pattern of a service that pays out to many users rather than a pool of one operator, so its high success should be read with that in mind.

\emph{Second hop.} The addresses that receive the first hop of the 126-wallet team forward only 33\% of what they receive within the window: the proceeds are parked. Community c7 shows the opposite: its first-hop addresses forward 6\% to very few destinations, mostly hubs, a single exit route. For c0, c3 and c1 the first-hop addresses are pass-through addresses that also carry other flows.

\emph{Do the teams share a cash-out infrastructure?} A naive count of destinations that receive SOL from two teams suggests that the 126-wallet team shares dozens of collectors with c3 and c0. That count is an artefact: these teams overlap in wallets, and every destination of a shared wallet is counted as shared. Recomputed on disjoint wallet sets, the 126-wallet team and c3 share 1 destination and the 126-wallet team and c0 share 6; the communities c0, c1 and c2 share almost only hubs, which exchanges, bridges and services produce without any common operator. One step further, c0 and c3 lead to a handful of common second-hop destinations that are reached through four first-hop addresses that both communities use and are not independent. By the cash-out the 126-wallet team is not linked to any other team.

\emph{One operator, or two?} For the two pairs that the transfers suggested, other evidence is stronger. Communities c0 and c3 (no common wallet) occur together in 3\,174 tokens: given c0, c3 is present in 31\% of the tokens (base rate 4\%). Their distributions of priority fees are close, they share funders and first-hop destinations, and c3 follows c0 (Section 8.7): tokens with c3 only reach 20 SOL in 15\% of cases against 55\% when both are present. We read c3 as a follower within the c0 ecosystem. Communities c1 and c2, in contrast, never occur together, have unlike behaviour and share only hubs and one transit address; they are different structures.

\emph{Wallet life cycle.} In most pumping teams the wallets stay for the whole window: few wallets of c7 (3\%), c3 (11\%) and c0 (13\%) are inactive in the last 24 hours, more of the 126-wallet team (19\%), and about half of c4 (49\%). In five of the seven teams a wallet whose last buy is followed by a sweep of at least 0.5 SOL to a non-team address is more likely to stop than one without a sweep ("sweep and leave"). Only c7 enlarges its pool: a batch of 23 wallets first appears together in the third half-day, most of them without a visible funding transfer. This is the closest we find to the planned rotation of wallets, and it is an expansion, not a replacement.

\emph{Reading.} On the trading side the pumping teams do not hide: their pools are stable, their wallets pay their own fees, their transactions sit in neighbouring positions of the same blocks, and the creator is not linked to them. The concealment that may exist is on the funding side (most wallets show no funding in the window) and at the exit (dedicated low-degree receiving addresses and parked proceeds for the 126-wallet team; a single route through hubs for c7; "sweep and leave" in several teams). Evidence for common operators behind different communities is weak at the level of cash-out and stronger at the level of behaviour (c0 and c3). These conclusions are bounded by an observation window of 88 hours that cannot show funding before it or the aging of wallets over weeks, by a coarse classification of exchange-like addresses, and by the fact that all communities are defined by co-occurrence of fast buys, not by proven ownership. We use \emph{concealment} only for the observed pattern of separation of flows; it does not imply intent.

\subsection{Which wallets join a team: graph methods}
If teams change wallets, the useful question is not only who is in a team now but who will be in it next. We test this with a prediction task on the first and second halves of the window and ask whether graph neural networks help. Graph-based learning has been used to trace the organisers of pump-and-dump schemes through information-diffusion graphs \cite{fu2025}, to detect rug pulls from token-flow graphs \cite{wu2025rphunter}, and to detect Sybil addresses in airdrops \cite{liu2025}, but on labelled data; we have no labels of team membership, so the target is a weak, behavioural one.

\emph{Task.} The kept merged teams of Section 8.6 at the middle of the window contain about 1\,414 wallets. Candidates are the fast wallets (at least 20 early buys) outside every kept team. A candidate is positive if in the second half it buys together with at least three team members among the first-minute buyers in at least ten tokens and these tokens are at least half of its early buys, so that a wallet that is simply everywhere does not count; 36\% of the candidates are positive. All features come from the first half: co-buys with team members, in-block adjacency to team members, similarity of the lag and priority-fee distributions to the team, and the activity of the wallet. Models are evaluated on a held-out half of the candidates, which are selected at random; wallets of one team can therefore appear on both sides, and the task measures who joins which kind of team, not generalisation to unseen teams.

\emph{Results} (Table~\ref{tab:h9join}). Single features rank candidates poorly (AUC between 0.53 and 0.61). A gradient-boosting model on the node features reaches 0.90. Smoothing the features over the graph (one and two hops, a linear graph convolution \cite{wu2019}) and feeding them to the same model gives 0.92, and two-layer graph neural networks trained end to end reach 0.87 (graph convolution \cite{kipf2017}) and 0.88 (GraphSAGE \cite{hamilton2017}). The neural networks therefore help relative to a model without graph information, but they do not beat simple smoothing followed by boosting.

\begin{table}[htbp]
\centering\small
\setlength{\tabcolsep}{3pt}
\caption{Prediction of the wallets that join a team in the second half of the window (candidates: fast wallets outside every kept team; positive: at least ten second-half tokens with at least three team members among the first-minute buyers that make up at least half of the wallet's early buys). AUC and precision at the top 50 and 100 candidates on a held-out half of the candidates; all features from the first half.}\label{tab:h9join}
\begin{tabular}{p{288pt}rrr}
\toprule
Method & AUC & \shortstack{Precision, \\[0pt] top 50} & \shortstack{Precision, \\[0pt] top 100} \\
\midrule
co-buy count with team members & 0.627 & 66\,\% & 61\,\% \\
distinct team members co-bought with & 0.682 & 58\,\% & 60\,\% \\
in-block adjacency to team members & 0.532 & 66\,\% & 63\,\% \\
lag-distribution similarity to the team & 0.611 & 60\,\% & 56\,\% \\
priority-fee distribution similarity to the team & 0.557 & 50\,\% & 63\,\% \\
LightGBM on node features & 0.902 & 98\,\% & 99\,\% \\
LightGBM without co-buy features & 0.816 & 94\,\% & 94\,\% \\
LightGBM on activity only (n\_buys, mean lag) & 0.729 & 94\,\% & 83\,\% \\
SGC-style smoothing + LightGBM & 0.919 & 100\,\% & 99\,\% \\
SGC-style smoothing + LightGBM, without own co-buy features & 0.888 & 100\,\% & 100\,\% \\
2-layer GCN (PyTorch) & 0.870 & 96\,\% & 93\,\% \\
2-layer GCN (PyTorch), without own co-buy features & 0.855 & 92\,\% & 92\,\% \\
2-layer GraphSAGE (PyTorch) & 0.882 & 96\,\% & 94\,\% \\
2-layer GraphSAGE (PyTorch), without own co-buy features & 0.871 & 98\,\% & 94\,\% \\
\bottomrule
\end{tabular}
\end{table}

\emph{Is this just activity?} A model on the activity of the wallet alone (number of early buys and mean lag) reaches an AUC of 0.73. Within terciles of activity the full node-feature model keeps a clearly higher AUC than activity alone, so the behavioural and neighbourhood features carry information beyond how active a wallet is.

\emph{Held-out teams.} The split above is random over candidates, so wallets of one team appear on both sides. We therefore repeated the experiment with whole teams held out: candidates are grouped by the kept team with which they co-bought most in the first half, half of the teams go to training and the other half to testing (Table~\ref{tab:h9heldout}; ten random splits). The picture changes. The models keep a clear signal for unseen teams (AUC about 0.72 for boosting on the node features and 0.74 with graph smoothing, against 0.90 and 0.92 above), but the margin is smaller and varies a lot between splits, because about nine in ten candidates are associated with one team, so a split puts that team either in training or in testing. Without the candidate's own co-buy features the AUC drops to between 0.57 and 0.68, and a model on activity alone is at chance (0.52), unlike in the random split (0.73): part of the earlier performance came from team-specific habits that the model had seen. Graph smoothing keeps its small advantage over boosting alone. The conclusion that graph neural networks do not beat simple smoothing is therefore unchanged, but the level of predictability for teams the model has not seen is considerably lower than the main text first suggested, and we did not repeat the neural networks on the team split.

Two cautions. The target is a behavioural proxy for membership, not an observed membership, and the first version of the label (any five tokens) was trivially satisfied by 62\% of the candidates, which is why the stricter definition is used. The networks were trained with default settings (two layers, 32 hidden units, 300 epochs), without tuning and without temporal models; the conclusion that they do not beat the simple baseline holds for this setting only.

\subsection{Final hold-out, signal timing and incremental value}
The rules of Sections 8.1 and 8.6 were developed on one window and selected from many candidates, so their figures are optimistic and, as the review of an earlier version of this paper pointed out, a signal that appears after the event cannot predict it. This section reports what remains under a stricter protocol (Section 4.8).

\emph{Frozen rules on an untouched block.} The last block of the window (about 36,500 complete tokens, of which 8.6\% reach 20 SOL) took no part in building teams, in fitting reputations and scouts, or in tuning parameters. Applied once to it, the rules keep most of their precision (Table~\ref{tab:holdout}). With parameters tuned on the middle block the reputation rule flags about 2,400 tokens of which half reach 20 SOL (lift 5.9), the agreement of the two layers flags about 900 tokens at 71\% (lift 8.2) and agreement with a scout buying first about 600 at 73\% (lift 8.5). The parameters of the first manuscript version give 55\%, 78\% and 85\% for the same three rules on fewer tokens (about 1,900, 500 and 350), so the earlier figure of about three quarters is reproduced on new data by the strict rules, and the tuning did not inflate it. The reach-60 share of the flagged tokens is between 12\% and 25\%, several times the base rate.

\begin{table}[htbp]
\centering\scriptsize
\setlength{\tabcolsep}{3pt}
\caption{Final hold-out. Teams, reputations and scouts are fitted on the first part of the window; the parameters are either tuned on a second block (B) or fixed at the values of the first manuscript version; the rules are then applied once to the last block (C), 36 532 complete tokens with a base rate of 8.6\%. Reach 20: share of signal tokens that ever reach 20 SOL; already at 20 SOL: signal tokens whose curve already held 20 SOL when the third team member bought; the last two columns restrict to signals that fire before the curve reaches 20 SOL and give the share of those that reach 20 SOL later and the base rate of the same kind of tokens.}\label{tab:holdout}
\begin{tabular}{p{76pt}rrrrrrrr}
\toprule
Rule & \shortstack{Signal \\[0pt] tokens} & Reach 20 & Lift & \shortstack{Top team \\[0pt] share} & \shortstack{Reach 20 without \\[0pt] top team} & \shortstack{Already at \\[0pt] 20 SOL} & \shortstack{Reach 20, \\[0pt] not yet at 20} & \shortstack{Base, \\[0pt] not yet at 20} \\
\midrule
reputation of the merged team (tuned on B) & 2\,359 & 50.4\,\% & 5.9 & 67\,\% & 50.1\,\% & 26\,\% & 33.3\,\% & 4.9\,\% \\
agreement of both layers (tuned on B) & 904 & 70.6\,\% & 8.2 & 70\,\% & 70.1\,\% & 37\,\% & 53.0\,\% & 4.6\,\% \\
agreement and scout first (tuned on B) & 615 & 73.0\,\% & 8.5 & 79\,\% & 76.7\,\% & 40\,\% & 54.9\,\% & 4.4\,\% \\
reputation of the merged team (first version) & 1\,884 & 55.5\,\% & 6.5 & 66\,\% & 54.3\,\% & 30\,\% & 36.4\,\% & 4.7\,\% \\
agreement of both layers (first version) & 499 & 78.4\,\% & 9.1 & 62\,\% & 73.2\,\% & 46\,\% & 59.9\,\% & 4.6\,\% \\
agreement and scout first (first version) & 354 & 84.7\,\% & 9.9 & 73\,\% & 85.6\,\% & 48\,\% & 70.8\,\% & 4.6\,\% \\
\bottomrule
\end{tabular}
\end{table}

\emph{One team dominates.} The signals are concentrated: the largest team supplies between 62\% and 79\% of the flagged tokens of every rule (the 126-wallet team of Section 8.3). Dropping each team in turn leaves the precision of the rules within a few points of the full value (for the tuned rules the lowest value over all teams is 49\% for the reputation rule and 69\% to 72\% for the agreement rules), and without the largest team the precision stays between 50\% and 86\%, but the number of flagged tokens falls to fewer than 800 and, for the strict rules, to about 100 to 270, so the claim "the rule works for many teams" rests on a small number of tokens outside the largest team.

\emph{Many signals arrive too late.} A team signal needs three members to have bought, at a median of two to three seconds after creation. In many cases the curve already held 20 SOL by then: between 26\% and 40\% of the signals of the tuned rules and between 30\% and 48\% of those of the first-version rules were already at 20 SOL when they fired, the more so the stricter the rule. For those tokens the "outcome" was already reached when the signal appeared. Among the signals that fire while the curve is still below 20 SOL the precision is 33\% for the reputation rule, 53\% for agreement and 55\% for agreement with a scout (36\%, 60\% and 71\% for the first-version parameters), against a base rate of 4.4\% to 4.9\% for tokens in the same position; the lift is still seven- to fifteen-fold, but the precision falls by a quarter to a third compared with the figures of the earlier sections when the signal has to precede the event. The pipeline's own earliest signals (Sections 5--7) are timelier, with only about 4\% of them at 20 SOL when detected, because they require far less of the token.

\emph{Equal budgets and the crowd.} At equal numbers of flagged tokens, naive counts of early buyers from a large co-buy population rank tokens better than the team layer (Section 5.1, Table~\ref{tab:equalbudget}), which prompts the question of what the team signal adds. We answer it for the strictly prospective outcome (the token is still below 20 SOL at 60 s and reaches 20 SOL later; about 34,000 tokens in the last block, 1.6\% of which succeed) by comparing logistic models fitted on the block before (Table~\ref{tab:incremental}). A model with only the curve SOL at 60 s reaches an AUC of 0.94, and a model with all first-minute features that carry no group information the same; adding the group features (membership in the best merged team and adjacency cluster, their shrunk reputations, agreement, scout first) does not improve it (change in AP $-0.015$, interval $-0.030$ to $+0.004$; change in AUC $0.000$). The group features alone give an AUC of 0.79, so they do carry information, but it is the information already contained in how far the curve has run.

Stratifying by the curve SOL at 60 s shows where the association between a team signal and the later outcome lives (Table~\ref{tab:strata}). In the lower four quintiles (median curve SOL at 60 s up to about 0.4 SOL) almost no token reaches 20 SOL later, whatever the signal. In the top quintile, tokens with a team signal reach it in 12\% of cases against 4\% without, and tokens flagged by the agreement rule in 27\% against 7\%; but the quintile is wide (its median curve is 2 SOL and it extends to just under 20 SOL), and the regression above finds no gain beyond the curve level itself.

\emph{Reading.} The team signals are real in the sense that matters for the descriptive questions of this paper: a small number of stable pools of wallets buys in the first seconds of a large share of the launches of one window and is associated with the tokens that run furthest, beyond what the covariates measured before the entry explain (Section 8.3). They are not shown to be informative beyond the early crowd for a trader at 60 s, a large part of the flagged tokens had already reached the level when the signal fired, and one team supplies most of the signals. The earlier statement that the layer raises the share of flagged tokens that reach 20 SOL from about one third to about three quarters should be read with these three qualifications; as a statement about association it holds for the final block, as a statement about forecasting it does not.

\section{Robustness}
\begin{figure}[tbp]
\centering
\includegraphics[width=\linewidth]{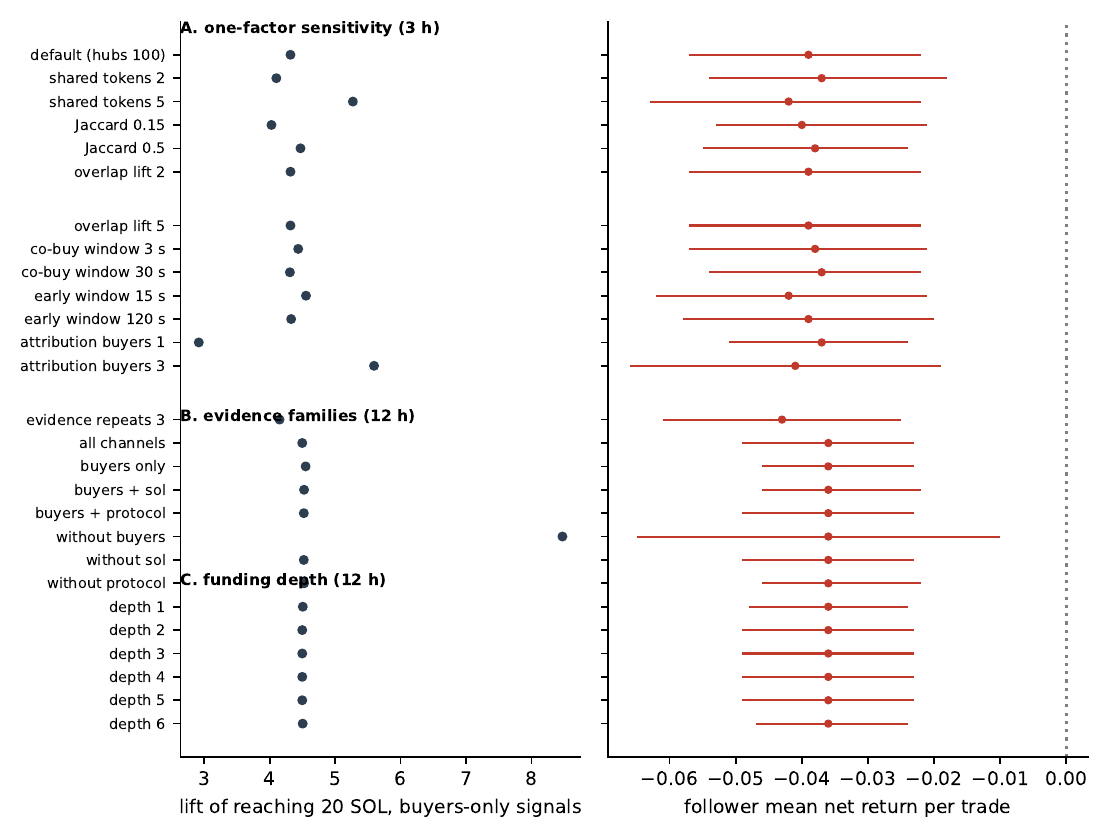}
\caption{Robustness of group discovery and of the follower result. Left: lift of reaching 20 SOL for buyers-only signals. Right: mean net return per trade of the follower with 95\% bootstrap intervals over groups; the interval lies below zero in all 27 settings. Panels A-C: parameters of the 3-hour slice, evidence families and funding depth of the 12-hour slice; values in Table~\ref{tab:robust}.}\label{fig:robust}
\end{figure}

The conclusions about group quality and follower returns are insensitive to discovery parameters. On a three-hour slice we varied the co-buy count, overlap threshold, overlap lift, coordination window, early-buy window, attribution rule and number of evidence repeats one at a time (Figure~\ref{fig:robust}). The accepted edge set is sensitive to the discovery rules, but the properties that matter for the conclusions are not: the lift of reaching 20 SOL for buyers-only signals stays within a narrow band (2.9 to 5.6), and the net return of a fixed follower rule stays between -4\% and -4\% with every bootstrap interval below zero. The attribution rule moves the precision--coverage trade-off: requiring one matching buyer instead of two yields more signals at a lower lift, requiring three yields fewer at a higher one. The two overlap-lift thresholds leave all results identical because the lift of accepted pairs against the uniform-overlap baseline is far above both thresholds on this slice; this baseline therefore does not discriminate, and a popularity-aware null model would be needed to make it useful.

\emph{Which evidence families matter.} On a 12-hour slice we removed evidence families from the discovery (Figure~\ref{fig:robust}). Co-buying and co-selling alone reproduce 87\% of the accepted edges of the full run and the same signals: signal coverage, the lift of reaching 20 SOL for buyers-only signals and the follower result are unchanged. Co-buying plus SOL-flow evidence retains 92\% and co-buying plus protocol evidence 95\%; token-transfer evidence changes nothing on this slice. Without the buyer family only 28\% of the edges remain; their lift is high but rests on few tokens. Every variant accepts only edges that the full run also accepts. The additional channels therefore enlarge and connect groups (and enable the campaign analysis) but, in this window, do not change the signal on which the follower and the dynamics results are based. This agrees with the link-prediction finding that non-trading links add no persistence beyond co-buy history.

\emph{Funding search depth.} The funding search follows paths of up to a maximum number of hops (default three). We varied this depth on the 12-hour slice (depths 1--6; Figure~\ref{fig:robust}) and on a 3-hour slice (depths 1--10, same pattern). From depth two on, the accepted edge set is practically unchanged, as are the number of groups, the signal coverage, the lift of buyers-only signals and the follower result (interval below zero at every depth). Only the campaign inventory depends on depth: the number of campaigns rises from 839 at depth 1 to 4\,738 at depth 3 and 7\,992 at depth 6 without saturating, and the share of campaigns with a consolidation falls from 29\% to 14\%, i.e. deeper searches add chains that are less often followed by a fan-in. The campaign counts in Section 5.3 are therefore a depth-three inventory and a lower bound on what a deeper search would report; the conclusions on groups, signals and followers do not depend on this choice.

\emph{Other robustness checks reported elsewhere.} The sensitivity of the price-path contrast to the covariate set, to overlap and to the resampling unit is in Section 6.4; the stability of the team rules over seeds and prior strengths in Section 8.6; the dependence of the team results on one team, on signal timing and on the choice of block in Section 8.10; and the latency comparison on identical trades in Section 7.1. \emph{Day by day.} Table~\ref{tab:daily} repeats three headline results on each of the four days without re-tuning anything. The matched price-path contrast with creation-time covariates is the same on every day (a rise of about seven to eight percentage points in the probability of a doubling and a fall of 13 to 15 points in the retention after the peak), the follower loses on every day under both cost conventions with intervals clear of zero, and three of the four fast-wallet communities keep their reach-20 share within a few points from day to day, far above the base rate of the day. The exception is community 4, whose share falls from about 60\% on the first two days to 49\% and then 30\% (on few tokens), in line with its low persistence of wallets (Section 8.4). The window is too short for a test across regimes, and an independent window is not available with these data.

\section{Discussion}
\textbf{What the data say.} Coordinated wallet groups are a real and concentrated feature of this launchpad: the top tenth of the groups (about two thousand) supply about four fifths of the attributable launches, the largest buyer rings persist for days, funding campaigns recycle a small pool of wallets, and members act earlier than other early buyers of the same token and leave with cash while late buyers hold the inventory. Tokens that attract known members have a pump-like path, but the data do not show that the members' presence adds to what the early activity of the token already predicts (Section 6.4). The signal that reveals a group is already priced when it becomes visible: a follower with realistic latency and costs earns less than a random entrant. The coexistence of a strong microstructure signature and no follower profit is what one expects from a venue in which informed early order flow is observable only after it has moved the price.

\textbf{Large teams.} The most consequential finding for the original question is Section 8: a handful of pools of tens to over a hundred wallets act in the first seconds of thousands of launches by thousands of creators and are associated with a reach-20 share two to four times that of comparable tokens, also within creators; they take cash out of the curve quickly and show hardly any rotation in the observed window. The group builder splits them; a merge on shared wallets recovers much of the signal. Their association stops at about 20 SOL, and followers still lose. The pools act after the creator: the creator's initial buy is the strongest single predictor, a bundle of creator-linked wallets buys in the first slot, and the large pools follow a few slots later (Section 8.7). The final hold-out (Section 8.10) tempers this: a large share of the flagged tokens had already reached 20 SOL when the signal fired, one team supplies most signals, and the team features do not improve the prediction of later success beyond how far the curve has run.

\textbf{Relation to earlier work.} Four points connect our results to the literature. (i) Joining coordinated schemes is not profitable on average for participants who follow: the Telegram-organised pumps of centralised exchanges attract participants despite negative expected returns \cite{dhawan2023}, and our follower, who sees the group only after its first buys, loses in the same way, here without any announcement channel. (ii) Early curve state is informative for outcomes on this launchpad: the SOL locked in a curve and behavioural covariates predict graduation \cite{marino2026}, and the creator's initial buy is the strongest single predictor of reaching 20 SOL in our data; early data can also flag fraudulent launches \cite{li2026rug}, but classifying a launch and finding a profitable entry after costs are different problems, and the cited model addresses the first. Our incremental-value result (Section 8.10) is in line with this: the information that group features carry is largely the information in the early curve. (iii) The fast-wallet teams behave like the sniper bots described on Ethereum and the BNB Smart Chain, which buy within seconds of a listing \cite{cernera2023,cernera2025}; on a bonding curve the target is the creation itself, and our teams add the scale (tens of wallets across thousands of creators) and the matched-control estimate of their association with the outcome. (iv) Our concealment results are compatible with the manipulation strategies found in a large measurement of pump.fun, which include creator-address obfuscation and coordinated selling \cite{szwajcok2026}, and with bundle traces that link accounts of one entity \cite{hu2026melt}: the wallets of the studied teams are almost never creators of the tokens they buy, and for the largest team the cash-out destinations are overwhelmingly specific to the team (Section 8.8). We measure cash flows and timing, not ownership, so these agreements are qualitative.

\textbf{What is new.} Compared with measurement studies that label manipulation strategies or fraud at the token level \cite{szwajcok2026,chen2026rug,li2026rug}, this paper attaches a validated group structure to price paths and to the cash flows of outside traders, evaluates it prospectively under an exact-reserve execution model, and reports negative and null results with the checks that produced them: that naive co-buy clustering ranks tokens as well as the pipeline at equal budgets, that group features add nothing to early curve state for the prospective outcome, and that the price-path contrast depends on comparability of controls. We consider these as findings and not as caveats: they delimit what group structure on this launchpad can and cannot be used for.

\textbf{Implications.} For platforms and researchers, the practical lever is on the creation and first-seconds side (rate limits, delays, visibility of early buys), not on after-the-fact detection. For detection research, label-free validation is feasible but must be run against a baseline at equal numbers of flagged items; a baseline compared at unequal coverage can mislead in either direction. For trading research, evaluation must keep the trader's own position in the reserve state, test selection rules on a held-out period, and check that the signal precedes the event it is said to predict.

\textbf{Limitations and threats to validity.} (1) \emph{One window.} 88 hours of one regime; results on groups' lifetimes are bounded by the window, and the outcome levels (20 and 60 SOL) are round numbers chosen in advance. (2) \emph{No ground truth.} Validation shows internal consistency and independent-link agreement but not ownership; the acceptance rules add no persistence beyond co-buy counts, and at equal budgets a naive co-buy score ranks tokens at least as well, so the group structure refines but does not outperform the simplest predictor. (3) \emph{Time resolution.} Block time is in seconds; latency is modelled at the second boundary; the order of transactions inside a block is observed, but submission order, tips and bundle membership are not, so real front-runners may exist that our model cannot represent. (4) \emph{Execution model.} Reserve-based quotes are an approximation; the trader's own position is added to the state at first order and later flow is not re-simulated. No real transactions were sent; the quote arithmetic is validated against observed trades but confirmation delay, priority fees and bundle inclusion are not. (5) \emph{Few large teams, one dominant.} One 126-wallet team supplies most large-team signals; the other candidates have fewer tokens and, for most, poorly balanced controls, so conclusions on large teams are case-based. (6) \emph{Residual confounding and overlap.} Covariate balance between group tokens and controls is incomplete, part of the covariates are partly post-treatment, and the price-path contrast does not survive restriction to the region of common support (Section 6.4); the dynamics results are conditional associations. (7) \emph{Single-wallet signals.} Most earliest signals come from one wallet that is also the creator; claims about coordination rest on the multi-wallet subset. (8) \emph{Exploratory analyses.} The team layer, its refinements, the case studies and the parameter sweeps were developed on the same window; the hold-out of Section 8.10 is a hold-out in time within that window, not an independent sample, and the teams and the dominant team are the same ones in every block. (9) \emph{Signal timing.} The team signal needs three members to have bought; a large part of the flagged tokens had reached the level by then. (10) \emph{Incomplete observation.} About a quarter of the launches are incomplete and excluded from outcomes; contrasts between group and control tokens are unaffected, but absolute base rates are conditional on recorded trading. (11) \emph{Concurrent work.} Other measurements of pump.fun \cite{szwajcok2026,hu2026melt,mongardini2025} use different samples, definitions and goals, and the directly comparable cohort preprint \cite{kamat2026} has withdrawn its numbers, so a head-to-head numerical comparison is not yet possible; audits of collection pipelines on the same launchpad \cite{kamat2026b,kamat2026c} show that coverage and labels differ between pipelines, which is why we report a completeness flag for every token.

\textbf{Ethics.} Addresses are public on-chain data, but the paper reports aggregates only and group identifiers are hashes. The working package keeps raw addresses so that every number can be recomputed; the public release replaces every wallet and mint address by a salted hash (the salt is not published) and omits the raw database, and a verification step recomputes the released tables from the hashed inputs. We name no individual and publish no list that could be used to copy or target specific wallets.

\section{Conclusion}
On a complete window of a major bonding-curve launchpad, coordinated wallet groups can be recovered prospectively (using only information available at each decision time) and checked without labels; they are highly concentrated, come in distinct types, recycle funded wallets across campaigns, and tokens with known members show a pump-like path and members who leave with cash while late buyers hold the inventory. The detectability of the groups does not translate into follower profit: after latency and realistic costs a follower loses money out of sample, and no entry-ranking model we trained recovers a positive return. The clearest coordinated actors are not the small groups but a few pools of tens to over a hundred wallets that buy within seconds of creation on thousands of launches; their tokens reach 20 SOL much more often than matched controls, 60 SOL somewhat more often and graduate no more often, and their buy is already in the price when it can be seen. A final hold-out shows that the team rules keep a large part of their precision on unseen data, that many of their signals arrive after the level has been reached, and that they add no information about later success beyond how far the curve has run. Extending the window to test wallet rotation over longer horizons and testing across market regimes are the natural next steps.

\appendix
\renewcommand{\thetable}{A\arabic{table}}\setcounter{table}{0}
\renewcommand{\thefigure}{A\arabic{figure}}\setcounter{figure}{0}
\section{Additional tables and figures}
This appendix collects the tables that support the results of the main text; the tables of the main text are numbered 1, 2, ..., those of the appendix A1, A2, .... The numbers behind every table and figure are also stored as CSV files in the replication package.

\FloatBarrier
\subsection{Data and validation}
\begin{table}[htbp]
\centering\small
\setlength{\tabcolsep}{3pt}
\caption{Reconciliation of the denominators used in the paper: tokens, trades and wallets.}\label{tab:reconcile}
\begin{tabular}{p{185pt}rp{185pt}}
\toprule
Quantity & Value & Note \\
\midrule
Launches observed (all) & 200,443 & All tokens created in the window, complete or not \\
Complete launches & 154,045 & Used for every outcome statistic; 76.9\% of launches \\
Incomplete launches & 46,398 & Median 1 recorded trade; reach 20 SOL in 1.1\% against 7.7\% of complete launches \\
Complete launches, second half & 82,248 & Population of the second-half tests; 7.4\% of them reach 20 SOL \\
Second-half launches with recorded early trades & 69,223 & Population of the simulation set; 8.8\% reach 20 SOL (launches with no trade in the first minute drop out) \\
Curve trades, source table & 15,854,121 & Trade events decoded from the protocol \\
Curve trades, derived table & 15,005,317 & Complete launches only; the first trade of each launch (the creation buy) is not used as a signal \\
Wallet memberships of groups & 91,764 & One row per wallet and group \\
Distinct wallets in groups & 48,162 & About 30\% of the wallets belong to more than one group \\
\bottomrule
\end{tabular}
\end{table}

\begin{table}[htbp]
\centering\scriptsize
\setlength{\tabcolsep}{3pt}
\caption{Basic token statistics. Panel A: stability between the two halves of the observation window. Panel B: complete tokens versus incomplete tokens (incomplete: the observation window of the token is cut by the end of the data or its early trades are missing; shares for incomplete tokens are biased downward by the truncated observation).}\label{tab:dataaux}
\begin{tabular}{lrrrrrr}
\toprule
\multicolumn{7}{l}{\emph{A. The two halves of the window}} \\
\midrule
Half & Tokens & Reach 20 & Reach 60 & Graduation & Median trades & Tokens per hour \\
\midrule
first half & 71\,797 & 8.1\,\% & 2.4\,\% & 2.8\,\% & 26 & 1\,627 \\
second half & 82\,248 & 7.4\,\% & 2.5\,\% & 2.8\,\% & 14 & 1\,864 \\
\bottomrule
\end{tabular}
\\[0.8ex]
\begin{tabular}{lrrrrrrr}
\toprule
\multicolumn{8}{l}{\emph{B. Complete and incomplete tokens}} \\
\midrule
Population & Tokens & Share & Median trades & Reach 20 & Reach 60 & Graduation & Median first-trade age (s) \\
\midrule
complete & 154\,045 & 76.9\,\% & 20 & 7.7\,\% & 2.4\,\% & 2.8\,\% & 5.0 \\
incomplete & 46\,398 & 23.1\,\% & 1 & 1.1\,\% & 0.2\,\% & 0.3\,\% &  \\
\bottomrule
\end{tabular}
\end{table}

\begin{table}[htbp]
\centering\small
\setlength{\tabcolsep}{3pt}
\caption{Naive baseline (connected components of the co-buy graph above a co-buy count threshold) versus groups from the discovery rules. The last row counts groups, members and accepted edges of the discovery output.}\label{tab:baseline}
\begin{tabular}{rrrrrrr}
\toprule
Min. co-buys & Edges & Wallets & Components & Largest & \shortstack{Share in \\[0pt] largest} & \shortstack{Components \\[0pt] $\ge$ 10} \\
\midrule
3 & 1\,903\,708 & 59\,212 & 1\,143 & 55\,872 & 94.4\,\% & 20 \\
5 & 814\,745 & 26\,917 & 784 & 24\,400 & 90.6\,\% & 16 \\
10 & 302\,465 & 13\,787 & 571 & 11\,981 & 86.9\,\% & 11 \\
20 & 120\,569 & 7\,437 & 383 & 6\,134 & 82.5\,\% & 13 \\
group rules & 228\,768 & 91\,764 & 23\,710 & 175 & 0.2\,\% & 1\,803 \\
\bottomrule
\end{tabular}
\end{table}

\begin{table}[htbp]
\centering\small
\setlength{\tabcolsep}{3pt}
\caption{Group-based signal on second-half tokens (at least two members of one group among the buyers of the first 60 s): groups built from first-half co-buy graphs (connected components and Louvain communities, K = minimum number of co-buys) versus discovery groups known before the middle of the window. Base rate of reaching 20 SOL among complete second-half tokens: 0.088.}\label{tab:basecomm}
\begin{tabular}{p{162pt}rrrrrr}
\toprule
Method & Groups & Largest & Signal tokens & \shortstack{Share of \\[0pt] tokens} & Reach 20 & Lift \\
\midrule
components, K=3 & 728 & 31\,612 & 45\,532 & 65.8\,\% & 11.2\,\% & 1.27 \\
components, K=5 & 491 & 14\,521 & 44\,384 & 64.1\,\% & 11.4\,\% & 1.30 \\
Louvain, K=5 & 658 & 5\,393 & 41\,729 & 60.3\,\% & 12.0\,\% & 1.36 \\
components, K=10 & 348 & 6\,962 & 41\,997 & 60.7\,\% & 12.0\,\% & 1.36 \\
discovery groups (known before mid-window) & 13\,020 & 103 & 16\,921 & 24.4\,\% & 22.9\,\% & 2.60 \\
discovery groups with $\ge$ 5 wallets & 2\,218 & 103 & 11\,469 & 16.6\,\% & 29.5\,\% & 3.35 \\
discovery only (not Louvain) &  &  & 179 &  & 1.1\,\% & 0.13 \\
Louvain only (not discovery) &  &  & 24\,987 &  & 4.5\,\% & 0.51 \\
\bottomrule
\end{tabular}
\end{table}

\begin{table}[htbp]
\centering\small
\setlength{\tabcolsep}{3pt}
\caption{Persistence of co-buying pairs on the second half of the window (share of pairs that co-buy again), by number of first-half co-buys. Panel A compares pairs accepted using only first-half evidence with all other co-buying pairs. Panel B compares pairs connected by an independent non-trading link (shared funding campaign, joint signing, transfers) with the remaining pairs. Ratio: persistence of the first group over that of the second.}\label{tab:persistence}
\begin{tabular}{lrrrrr}
\toprule
\multicolumn{6}{l}{\emph{A. Pairs accepted with first-half evidence versus other co-buying pairs}} \\
\midrule
\shortstack{Co-buys in \\[0pt] first half} & \shortstack{Accepted \\[0pt] pairs} & Persist. & \shortstack{Other co-buying \\[0pt] pairs} & Persist. & Ratio \\
\midrule
0 (none) & 2\,991 & 5.5\,\% &  &  &  \\
1 & 7\,333 & 6.7\,\% & 13\,374\,024 & 4.3\,\% & 1.57 \\
2 & 9\,261 & 16.9\,\% & 1\,432\,714 & 15.5\,\% & 1.09 \\
3 & 40\,492 & 21.5\,\% & 349\,374 & 31.3\,\% & 0.69 \\
4-5 & 19\,088 & 41.5\,\% & 260\,859 & 41.9\,\% & 0.99 \\
6-10 & 14\,362 & 61.2\,\% & 169\,782 & 54.2\,\% & 1.13 \\
11+ & 15\,118 & 79.2\,\% & 108\,736 & 71.9\,\% & 1.10 \\
\bottomrule
\end{tabular}
\\[0.8ex]
\begin{tabular}{lrrrrr}
\toprule
\multicolumn{6}{l}{\emph{B. Pairs with an independent non-trading link versus pairs without}} \\
\midrule
\shortstack{Co-buys in \\[0pt] first half} & \shortstack{With hard \\[0pt] link} & Persist. & \shortstack{Without hard \\[0pt] link} & Persist. & Ratio \\
\midrule
0 (none) & 2\,972 & 5.5\,\% &  &  &  \\
1 & 7\,092 & 5.3\,\% & 13\,374\,265 & 4.3\,\% & 1.23 \\
2 & 8\,099 & 12.0\,\% & 1\,433\,876 & 15.5\,\% & 0.77 \\
3 & 4\,496 & 22.0\,\% & 385\,370 & 30.4\,\% & 0.72 \\
4-5 & 3\,547 & 42.7\,\% & 276\,400 & 41.9\,\% & 1.02 \\
6-10 & 2\,262 & 64.1\,\% & 181\,882 & 54.6\,\% & 1.17 \\
11+ & 4\,768 & 73.4\,\% & 119\,086 & 72.8\,\% & 1.01 \\
\bottomrule
\end{tabular}
\end{table}

\begin{table}[htbp]
\centering\small
\setlength{\tabcolsep}{3pt}
\caption{Sensitivity of discovered edges and signals to the transfer-hub threshold (hub threshold). Jaccard is taken over accepted edge sets against the default threshold of 100; ``same class'' is the share of default-run signal tokens whose earliest signal has the same creator link and group size.}\label{tab:hubs}
\begin{tabular}{lrrrrrrr}
\toprule
Slice & \shortstack{Hub \\[0pt] threshold} & Hubs & Groups & \shortstack{Accepted \\[0pt] edges} & \shortstack{Edge Jaccard \\[0pt] vs. 100} & \shortstack{Signal \\[0pt] tokens} & \shortstack{Same signal \\[0pt] class} \\
\midrule
3 h & 15 & 1\,678 & 1\,162 & 7\,766 & 0.930 & 3\,518 & 98.5\,\% \\
3 h & 30 & 578 & 1\,163 & 7\,876 & 0.943 & 3\,518 & 99.2\,\% \\
3 h & 100 & 312 & 1\,157 & 8\,354 & 1.000 & 3\,519 & 100.0\,\% \\
3 h & 300 & 127 & 1\,170 & 8\,550 & 0.977 & 3\,520 & 99.8\,\% \\
3 h & 1\,000 & 27 & 1\,180 & 8\,766 & 0.953 & 3\,520 & 99.8\,\% \\
3 h & 3\,000 & 13 & 1\,185 & 8\,910 & 0.938 & 3\,521 & 99.5\,\% \\
3 h & 1\,000\,000 & 8 & 1\,185 & 8\,910 & 0.938 & 3\,521 & 99.5\,\% \\
12 h & 100 & 1\,088 & 2\,666 & 15\,107 & 1.000 & 11\,346 & 100.0\,\% \\
12 h & 1\,000 & 99 & 2\,731 & 16\,693 & 0.905 & 11\,353 & 99.4\,\% \\
\bottomrule
\end{tabular}
\end{table}

\FloatBarrier
\subsection{Groups, types and signal quality}
\begin{table}[htbp]
\centering\small
\setlength{\tabcolsep}{3pt}
\caption{Data-driven group types (k-means, k=10, chosen by silhouette; mean adjusted Rand index over 25 80\% subsamples = 0.97). Wallets and lifetime are medians over groups; creator and buyer-only shares are means of the group's token shares; graduation is the mean graduation rate of the group's tokens; signal tokens, reach 20 and reach 60 refer to the group's earliest-signal tokens (share whose curve reached 20 / 60 SOL).}\label{tab:typology}
\begin{tabular}{rrrrrrrrrr}
\toprule
Type & Groups & Wallets & \shortstack{Creator \\[0pt] share} & \shortstack{Buyer-only \\[0pt] share} & \shortstack{Lifetime \\[0pt] (h)} & \shortstack{Gradua- \\[0pt] tion} & \shortstack{Signal \\[0pt] tokens} & Reach 20 & Reach 60 \\
\midrule
0 & 3\,773 & 1.0 & 99.7\,\% & 0.3\,\% & 27.5 & 2.8\,\% & 15\,674 & 3.4\,\% & 0.7\,\% \\
1 & 4\,535 & 2.0 & 0.2\,\% & 99.8\,\% & 0.3 & 10.7\,\% & 680 & 39.7\,\% & 17.4\,\% \\
2 & 3\,371 & 1.0 & 99.9\,\% & 0.1\,\% & 0.7 & 1.9\,\% & 8\,714 & 2.1\,\% & 0.4\,\% \\
3 & 1\,169 & 16.0 & 16.8\,\% & 83.2\,\% & 65.1 & 8.6\,\% & 11\,875 & 34.0\,\% & 7.7\,\% \\
4 & 2\,663 & 3.0 & 3.1\,\% & 96.9\,\% & 48.5 & 14.2\,\% & 3\,572 & 23.8\,\% & 6.8\,\% \\
5 & 1\,208 & 3.0 & 18.9\,\% & 81.1\,\% & 45.9 & 10.4\,\% & 3\,132 & 13.5\,\% & 2.4\,\% \\
6 & 2\,164 & 2.0 & 2.8\,\% & 97.2\,\% & 0.2 & 7.3\,\% & 769 & 25.2\,\% & 7.2\,\% \\
7 & 1\,059 & 1.0 & 91.6\,\% & 8.4\,\% & 63.9 & 2.2\,\% & 70\,823 & 4.1\,\% & 0.7\,\% \\
8 & 1\,055 & 7.0 & 0.7\,\% & 99.3\,\% & 0.4 & 5.8\,\% & 654 & 37.8\,\% & 8.3\,\% \\
9 & 2\,713 & 1.0 & 100.0\,\% & 0.0\,\% & 0.0 & 0.6\,\% & 7\,859 & 1.2\,\% & 0.4\,\% \\
\bottomrule
\end{tabular}
\end{table}

\begin{table}[htbp]
\centering\small
\setlength{\tabcolsep}{3pt}
\caption{Share of earliest-signal tokens whose bonding curve reached 20 and 60 SOL, by group size at detection (Panel A) and by day of creation (Panel B), with lift relative to all complete tokens (Panel A) or to the base rate of the day (Panel B); 95\% Wilson intervals. Figures~\ref{fig:dose} and~\ref{fig:daystab} plot these values.}\label{tab:doseday}
\begin{tabular}{llrrrrrr}
\toprule
\multicolumn{8}{l}{\emph{A. By group size at detection}} \\
\midrule
Signal kind & Members & Tokens & Reach 20 & 95\% interval & Lift 20 & Reach 60 & Lift 60 \\
\midrule
buyers only & 2 & 2\,202 & 20.3\,\% & [0.186,0.220] & 2.62 & 4.0\,\% & 1.62 \\
buyers only & 3-5 & 9\,003 & 19.3\,\% & [0.185,0.201] & 2.50 & 5.3\,\% & 2.15 \\
buyers only & 6-12 & 3\,414 & 32.5\,\% & [0.310,0.341] & 4.21 & 9.4\,\% & 3.84 \\
buyers only & 13-29 & 4\,414 & 44.2\,\% & [0.428,0.457] & 5.73 & 11.5\,\% & 4.71 \\
buyers only & 30+ & 832 & 48.3\,\% & [0.449,0.517] & 6.26 & 10.8\,\% & 4.43 \\
creator-linked & 1 & 91\,346 & 2.5\,\% & [0.024,0.026] & 0.33 & 0.4\,\% & 0.17 \\
creator-linked & 2 & 3\,813 & 1.5\,\% & [0.011,0.019] & 0.19 & 0.2\,\% & 0.06 \\
creator-linked & 3-5 & 2\,639 & 8.2\,\% & [0.072,0.093] & 1.07 & 0.9\,\% & 0.39 \\
creator-linked & 6-12 & 3\,676 & 20.6\,\% & [0.193,0.219] & 2.67 & 3.4\,\% & 1.40 \\
creator-linked & 13-29 & 2\,167 & 29.5\,\% & [0.277,0.315] & 3.83 & 3.4\,\% & 1.40 \\
creator-linked & 30+ & 246 & 35.4\,\% & [0.297,0.415] & 4.58 & 4.5\,\% & 1.83 \\
\bottomrule
\end{tabular}
\\[0.8ex]
\begin{tabular}{lrrrrr}
\toprule
\multicolumn{6}{l}{\emph{B. By day of creation (UTC; day 0 is the first observed day)}} \\
\midrule
Signal kind & Day & Tokens & Reach 20 [95\% interval] & Day base rate & Lift \\
\midrule
all signals & 0 & 29\,285 & 9.4\,\% [9.1, 9.7] & 8.7\,\% & 1.09 \\
all signals & 1 & 34\,269 & 8.1\,\% [7.8, 8.4] & 8.0\,\% & 1.02 \\
all signals & 2 & 33\,859 & 7.5\,\% [7.2, 7.7] & 7.4\,\% & 1.01 \\
all signals & 3 & 26\,339 & 6.2\,\% [5.9, 6.5] & 6.7\,\% & 0.93 \\
buyers only & 0 & 5\,705 & 28.4\,\% [27.3, 29.6] & 8.7\,\% & 3.29 \\
buyers only & 1 & 5\,689 & 29.1\,\% [27.9, 30.2] & 8.0\,\% & 3.63 \\
buyers only & 2 & 5\,453 & 28.9\,\% [27.7, 30.1] & 7.4\,\% & 3.90 \\
buyers only & 3 & 3\,018 & 26.4\,\% [24.9, 28.0] & 6.7\,\% & 3.96 \\
creator-linked & 0 & 23\,580 & 4.8\,\% [4.5, 5.1] & 8.7\,\% & 0.55 \\
creator-linked & 1 & 28\,580 & 4.0\,\% [3.7, 4.2] & 8.0\,\% & 0.49 \\
creator-linked & 2 & 28\,406 & 3.4\,\% [3.1, 3.6] & 7.4\,\% & 0.45 \\
creator-linked & 3 & 23\,321 & 3.6\,\% [3.4, 3.9] & 6.7\,\% & 0.54 \\
\bottomrule
\end{tabular}
\end{table}

\begin{table}[htbp]
\centering\small
\setlength{\tabcolsep}{3pt}
\caption{Logistic regressions of reaching 20 SOL or 60 SOL on group size at detection, creator link, high attribution score, initial buy and time-of-day terms; odds ratios with 95\% intervals, standard errors clustered by group. Variable definitions are given in the first column.}\label{tab:logit}
\begin{tabular}{lp{253pt}rr}
\toprule
Outcome & Variable & Odds ratio [95\% CI] & p \\
\midrule
reach20 & log group size (natural log of members at detection) & 1.38 [1.23, 1.56] & $<$0.001 \\
reach20 & creator-linked signal (1/0) & 0.21 [0.17, 0.26] & $<$0.001 \\
reach20 & attribution score $\ge$ 80 (1/0) & 2.21 [1.80, 2.71] & $<$0.001 \\
reach20 & log initial buy (natural log of SOL, floor 0.01) & 1.92 [1.73, 2.12] & $<$0.001 \\
reach20 & sin(2 pi hour/24) & 0.99 [0.93, 1.05] & 0.679 \\
reach20 & cos(2 pi hour/24) & 1.20 [1.12, 1.29] & $<$0.001 \\
reach60 & log group size (natural log of members at detection) & 1.39 [1.15, 1.68] & $<$0.001 \\
reach60 & creator-linked signal (1/0) & 0.19 [0.15, 0.24] & $<$0.001 \\
reach60 & attribution score $\ge$ 80 (1/0) & 1.44 [1.15, 1.80] & 0.001 \\
reach60 & log initial buy (natural log of SOL, floor 0.01) & 1.46 [1.17, 1.81] & $<$0.001 \\
reach60 & sin(2 pi hour/24) & 1.04 [0.94, 1.14] & 0.455 \\
reach60 & cos(2 pi hour/24) & 1.22 [1.11, 1.34] & $<$0.001 \\
\bottomrule
\end{tabular}
\end{table}

\begin{table}[htbp]
\centering\small
\setlength{\tabcolsep}{3pt}
\caption{Graduation (migration to PumpSwap) rate by earliest-signal group type; 95\% Wilson interval.}\label{tab:grad}
\begin{tabular}{lrrrr}
\toprule
Group & Tokens & Graduated & Rate [95\% CI] & Lift \\
\midrule
all complete tokens & 154\,045 & 4\,250 & 2.8\,\% [2.7, 2.8] & 1.00 \\
buyers only, 1-2 & 2\,202 & 104 & 4.7\,\% [3.9, 5.7] & 1.71 \\
buyers only, 13+ & 5\,246 & 371 & 7.1\,\% [6.4, 7.8] & 2.56 \\
buyers only, 3-12 & 12\,417 & 693 & 5.6\,\% [5.2, 6.0] & 2.02 \\
creator-linked, 1-2 & 95\,159 & 1\,076 & 1.1\,\% [1.1, 1.2] & 0.41 \\
creator-linked, 13+ & 2\,413 & 62 & 2.6\,\% [2.0, 3.3] & 0.93 \\
creator-linked, 3-12 & 6\,315 & 108 & 1.7\,\% [1.4, 2.1] & 0.62 \\
no group signal & 30\,293 & 1\,836 & 6.1\,\% [5.8, 6.3] & 2.20 \\
\bottomrule
\end{tabular}
\end{table}

\FloatBarrier
\subsection{Price paths}
\begin{table}[htbp]
\centering\small
\setlength{\tabcolsep}{3pt}
\caption{Price-path measures after the 60-s reference time for tokens with known group members versus stratified controls (curve SOL at the reference time by time-of-day block). Means for the group tokens and for their matched controls (same strata), difference and 95\% bootstrap interval; times in seconds, other measures dimensionless.}\label{tab:h5}
\begin{tabular}{p{173pt}rrr}
\toprule
Measure & Group tokens & Matched controls & Difference [95\% CI] \\
\midrule
active after reference (share of tokens with any later trade) & 85.6\,\% & 80.5\,\% & +5.2 [+4.9, +5.4] pp \\
post-reference peak multiple (max price after reference / price at reference) & 1.46 & 1.12 & +0.34 [+0.25, +0.52] \\
time to peak (seconds after the reference time) & 47.2 & 69.8 & $-$22.6 [$-$23.6, $-$21.8] \\
price 60 s after the peak / peak price & 0.83 & 0.91 & $-$0.09 [$-$0.09, $-$0.09] \\
price at 10 min after creation / price at reference & 0.99 & 0.96 & +0.03 [$-$0.03, +0.17] \\
share of tokens with peak multiple $\ge$ 2 & 8.6\,\% & 3.2\,\% & +5.4 [+5.1, +5.6] pp \\
share of tokens with peak multiple $\ge$ 5 & 1.7\,\% & 0.4\,\% & +1.3 [+1.2, +1.4] pp \\
\bottomrule
\end{tabular}
\end{table}

\begin{table}[htbp]
\centering\small
\setlength{\tabcolsep}{3pt}
\caption{Price-path measures after the 60-s reference time for tokens with at least three (upper panel) or five (lower panel) known group members versus inverse-probability-weighted controls; the corresponding results for at least two members are in the main text. Differences in percentage points for shares, otherwise in the unit of the measure; 95\% bootstrap intervals over groups.}\label{tab:h5matchedall}
\begin{tabular}{p{153pt}rrr}
\toprule
\multicolumn{4}{l}{\emph{Tokens with at least 3 known members}} \\
\midrule
Measure & Group tokens & Weighted controls & Difference [95\% CI] \\
\midrule
active after reference (share) & 88.2\,\% & 71.8\,\% & +16.4 [+13.8, +19.2] pp \\
peak multiple $\ge$ 2 (share) & 7.9\,\% & 3.1\,\% & +4.9 [+3.5, +6.3] pp \\
peak multiple $\ge$ 5 (share) & 1.3\,\% & 0.4\,\% & +0.9 [+0.1, +1.3] pp \\
peak multiple (price / reference price) & 1.40 & 1.12 & +0.28 [+0.17, +0.45] \\
time to peak (s) & 51.3 & 49.5 & +1.8 [$-$5.2, +7.8] \\
price 60 s after peak / peak & 0.84 & 0.91 & $-$0.07 [$-$0.09, $-$0.06] \\
price at 10 min / reference price & 0.99 & 0.95 & +0.03 [$-$0.05, +0.16] \\
\bottomrule
\end{tabular}
\\[0.8ex]
\begin{tabular}{p{153pt}rrr}
\toprule
\multicolumn{4}{l}{\emph{Tokens with at least 5 known members}} \\
\midrule
Measure & Group tokens & Weighted controls & Difference [95\% CI] \\
\midrule
active after reference (share) & 88.9\,\% & 71.9\,\% & +17.0 [+13.6, +20.8] pp \\
peak multiple $\ge$ 2 (share) & 8.8\,\% & 3.5\,\% & +5.4 [+3.5, +7.2] pp \\
peak multiple $\ge$ 5 (share) & 1.4\,\% & 0.5\,\% & +0.9 [+0.1, +1.4] pp \\
peak multiple (price / reference price) & 1.45 & 1.14 & +0.31 [+0.18, +0.57] \\
time to peak (s) & 53.3 & 46.5 & +6.8 [$-$3.3, +14.8] \\
price 60 s after peak / peak & 0.82 & 0.90 & $-$0.07 [$-$0.09, $-$0.05] \\
price at 10 min / reference price & 1.00 & 0.95 & +0.05 [$-$0.06, +0.23] \\
\bottomrule
\end{tabular}
\end{table}

\FloatBarrier
\subsection{Execution model and learned rankings}
\begin{table}[htbp]
\centering\small
\setlength{\tabcolsep}{3pt}
\caption{Check of the constant-product quote arithmetic against observed trades: relative error between the amount implied by the pre-trade reserves and the observed amount (SOL amounts are net of the protocol fee as recorded in the event).}\label{tab:quotecheck}
\begin{tabular}{lrrrr}
\toprule
Direction & Trades & Median relative error & 99th percentile & Within 10$^{-}$$^{4}$ \\
\midrule
buy: tokens out & 7\,739\,764 & 1.00$\times$10$^{-}$$^{9}$ & 1.27$\times$10$^{-}$$^{5}$ & 99.63\,\% \\
sell: SOL out & 7\,088\,067 & 7.29$\times$10$^{-}$$^{9}$ & 9.79$\times$10$^{-}$$^{4}$ & 98.07\,\% \\
\bottomrule
\end{tabular}
\end{table}

\begin{table}[htbp]
\centering\footnotesize
\setlength{\tabcolsep}{3pt}
\caption{Gradient-boosting ranking of entry moments: validation and out-of-sample (last 30\% in time) mean net return of the top-ranked decisions (60-s hold). Label y\_p: project cost model; y\_e: event fees and 0.3\% slippage. Intervals: bootstrap over tokens.}\label{tab:ml}
\begin{tabular}{llrrrr}
\toprule
Label & Features & Top share & Validation mean & Test decisions & Test mean [95\% CI] \\
\midrule
y\_p60 & flow & 5\,\% & $-$0.056 & 5\,267 & $-$0.058 [$-$0.069, $-$0.046] \\
y\_p60 & flow & 1\,\% & $-$0.051 & 1\,263 & $-$0.056 [$-$0.084, $-$0.024] \\
y\_p60 & flow+reputation & 5\,\% & $-$0.058 & 4\,989 & $-$0.051 [$-$0.067, $-$0.034] \\
y\_p60 & flow+reputation & 1\,\% & $-$0.031 & 1\,010 & $-$0.039 [$-$0.087, +0.007] \\
y\_e60 & flow & 5\,\% & $-$0.020 & 5\,241 & $-$0.021 [$-$0.034, $-$0.009] \\
y\_e60 & flow & 1\,\% & $-$0.034 & 1\,257 & $-$0.012 [$-$0.045, +0.024] \\
y\_e60 & flow+reputation & 5\,\% & $-$0.024 & 4\,995 & $-$0.016 [$-$0.034, +0.002] \\
y\_e60 & flow+reputation & 1\,\% & +0.008 & 1\,075 & +0.003 [$-$0.043, +0.051] \\
\bottomrule
\end{tabular}
\end{table}

\begin{table}[htbp]
\centering\small
\setlength{\tabcolsep}{3pt}
\caption{Hyperparameters of the gradient-boosting model (LightGBM) and of the decision set.}\label{tab:mlhp}
\begin{tabular}{lr}
\toprule
Parameter & Value \\
\midrule
n\_estimators & 400 \\
learning\_rate & 0.03 \\
num\_leaves & 31 \\
min\_child\_samples & 300 \\
subsample & 0.8 \\
subsample\_freq & 1 \\
colsample\_bytree & 0.8 \\
reg\_lambda & 10 \\
early stopping rounds & 40 \\
objective & squared error \\
label & net return of a 0.05 SOL position, 60-s hold, latency 1 s \\
decision ages (s) & 4, 8, 15, 30, 60, 120, 240 \\
\bottomrule
\end{tabular}
\end{table}

\begin{figure}[tbp]
\centering
\includegraphics[width=\linewidth]{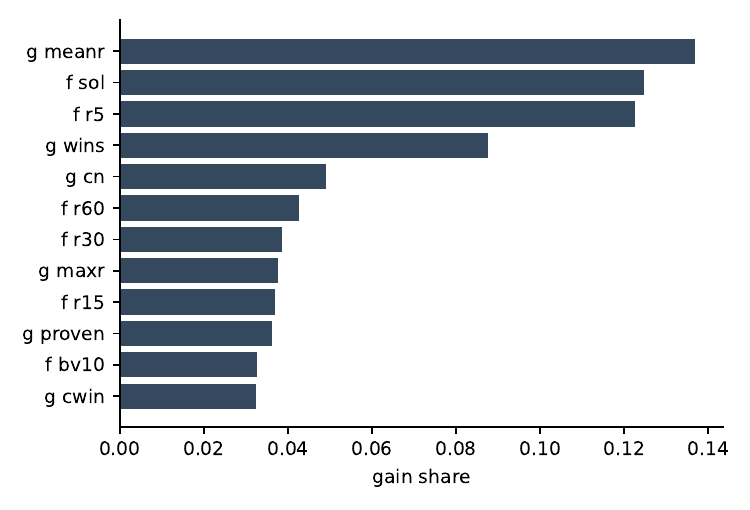}
\caption{Gain share of the features of the entry-ranking model (flow plus reputation features, project cost label, 60-s hold).}\label{fig:mlimportance}
\end{figure}

\FloatBarrier
\subsection{Teams}
\begin{table}[htbp]
\centering\small
\setlength{\tabcolsep}{3pt}
\caption{Cards of three sniper communities (tokens with at least three members among the buyers of the first 5 s): wallets, tokens, distinct creators, tokens per hour, median first-buy lag (s), median members per token, share of tokens reaching 20 and 60 SOL and graduating, members' total cash flow and liquidation-basis result (SOL), rotation (new share and retention across 12-hour blocks).}\label{tab:cards}
\begin{tabular}{lrrr}
\toprule
 & 4 & 7 & 3 \\
\midrule
Wallets & 45 & 38 & 76 \\
Tokens & 1\,656 & 886 & 5\,217 \\
Distinct creators & 1\,145 & 567 & 2\,973 \\
Tokens per hour & 19.7 & 10.1 & 59.2 \\
Median first-buy lag (s) & 0.0 & 1.0 & 1.0 \\
Median members per token & 4.0 & 3.0 & 5.0 \\
Reach 20 & 57.8\,\% & 67.5\,\% & 39.1\,\% \\
Reach 60 & 14.1\,\% & 20.2\,\% & 8.1\,\% \\
Graduation & 7.4\,\% & 10.8\,\% & 5.1\,\% \\
Cash flow (SOL) & 345.0 & 1802.3 & 1689.8 \\
Liquidation basis (SOL) & 547.6 & 2166.4 & 1865.8 \\
Rotation: new share & 0.3\,\% & 14.3\,\% & 2.6\,\% \\
Rotation: retention & 80.8\,\% & 95.4\,\% & 92.1\,\% \\
\bottomrule
\end{tabular}
\end{table}

\begin{table}[htbp]
\centering\footnotesize
\setlength{\tabcolsep}{3pt}
\caption{Sniper communities versus weighted controls (no member of any candidate community or large team among first-minute buyers): reach-20 shares, difference with 95\% interval, within-creator difference, differences in reach-60 and graduation shares, and the largest absolute standardised mean difference of the covariates after weighting (values above 0.5 mean poor balance).}\label{tab:candcontrols}
\begin{tabular}{rrrrrrrrrr}
\toprule
Comm. & Wallets & Tokens & Team & Ctrl. & Reach-20 diff., pp [95\% CI] & \shortstack{Within \\[0pt] creator} & \shortstack{Reach-60 \\[0pt] diff.} & \shortstack{Grad. \\[0pt] diff.} & \shortstack{Max. \\[0pt] SMD} \\
\midrule
0 & 294 & 10\,099 & 46.7\,\% & 27.6\,\% & +19.1 [+13.7, +23.9] & +20.2 & $-$16.2 & $-$20.4 & 0.93 \\
3 & 76 & 5\,217 & 39.1\,\% & 16.9\,\% & +22.2 [+19.4, +24.5] & +11.1 & $-$8.1 & $-$11.2 & 0.49 \\
2 & 85 & 4\,368 & 21.9\,\% & 20.0\,\% & +1.9 [$-$1.7, +5.1] & $-$0.0 & $-$15.8 & $-$17.4 & 1.24 \\
1 & 133 & 3\,426 & 0.8\,\% & 0.6\,\% & +0.2 [$-$0.1, +0.5] & +0.6 & $-$0.3 & +2.5 & 0.34 \\
5 & 44 & 2\,119 & 16.6\,\% & 12.4\,\% & +4.2 [+1.3, +6.8] & +3.2 & $-$9.7 & $-$10.4 & 1.59 \\
4 & 45 & 1\,642 & 57.8\,\% & 16.7\,\% & +41.1 [+37.3, +45.4] & +31.2 & $-$1.9 & $-$8.5 & 0.14 \\
6 & 41 & 957 & 33.1\,\% & 11.5\,\% & +21.7 [+14.3, +27.4] & +13.7 & $-$5.7 & $-$7.2 & 2.38 \\
22 & 7 & 941 & 6.4\,\% & 8.8\,\% & $-$2.4 [$-$3.9, $-$0.7] & $-$0.7 & $-$6.8 & $-$7.5 & 0.06 \\
9 & 29 & 913 & 34.5\,\% & 9.4\,\% & +25.1 [+13.0, +33.0] &  & +1.5 & $-$0.6 & 5.04 \\
7 & 38 & 870 & 67.6\,\% & 17.7\,\% & +49.9 [+46.2, +53.6] & +25.7 & +3.3 & $-$6.2 & 0.54 \\
8 & 35 & 777 & 44.5\,\% & 24.3\,\% & +20.3 [+11.2, +29.6] & +6.0 & $-$14.6 & $-$18.7 & 1.43 \\
13 & 18 & 595 & 19.3\,\% & 11.0\,\% & +8.3 [+5.3, +11.4] & $-$8.4 & $-$6.8 & $-$8.0 & 0.15 \\
10 & 28 & 475 & 19.4\,\% & 21.2\,\% & $-$1.8 [$-$15.1, +7.0] & +4.3 & $-$18.1 & $-$19.0 & 1.24 \\
\bottomrule
\end{tabular}
\end{table}

\begin{table}[htbp]
\centering\small
\setlength{\tabcolsep}{3pt}
\caption{Selection of teams by their own first-half success (share of first-half signal tokens reaching 20 SOL, at least 20 tokens), evaluated on second-half tokens: pipeline groups merged by overlap and fine clusters of the in-block adjacency graph (Louvain, resolution 10); threshold 0 = no selection.}\label{tab:repoot}
\begin{tabular}{p{147pt}rrrrrr}
\toprule
Method & \shortstack{Min. first-half \\[0pt] reach 20} & Teams & \shortstack{Signal \\[0pt] tokens} & Reach 20 & Lift 20 & Reach 60 \\
\midrule
pipeline groups merged by overlap ($\ge$10 wallets) & 0\,\% & 97 & 7\,105 & 36.1\,\% & 4.09 & 7.7\,\% \\
pipeline groups merged by overlap ($\ge$10 wallets) & 40\,\% & 31 & 3\,973 & 53.3\,\% & 6.05 & 12.5\,\% \\
adjacency Louvain res 10 ($\ge$5 wallets) & 0\,\% & 52 & 8\,962 & 32.4\,\% & 3.68 & 7.1\,\% \\
adjacency Louvain res 10 ($\ge$5 wallets) & 60\,\% & 11 & 1\,933 & 66.6\,\% & 7.56 & 16.0\,\% \\
\bottomrule
\end{tabular}
\end{table}

\begin{table}[htbp]
\centering\scriptsize
\setlength{\tabcolsep}{3pt}
\caption{Scout test: reach-20 share of second-half team tokens in which a first-half scout (top quartile by first-half success as first mover) enters first, against tokens where another member enters first; difference and one-sided permutation p-value (first-mover labels shuffled within hour).}\label{tab:h7roles}
\begin{tabular}{lrrrrrrrr}
\toprule
Team & Wallets & Tokens & \shortstack{Scout \\[0pt] candidates} & \shortstack{Top 10\% wallets: \\[0pt] share of first-mover tokens} & \shortstack{Reach 20, \\[0pt] scout first} & \shortstack{Reach 20, \\[0pt] other first} & Diff., pp & Perm. p \\
\midrule
adjacency sub-team 0 & 287 & 9\,166 & 54 & 60\,\% & 60.1\,\% & 32.1\,\% & +27.0 & 0.000 \\
adjacency sub-team 2 & 73 & 1\,824 & 9 & 74\,\% & 77.1\,\% & 31.5\,\% & +37.5 & 0.000 \\
adjacency sub-team 4 & 68 & 5\,161 & 40 & 38\,\% & 66.7\,\% & 49.9\,\% & +16.4 & 0.000 \\
adjacency sub-team 5 & 61 & 5\,281 & 30 & 49\,\% & 64.0\,\% & 43.1\,\% & +18.9 & 0.000 \\
adjacency sub-team 7 & 49 & 1\,757 & 16 & 39\,\% & 58.0\,\% & 53.9\,\% & +5.8 & 0.357 \\
adjacency sub-team 9 & 24 & 1\,523 & 12 & 36\,\% & 33.3\,\% & 35.4\,\% & $-$1.1 & 0.467 \\
sniper community 4 & 45 & 1\,656 & 13 & 37\,\% & 75.9\,\% & 34.5\,\% & +37.6 & 0.000 \\
sniper community 7 & 38 & 886 & 9 & 32\,\% & 80.3\,\% & 53.9\,\% & +16.8 & 0.000 \\
sniper community 3 & 76 & 5\,217 & 23 & 41\,\% & 58.2\,\% & 35.3\,\% & +20.9 & 0.000 \\
\bottomrule
\end{tabular}
\end{table}

\begin{table}[htbp]
\centering\small
\setlength{\tabcolsep}{3pt}
\caption{Stability of the soft-reputation rules. Panel A: range of the number of flagged tokens and of their reach-20 share over eight seeds of the clustering (one seed gives a borderline cluster that the hard reputation threshold rejects). Panel B: choice of the shrinkage strength by mean precision on a validation block that excludes the second half.}\label{tab:softprior}
\begin{tabular}{p{166pt}rrrr}
\toprule
\multicolumn{5}{l}{\emph{A. Stability over eight random seeds of the clustering}} \\
\midrule
Selection & Tokens, min & Tokens, max & Reach 20, min & Reach 20, max \\
\midrule
S4 agreement score: score $\ge$ 0.5 & 208 & 221 & 78.4\,\% & 79.2\,\% \\
S4 agreement score: top 1000 & 1\,000 & 1\,000 & 70.7\,\% & 74.6\,\% \\
S4 agreement score: top 1500 & 1\,500 & 1\,500 & 70.2\,\% & 72.6\,\% \\
S4 agreement score: top 500 & 500 & 500 & 73.0\,\% & 76.4\,\% \\
S5 agreement score, scout first: top 1000 & 1\,000 & 1\,000 & 73.8\,\% & 75.2\,\% \\
hard R4 (kept teams) & 369 & 1\,248 & 74.0\,\% & 78.6\,\% \\
\bottomrule
\end{tabular}
\\[0.8ex]
\begin{tabular}{rrrrr}
\toprule
\multicolumn{5}{l}{\emph{B. Strength of the shrinkage prior of the soft reputation}} \\
\midrule
Prior (pseudo-tokens) & Reach 20 (mean over seeds) & min & max & Validation precision \\
\midrule
0 & 67.2\,\% & 66.8\,\% & 68.1\,\% & 65.1\,\% \\
2 & 68.1\,\% & 67.2\,\% & 70.0\,\% & 65.2\,\% \\
5 & 69.4\,\% & 69.1\,\% & 70.0\,\% & 68.1\,\% \\
10 & 69.3\,\% & 68.5\,\% & 70.9\,\% & 68.3\,\% \\
20 & 71.5\,\% & 70.7\,\% & 71.9\,\% & 67.9\,\% \\
50 & 74.7\,\% & 72.7\,\% & 75.7\,\% & 67.5\,\% \\
100 & 72.8\,\% & 71.8\,\% & 73.3\,\% & 66.2\,\% \\
\bottomrule
\end{tabular}
\end{table}

\begin{table}[htbp]
\centering\small
\setlength{\tabcolsep}{3pt}
\caption{Who enters first when two sniper communities both have at least three members among the buyers of the first 10 s: share of tokens led by the first community, median gap (s) between the two first buys, share where the second community's first buy follows the third member of the first (reaction) and vice versa, and the reach-20 / reach-60 shares by leader.}\label{tab:leadlag}
\begin{tabular}{llrrrrr}
\toprule
\multicolumn{7}{l}{\emph{A. Timing}} \\
\midrule
A & B & Tokens & A leads & \shortstack{Median gap \\[0pt] B$-$A (s)} & \shortstack{B first after \\[0pt] A's third} & \shortstack{A first after \\[0pt] B's third} \\
\midrule
c0 & c3 & 4\,337 & 62\,\% & 0.0 & 51\,\% & 33\,\% \\
c0 & c2 & 2\,135 & 17\,\% & 0.0 & 18\,\% & 71\,\% \\
c0 & c4 & 1\,251 & 60\,\% & 0.0 & 47\,\% & 37\,\% \\
c3 & c4 & 662 & 44\,\% & 0.0 & 35\,\% & 40\,\% \\
c2 & c3 & 788 & 90\,\% & 1.0 & 69\,\% & 21\,\% \\
c0 & c7 & 928 & 67\,\% & 0.0 & 57\,\% & 31\,\% \\
c3 & c7 & 511 & 60\,\% & 0.0 & 48\,\% & 30\,\% \\
c4 & c7 & 251 & 61\,\% & 0.0 & 41\,\% & 30\,\% \\
\bottomrule
\end{tabular}
\\[0.8ex]
\begin{tabular}{llrrrrrrr}
\toprule
\multicolumn{9}{l}{\emph{B. Outcomes}} \\
\midrule
A & B & \shortstack{Reach 20, \\[0pt] all} & A leads & B leads & \shortstack{Reach 60, \\[0pt] A leads} & B leads & \shortstack{Tokens, \\[0pt] A leads} & B leads \\
\midrule
c0 & c3 & 55.2\,\% & 59.9\,\% & 47.3\,\% & 12.3\,\% & 10.3\,\% & 2\,701 & 1\,636 \\
c0 & c2 & 39.3\,\% & 41.6\,\% & 38.9\,\% & 6.8\,\% & 6.5\,\% & 370 & 1\,765 \\
c0 & c4 & 75.1\,\% & 84.0\,\% & 61.8\,\% & 25.4\,\% & 14.6\,\% & 751 & 500 \\
c3 & c4 & 77.9\,\% & 86.3\,\% & 71.4\,\% & 26.7\,\% & 20.3\,\% & 292 & 370 \\
c2 & c3 & 44.9\,\% & 45.5\,\% & 39.5\,\% & 9.2\,\% & 8.6\,\% & 707 & 81 \\
c0 & c7 & 77.3\,\% & 80.0\,\% & 71.6\,\% & 25.1\,\% & 22.8\,\% & 625 & 303 \\
c3 & c7 & 77.9\,\% & 78.2\,\% & 77.5\,\% & 27.0\,\% & 17.2\,\% & 307 & 204 \\
c4 & c7 & 89.6\,\% & 86.4\,\% & 94.8\,\% & 30.5\,\% & 33.0\,\% & 154 & 97 \\
\bottomrule
\end{tabular}
\end{table}

\begin{table}[htbp]
\centering\footnotesize
\setlength{\tabcolsep}{3pt}
\caption{Creator, bundle and snipers. Panel A: presence of three sniper communities (at least three members among the buyers of the first 5 s) by the creator's initial buy. Panel B: reach-20 share with and without c0/c3 snipers within strata of the creator's initial buy (stratum 0 is the lowest).}\label{tab:chaincreator}
\begin{tabular}{lrrrrr}
\toprule
\multicolumn{6}{l}{\emph{A. Presence of the communities c2, c0 and c3 by the creator's initial buy}} \\
\midrule
Creator's initial buy (SOL) & Tokens & Share with c2 & Share with c0 & Share with c3 & Reach 20 \\
\midrule
none & 6\,413 & 5.4\,\% & 9.1\,\% & 1.4\,\% & 5.3\,\% \\
0-0.5 & 73\,138 & 0.9\,\% & 2.3\,\% & 1.0\,\% & 2.1\,\% \\
0.5-1 & 16\,989 & 5.1\,\% & 11.9\,\% & 5.1\,\% & 13.4\,\% \\
1-2 & 13\,707 & 9.6\,\% & 15.7\,\% & 7.7\,\% & 10.6\,\% \\
2-4 & 11\,995 & 8.9\,\% & 29.7\,\% & 15.5\,\% & 17.8\,\% \\
$>$4 & 10\,221 & 6.1\,\% & 26.9\,\% & 16.0\,\% & 40.5\,\% \\
\bottomrule
\end{tabular}
\\[0.8ex]
\begin{tabular}{rrrrr}
\toprule
\multicolumn{5}{l}{\emph{B. Reach-20 share with and without c0/c3 snipers within strata of the initial buy}} \\
\midrule
Initial-buy stratum & Tokens & \shortstack{With c0/c3 \\[0pt] snipers} & \shortstack{Reach 20, \\[0pt] with} & \shortstack{Reach 20, \\[0pt] without} \\
\midrule
0 & 6\,413 & 611 & 30.0\,\% & 2.7\,\% \\
1 & 63\,701 & 1\,540 & 36.4\,\% & 1.0\,\% \\
2 & 9\,437 & 538 & 36.4\,\% & 1.6\,\% \\
3 & 16\,989 & 2\,501 & 37.9\,\% & 9.2\,\% \\
4 & 13\,707 & 2\,697 & 31.7\,\% & 5.4\,\% \\
5 & 6\,696 & 1\,951 & 35.0\,\% & 8.0\,\% \\
6 & 8\,192 & 3\,033 & 45.3\,\% & 10.2\,\% \\
7 & 3\,573 & 1\,400 & 53.7\,\% & 8.7\,\% \\
8 & 3\,755 & 692 & 78.9\,\% & 59.8\,\% \\
\bottomrule
\end{tabular}
\end{table}

\begin{table}[htbp]
\centering\small
\setlength{\tabcolsep}{3pt}
\caption{Do group features add information beyond the first minute? Prospective outcome (token still below 20 SOL at 60 s reaches 20 SOL later), last block of the window, models fitted on the preceding block. First-minute features: curve SOL, trades, buyers, sellers, buy volume, share of the largest buyer, creator's initial buy. Group features: members of the best merged team and adjacency cluster, shrunk reputations, agreement, scout first. Adding the group features changes AP by -0.015 (hour-block bootstrap interval -0.030 to +0.004) and AUC by 0.000.}\label{tab:incremental}
\begin{tabular}{lrrrrr}
\toprule
Model (logistic regression) & AP & AUC & P@100 & P@500 & P@1000 \\
\midrule
curve SOL at 60 s only & 0.364 & 0.939 & 65.0\,\% & 41.6\,\% & 31.5\,\% \\
first-minute features (no group information) & 0.372 & 0.938 & 61.0\,\% & 42.6\,\% & 31.2\,\% \\
first-minute features and group features & 0.357 & 0.938 & 59.0\,\% & 43.0\,\% & 31.0\,\% \\
group features only & 0.099 & 0.793 & 21.0\,\% & 18.4\,\% & 14.3\,\% \\
\bottomrule
\end{tabular}
\end{table}

\begin{table}[htbp]
\centering\footnotesize
\setlength{\tabcolsep}{3pt}
\caption{Prospective outcome within quintiles of the curve SOL at 60 s (last block of the window, tokens still below 20 SOL at 60 s). Within the lowest four quintiles almost no token reaches 20 SOL later whatever the signal; the contrast is concentrated in the top quintile, where the curve is already at a higher level at 60 s.}\label{tab:strata}
\begin{tabular}{rrlrrrr}
\toprule
\shortstack{Curve-SOL \\[0pt] quintile} & \shortstack{Median curve \\[0pt] SOL at 60 s} & Signal & \shortstack{Tokens with \\[0pt] signal} & \shortstack{Tokens \\[0pt] without} & \shortstack{Later reach 20, \\[0pt] with signal} & \shortstack{Later reach 20, \\[0pt] without} \\
\midrule
1 & 0.00 & any team signal & 71 & 6\,775 & 0.0\,\% & 0.0\,\% \\
1 & 0.00 & agreement rule R4 & 0 & 6\,846 &  & 0.0\,\% \\
2 & 0.00 & any team signal & 683 & 6\,027 & 0.3\,\% & 0.2\,\% \\
2 & 0.00 & agreement rule R4 & 7 & 6\,703 & 0.0\,\% & 0.2\,\% \\
3 & 0.05 & any team signal & 955 & 5\,774 & 0.2\,\% & 0.2\,\% \\
3 & 0.05 & agreement rule R4 & 7 & 6\,722 & 0.0\,\% & 0.2\,\% \\
4 & 0.36 & any team signal & 796 & 5\,966 & 0.4\,\% & 0.3\,\% \\
4 & 0.36 & agreement rule R4 & 13 & 6\,749 & 7.7\,\% & 0.3\,\% \\
5 & 1.99 & any team signal & 2\,652 & 4\,110 & 12.4\,\% & 4.1\,\% \\
5 & 1.99 & agreement rule R4 & 113 & 6\,649 & 27.4\,\% & 7.0\,\% \\
\bottomrule
\end{tabular}
\end{table}

\FloatBarrier
\subsection{Concealment and cash-out}
\begin{table}[htbp]
\centering\small
\setlength{\tabcolsep}{3pt}
\caption{Concealment and cash-out profile of the studied teams (native SOL transfers outside protocol transactions). A: share of the creators of the team's tokens that are themselves team wallets (tokens with at least three team wallets among the buyers of the first minute). B: team wallets with at least 20 early buys and visible inbound SOL. C: external SOL in and out, concentration of funders and destinations. D: class of destinations by number of distinct senders (hub, mid-sized, specific). E: share of wallets inactive at the end of the window, newcomers, and retirement after sweeping.}\label{tab:teamprofile}
\begin{tabular}{lrrrrrrr}
\toprule
\multicolumn{8}{l}{\emph{A. Creators}} \\
\midrule
 & team126 & c4 & c7 & c3 & c0 & c1 & c2 \\
\midrule
Tokens & 5\,635 & 1\,656 & 886 & 5\,217 & 10\,099 & 3\,445 & 4\,405 \\
Distinct creators & 3\,208 & 1\,145 & 567 & 2\,973 & 5\,580 & 217 & 2\,454 \\
Creators linked to a team wallet & 0.03\,\% & 0.44\,\% & 0.18\,\% & 0.00\,\% & 0.07\,\% & 5.99\,\% & 0.37\,\% \\
Same, controls & 0.04\,\% & 0.61\,\% & 0.00\,\% & 0.00\,\% & 0.01\,\% & 0.31\,\% & 0.01\,\% \\
Tokens with linked creator & 0.04\,\% & 0.30\,\% & 7.90\,\% & 0.00\,\% & 0.10\,\% & 19.71\,\% & 3.81\,\% \\
Reach 20, unlinked tokens & 50.9\,\% & 57.7\,\% & 69.1\,\% & 39.1\,\% & 46.7\,\% & 1.0\,\% & 21.9\,\% \\
\bottomrule
\end{tabular}
\\[0.8ex]
\begin{tabular}{p{180pt}rrrrrrr}
\toprule
\multicolumn{8}{l}{\emph{B. Funding visibility}} \\
\midrule
 & team126 & c4 & c7 & c3 & c0 & c1 & c2 \\
\midrule
Wallets with at least 20 early buys & 113 & 45 & 38 & 76 & 294 & 133 & 85 \\
With any inbound SOL & 35\,\% & 53\,\% & 21\,\% & 46\,\% & 47\,\% & 92\,\% & 56\,\% \\
With inbound SOL before first early buy & 1.8\,\% & 0.0\,\% & 7.9\,\% & 10.5\,\% & 14.6\,\% & 45.1\,\% & 24.7\,\% \\
\bottomrule
\end{tabular}
\\[0.8ex]
\begin{tabular}{lrrrrrrr}
\toprule
\multicolumn{8}{l}{\emph{C. Funding and cash-out}} \\
\midrule
 & team126 & c4 & c7 & c3 & c0 & c1 & c2 \\
\midrule
External SOL in & 418 & 1\,251 & 65 & 559 & 2\,588 & 4\,914 & 2\,500 \\
Inflow from shared funders & 33\,\% & 0\,\% & 0\,\% & 1\,\% & 23\,\% & 30\,\% & 11\,\% \\
Funders & 108 & 526 & 10 & 50 & 300 & 285 & 117 \\
Top-5 funder share & 63\,\% & 30\,\% & 96\,\% & 53\,\% & 40\,\% & 45\,\% & 65\,\% \\
External SOL out & 2\,664 & 97\,440 & 5\,739 & 2\,303 & 16\,524 & 5\,360 & 6\,428 \\
Outflow to shared collectors & 7.0\,\% & 0.0\,\% & 0.0\,\% & 32.4\,\% & 1.9\,\% & 3.9\,\% & 3.9\,\% \\
Destinations & 119 & 24\,418 & 549 & 114 & 331 & 368 & 299 \\
Top-5 destination share & 63\,\% & 6\,\% & 24\,\% & 58\,\% & 60\,\% & 54\,\% & 42\,\% \\
Internal share of outflow & 12.4\,\% & 0.1\,\% & 7.9\,\% & 1.5\,\% & 0.3\,\% & 1.9\,\% & 1.1\,\% \\
Out/in ratio & 6.4 & 77.9 & 88.7 & 4.1 & 6.4 & 1.1 & 2.6 \\
\bottomrule
\end{tabular}
\\[0.8ex]
\begin{tabular}{lrrrrrrr}
\toprule
\multicolumn{8}{l}{\emph{D. Class of cash-out destinations}} \\
\midrule
 & team126 & c4 & c7 & c3 & c0 & c1 & c2 \\
\midrule
Destinations: hubs & 1.3\,\% & 0.1\,\% & 13.1\,\% & 0.4\,\% & 1.5\,\% & 48.8\,\% & 3.5\,\% \\
Destinations: mid-sized & 1.7\,\% & 0.0\,\% & 0.0\,\% & 0.0\,\% & 1.1\,\% & 1.1\,\% & 2.2\,\% \\
Destinations: specific & 97.0\,\% & 42.1\,\% & 81.7\,\% & 99.5\,\% & 97.4\,\% & 49.8\,\% & 94.3\,\% \\
Destinations: unknown & 0.0\,\% & 57.8\,\% & 5.2\,\% & 0.0\,\% & 0.0\,\% & 0.3\,\% & 0.0\,\% \\
\bottomrule
\end{tabular}
\\[0.8ex]
\begin{tabular}{lrrrrrrr}
\toprule
\multicolumn{8}{l}{\emph{E. Wallet lifetimes}} \\
\midrule
 & team126 & c4 & c7 & c3 & c0 & c1 & c2 \\
\midrule
Wallets retired & 19\,\% & 49\,\% & 3\,\% & 11\,\% & 13\,\% & 29\,\% & 8\,\% \\
Newcomers & 1.8\,\% & 0.0\,\% & 63.2\,\% & 10.5\,\% & 12.6\,\% & 24.1\,\% & 14.1\,\% \\
Retired, given a sweep & 50\,\% & 56\,\% & 5\,\% & 18\,\% & 38\,\% & 37\,\% & 3\,\% \\
Retired, no sweep & 15\,\% & 22\,\% & 0\,\% & 8\,\% & 10\,\% & 26\,\% & 12\,\% \\
Wallets that sweep & 11\,\% & 80\,\% & 50\,\% & 22\,\% & 13\,\% & 23\,\% & 44\,\% \\
Median active hours & 87 & 68 & 60 & 87 & 85 & 67 & 80 \\
\bottomrule
\end{tabular}
\end{table}

\begin{table}[htbp]
\centering\footnotesize
\setlength{\tabcolsep}{3pt}
\caption{Second hop of the cash-out chains: first-hop addresses (non-hub destinations of a team's own wallets) with SOL received, SOL they forward in the window, ratio, number of second-hop destinations and the shares of the forwarded SOL going to hubs, mid and specific addresses (by number of distinct senders in the whole window).}\label{tab:chain2}
\begin{tabular}{lrrrrrrrrr}
\toprule
Team & \shortstack{First-hop \\[0pt] addresses} & \shortstack{First-hop \\[0pt] SOL} & \shortstack{Forwarded \\[0pt] SOL} & \shortstack{Forwarded \\[0pt] ratio} & \shortstack{Second-hop \\[0pt] destinations} & Hub & Mid & Specific & Unknown \\
\midrule
team126 & 48 & 1\,073 & 352 & 0.33 & 114 & 17\,\% & 20\,\% & 63\,\% & 0\,\% \\
c3 & 63 & 1\,958 & 4\,536 & 2.32 & 1\,494 & 38\,\% & 15\,\% & 45\,\% & 1\,\% \\
c0 & 228 & 15\,029 & 15\,179 & 1.01 & 1\,094 & 15\,\% & 12\,\% & 72\,\% & 1\,\% \\
c7 & 298 & 4\,684 & 270 & 0.06 & 3 & 85\,\% & 0\,\% & 15\,\% & 0\,\% \\
c2 & 245 & 6\,187 & 4\,758 & 0.77 & 416 & 18\,\% & 3\,\% & 78\,\% & 1\,\% \\
c1 & 275 & 2\,714 & 9\,495 & 3.50 & 1\,492 & 18\,\% & 13\,\% & 66\,\% & 3\,\% \\
\bottomrule
\end{tabular}
\end{table}

\begin{table}[htbp]
\centering\small
\setlength{\tabcolsep}{3pt}
\caption{Overlap of the cash-out destinations of pairs of teams. Direct destinations are computed on disjoint wallet sets (wallets of each team that are not in the other team); the columns hub, mid and specific split the shared destinations by their number of distinct senders. The last three columns refer to the first and second hop of the cash-out chains; a blank cell means that the pair was not part of that computation.}\label{tab:pairs}
\begin{tabular}{llrrrrrrrrr}
\toprule
A & B & \shortstack{Wallets \\[0pt] A only} & \shortstack{Wallets \\[0pt] B only} & \shortstack{Shared \\[0pt] destin.} & hub & mid & specific & \shortstack{Shared \\[0pt] first hop} & \shortstack{Shared \\[0pt] second hop} & \shortstack{of which \\[0pt] specific} \\
\midrule
c0 & c1 & 294 & 133 & 10 & 9 & 1 & 0 & 0 & 63 & 0 \\
c0 & c2 & 294 & 85 & 13 & 12 & 1 & 0 & 1 & 45 & 20 \\
c0 & c3 & 294 & 76 & 8 & 2 & 2 & 4 & 4 & 85 & 7 \\
c0 & c4 & 294 & 45 & 3 & 2 & 1 & 0 &  &  &  \\
c0 & c7 & 294 & 38 & 1 & 1 & 0 & 0 & 0 & 2 & 0 \\
c0 & team126 & 262 & 94 & 6 & 4 & 1 & 1 & 0 & 3 & 0 \\
c1 & c2 & 133 & 85 & 9 & 8 & 1 & 0 & 1 & 18 & 1 \\
c1 & c3 & 133 & 76 & 1 & 1 & 0 & 0 & 0 & 37 & 1 \\
c1 & c4 & 133 & 45 & 1 & 1 & 0 & 0 &  &  &  \\
c1 & c7 & 133 & 38 & 1 & 1 & 0 & 0 & 0 & 1 & 0 \\
c1 & team126 & 133 & 126 & 1 & 1 & 0 & 0 & 0 & 3 & 0 \\
c2 & c3 & 85 & 76 & 1 & 1 & 0 & 0 & 0 & 16 & 1 \\
c2 & c4 & 85 & 45 & 4 & 3 & 1 & 0 &  &  &  \\
c2 & c7 & 85 & 38 & 1 & 1 & 0 & 0 & 0 & 2 & 0 \\
c2 & team126 & 85 & 126 & 1 & 1 & 0 & 0 & 0 & 2 & 0 \\
c3 & c4 & 76 & 45 & 1 & 0 & 0 & 1 &  &  &  \\
c3 & c7 & 76 & 38 & 0 & 0 & 0 & 0 & 0 & 2 & 0 \\
c3 & team126 & 62 & 112 & 1 & 1 & 0 & 0 & 0 & 5 & 0 \\
c4 & c7 & 45 & 38 & 0 & 0 & 0 & 0 &  &  &  \\
c4 & team126 & 45 & 126 & 0 & 0 & 0 & 0 &  &  &  \\
c7 & team126 & 35 & 123 & 0 & 0 & 0 & 0 & 0 & 0 & 0 \\
\bottomrule
\end{tabular}
\end{table}

\FloatBarrier
\subsection{Robustness}
\begin{table}[htbp]
\centering\scriptsize
\setlength{\tabcolsep}{3pt}
\caption{Robustness of group discovery and of the follower result. Panel A varies one parameter at a time on a 3-hour slice; Panel B removes evidence families on a 12-hour slice; Panel C varies the depth of the funding search. Lift 20: reach-20 share of buyers-only signals relative to the base rate. Follower return: mean net return of the follower of Section~\ref{sec:followers} with a bootstrap interval over groups. Figure~\ref{fig:robust} plots the lift and the follower return.}\label{tab:robust}
\begin{tabular}{lrrrrrrr}
\toprule
\multicolumn{8}{l}{\emph{A. One-factor sensitivity of group discovery (3-hour slice; baseline reach 20 of 0.062)}} \\
\midrule
Setting & Groups & Edges & \shortstack{Signal \\[0pt] share} & \shortstack{Buyers-only \\[0pt] signals} & Lift 20 & \shortstack{Edge \\[0pt] Jaccard} & Follower return [95\% CI] \\
\midrule
default (hubs 100) & 1\,157 & 8\,354 & 41.5\,\% & 861 & 4.32 & 1.00 & -0.039 [-0.057, -0.022] \\
shared tokens 2 & 1\,676 & 15\,783 & 42.1\,\% & 918 & 4.10 & 0.53 & -0.037 [-0.054, -0.018] \\
shared tokens 5 & 874 & 4\,869 & 39.6\,\% & 690 & 5.27 & 0.58 & -0.042 [-0.063, -0.022] \\
Jaccard 0.15 & 1\,274 & 13\,325 & 42.3\,\% & 951 & 4.03 & 0.63 & -0.040 [-0.053, -0.021] \\
Jaccard 0.5 & 1\,076 & 6\,765 & 41.2\,\% & 810 & 4.47 & 0.81 & -0.038 [-0.055, -0.024] \\
overlap lift 2 & 1\,157 & 8\,354 & 41.5\,\% & 861 & 4.32 & 1.00 & -0.039 [-0.057, -0.022] \\
overlap lift 5 & 1\,157 & 8\,354 & 41.5\,\% & 861 & 4.32 & 1.00 & -0.039 [-0.057, -0.022] \\
co-buy window 3 s & 1\,101 & 6\,100 & 41.3\,\% & 820 & 4.44 & 0.73 & -0.038 [-0.057, -0.021] \\
co-buy window 30 s & 1\,223 & 11\,431 & 41.5\,\% & 859 & 4.31 & 0.73 & -0.037 [-0.054, -0.022] \\
early window 15 s & 924 & 6\,773 & 40.6\,\% & 774 & 4.55 & 0.71 & -0.042 [-0.062, -0.021] \\
early window 120 s & 1\,224 & 8\,716 & 41.5\,\% & 859 & 4.33 & 0.90 & -0.039 [-0.058, -0.020] \\
attribution buyers 1 & 1\,157 & 8\,354 & 47.3\,\% & 1\,346 & 2.92 & 1.00 & -0.037 [-0.051, -0.024] \\
attribution buyers 3 & 1\,157 & 8\,354 & 38.6\,\% & 610 & 5.59 & 1.00 & -0.041 [-0.066, -0.019] \\
evidence repeats 3 & 879 & 7\,980 & 36.9\,\% & 996 & 4.15 & 0.96 & -0.043 [-0.061, -0.025] \\
\bottomrule
\end{tabular}
\\[0.8ex]
\begin{tabular}{lrrrrrrrrr}
\toprule
\multicolumn{10}{l}{\emph{B. Ablation of evidence families (12-hour slice)}} \\
\midrule
Variant & Groups & Edges & \shortstack{Edge \\[0pt] Jaccard} & \shortstack{Edges \\[0pt] retained} & \shortstack{Signal \\[0pt] share} & \shortstack{Buyers-only \\[0pt] signals} & Lift 20 & \shortstack{Lift 20, \\[0pt] all} & \shortstack{Follower return \\[0pt] [95\% CI]} \\
\midrule
all channels & 2\,666 & 15\,107 & 1.00 & 100\,\% & 46.1\,\% & 2\,052 & 4.50 & 1.31 & -0.036 [-0.049, -0.023] \\
buyers only & 2\,601 & 13\,100 & 0.87 & 87\,\% & 46.1\,\% & 2\,048 & 4.55 & 1.31 & -0.036 [-0.046, -0.023] \\
buyers + sol & 2\,618 & 13\,923 & 0.92 & 92\,\% & 46.1\,\% & 2\,051 & 4.52 & 1.31 & -0.036 [-0.046, -0.022] \\
buyers + protocol & 2\,652 & 14\,287 & 0.95 & 95\,\% & 46.1\,\% & 2\,049 & 4.52 & 1.31 & -0.036 [-0.049, -0.023] \\
without buyers & 1\,740 & 4\,247 & 0.28 & 28\,\% & 40.4\,\% & 460 & 8.47 & 1.02 & -0.036 [-0.065, -0.010] \\
without sol & 2\,652 & 14\,287 & 0.95 & 95\,\% & 46.1\,\% & 2\,049 & 4.52 & 1.31 & -0.036 [-0.049, -0.023] \\
without protocol & 2\,618 & 13\,923 & 0.92 & 92\,\% & 46.1\,\% & 2\,051 & 4.52 & 1.31 & -0.036 [-0.046, -0.022] \\
\bottomrule
\end{tabular}
\\[0.8ex]
\begin{tabular}{rrrrrrrrr}
\toprule
\multicolumn{9}{l}{\emph{C. Funding search depth (12-hour slice; depth 3 is the default)}} \\
\midrule
Depth & Run time (s) & Groups & Edges & Campaigns & \shortstack{Consolidation \\[0pt] share} & \shortstack{Edge \\[0pt] Jaccard} & Lift 20 & Follower return [95\% CI] \\
\midrule
1 & 703 & 2\,668 & 14\,701 & 839 & 28.8\,\% & 0.97 & 4.51 & -0.036 [-0.048, -0.024] \\
2 & 760 & 2\,667 & 15\,093 & 3\,310 & 25.6\,\% & 1.00 & 4.50 & -0.036 [-0.049, -0.023] \\
3 & 767 & 2\,666 & 15\,107 & 4\,738 & 20.7\,\% & 1.00 & 4.50 & -0.036 [-0.049, -0.023] \\
4 & 876 & 2\,665 & 15\,109 & 5\,878 & 17.4\,\% & 1.00 & 4.50 & -0.036 [-0.049, -0.023] \\
5 & 1\,022 & 2\,665 & 15\,109 & 6\,882 & 15.3\,\% & 1.00 & 4.50 & -0.036 [-0.049, -0.023] \\
6 & 3\,424 & 2\,668 & 15\,114 & 7\,992 & 13.6\,\% & 1.00 & 4.50 & -0.036 [-0.047, -0.024] \\
\bottomrule
\end{tabular}
\end{table}

\FloatBarrier
\subsection{Further checks and the ledger of tests}
\begin{table}[htbp]
\centering\scriptsize
\setlength{\tabcolsep}{3pt}
\caption{Day-by-day replication (day 1 starts at the beginning of the window; days 1 and 4 are partial). Matched contrast: tokens with at least two known group members against weighted controls with covariates fixed at creation, difference in percentage points with interval over creators. Follower: mean net return per trade in percent (latency 1 s, take-profit 30\%, stop 30\%, 60 s), interval over groups. Communities: share of the tokens with at least three community members among the buyers of the first 5 s that reach 20 SOL, Wilson interval and number of tokens. Nothing is re-tuned from day to day.}\label{tab:daily}
\begin{tabular}{p{72pt}rrrr}
\toprule
Measure & Day 1 & Day 2 & Day 3 & Day 4 \\
\midrule
matched contrast, creation covariates: peak at least 2x, difference in pp & +7.4 [+5.8, +8.8] & +7.4 [+6.6, +8.2] & +7.5 [+6.4, +8.5] & +8.4 [+6.7, +10.0] \\
matched contrast, creation covariates: retention after peak, difference in pp & $-$13.0 [$-$14.4, $-$11.3] & $-$13.5 [$-$15.0, $-$12.1] & $-$12.9 [$-$14.3, $-$11.3] & $-$14.7 [$-$16.6, $-$12.9] \\
follower, all signals, event costs & $-$4.5 [$-$5.1, $-$4.0] & $-$4.3 [$-$4.8, $-$3.8] & $-$4.1 [$-$4.6, $-$3.6] & $-$4.5 [$-$5.4, $-$3.8] \\
follower, buyers only, event costs & $-$3.1 [$-$4.5, $-$2.1] & $-$3.9 [$-$5.1, $-$2.9] & $-$4.3 [$-$5.8, $-$2.9] & $-$4.0 [$-$5.7, $-$2.7] \\
follower, all signals, project costs & $-$8.8 [$-$9.5, $-$8.3] & $-$8.6 [$-$9.1, $-$8.2] & $-$8.6 [$-$9.0, $-$8.0] & $-$9.0 [$-$9.8, $-$8.2] \\
follower, buyers only, project costs & $-$7.0 [$-$8.3, $-$6.0] & $-$7.7 [$-$8.8, $-$6.7] & $-$8.1 [$-$9.5, $-$6.7] & $-$7.8 [$-$9.4, $-$6.5] \\
community 0: reach-20 share of signal tokens & 47\% [45, 49] (n=2870) & 46\% [44, 47] (n=3060) & 47\% [45, 49] (n=2644) & 48\% [45, 50] (n=1525) \\
community 3: reach-20 share of signal tokens & 40\% [37, 42] (n=1588) & 39\% [37, 42] (n=1388) & 38\% [35, 40] (n=1416) & 40\% [37, 44] (n=825) \\
community 4: reach-20 share of signal tokens & 61\% [58, 65] (n=802) & 58\% [54, 62] (n=643) & 49\% [41, 57] (n=157) & 30\% [19, 43] (n=54) \\
community 7: reach-20 share of signal tokens & 66\% [61, 72] (n=286) & 67\% [61, 73] (n=221) & 68\% [62, 73] (n=253) & 70\% [61, 77] (n=126) \\
base rate of reaching 20 SOL (complete tokens of the day) & 9.7\% & 9.3\% & 8.7\% & 8.0\% \\
\bottomrule
\end{tabular}
\end{table}

\begin{table}[htbp]
\centering\small
\setlength{\tabcolsep}{3pt}
\caption{Conditional information of each evidence channel: odds ratio of a pair co-buying again in the second half when the channel is present, at the same first-half co-buy count (logistic regression on all first-half co-buying pairs; intervals clustered on the first wallet). Adding the wallets' first-half activity changes the sign of several channels, so part of what a channel measures is how busy its wallets are.}\label{tab:channelinfo}
\begin{tabular}{p{114pt}p{114pt}rr}
\toprule
Model (controls) & Evidence channel & Odds ratio [95\% CI] & Pairs with channel \\
\midrule
count and acceptance & accepted by the discovery rules & 0.94 [0.87, 1.03] & 105\,654 \\
count and each reason & recurrent campaign funding & 0.81 [0.70, 0.93] & 25\,650 \\
count and each reason & recurrent entry and exit coordination & 3.15 [2.84, 3.49] & 68\,144 \\
count and each reason & recurrent joint authorization & 0.30 [0.27, 0.34] & 4\,627 \\
count and each reason & recurrent specific co-buy & 0.40 [0.36, 0.44] & 87\,303 \\
count and each reason & co-buy and consolidation & 3.77 [2.96, 4.79] & 2\,447 \\
count and each reason & transfer and trade & 2.43 [1.98, 3.00] & 2\,090 \\
count, each reason and wallet activity & recurrent campaign funding & 2.15 [1.79, 2.59] & 25\,650 \\
count, each reason and wallet activity & recurrent entry and exit coordination & 4.41 [3.93, 4.94] & 68\,144 \\
count, each reason and wallet activity & recurrent joint authorization & 0.42 [0.36, 0.49] & 4\,627 \\
count, each reason and wallet activity & recurrent specific co-buy & 3.17 [2.86, 3.52] & 87\,303 \\
count, each reason and wallet activity & co-buy and consolidation & 3.19 [2.39, 4.28] & 2\,447 \\
count, each reason and wallet activity & transfer and trade & 3.32 [2.59, 4.26] & 2\,090 \\
\bottomrule
\end{tabular}
\end{table}

\begin{table}[htbp]
\centering\small
\setlength{\tabcolsep}{3pt}
\caption{Incomplete observation. Panel A: the incomplete flag is almost a function of the creator's initial buy (a launch with an initial buy has a recorded trade by construction). Panel B: how much a lift computed against the complete tokens would change if the base rate were computed over all launches.}\label{tab:incompleteness}
\begin{tabular}{lrr}
\toprule
\multicolumn{3}{l}{\emph{A. Share of incomplete launches by characteristics fixed at creation}} \\
\midrule
Launches & Number & Incomplete \\
\midrule
all launches & 200\,443 & 23.1\,\% \\
initial buy = 0 & 74\,393 & 62.4\,\% \\
initial buy $>$ 0 & 126\,050 & 0.0\,\% \\
first token of the creator & 64\,531 & 26.3\,\% \\
creator with 10+ earlier tokens & 84\,431 & 20.4\,\% \\
\bottomrule
\end{tabular}
\\[0.8ex]
\begin{tabular}{p{178pt}rrr}
\toprule
\multicolumn{4}{l}{\emph{B. Effect on the base rate and the lift if incomplete launches were counted}} \\
\midrule
Assumption about the incomplete launches & Reach-20 rate & Base rate, all launches & Lift factor \\
\midrule
incomplete launches never reach 20 SOL & 0.0\,\% & 5.9\,\% & 1.30 \\
as recorded & 1.1\,\% & 6.2\,\% & 1.25 \\
at the rate of complete launches & 7.7\,\% & 7.7\,\% & 1.00 \\
\bottomrule
\end{tabular}
\end{table}

\begin{table}[htbp]
\centering\small
\setlength{\tabcolsep}{3pt}
\caption{Prediction of which wallets join a team when whole teams are held out. Candidates are grouped by the kept team with which they co-bought most in the first half; half of the teams (and of the candidates without any contact) are used for training and the other half for testing, repeated over ten random splits. About 90\% of the candidates are associated with one team, so splits differ strongly. Compare with the random split of the main text.}\label{tab:h9heldout}
\begin{tabular}{p{243pt}rrr}
\toprule
Method & AUC, mean & SD over splits & Worst split \\
\midrule
co-buy count with team members & 0.589 & 0.040 & 0.533 \\
in-block adjacency to team members & 0.533 & 0.011 & 0.518 \\
lag-distribution similarity to the team & 0.507 & 0.153 & 0.295 \\
LightGBM on node features & 0.723 & 0.050 & 0.637 \\
LightGBM without co-buy features & 0.568 & 0.068 & 0.475 \\
LightGBM on activity only & 0.522 & 0.036 & 0.457 \\
SGC-style smoothing + LightGBM & 0.735 & 0.063 & 0.633 \\
SGC-style smoothing + LightGBM, without own co-buy features & 0.678 & 0.087 & 0.564 \\
\bottomrule
\end{tabular}
\end{table}

\begin{table}[htbp]
\centering\scriptsize
\setlength{\tabcolsep}{3pt}
\caption{Ledger of the hypotheses and tests of the paper, part A (questions 1 to 5). Status: fixed in advance (the design of the validation and of the follower simulation, and the primary contrasts), developed on the same window (exploratory), or added in response to review. No multiplicity correction is applied; Outcome states what each test shows and Data used says whether the figure comes from data that played a part in selecting the rule.}\label{tab:ledgerA (questions 1 to 5)}
\begin{tabular}{lp{84pt}llp{84pt}p{84pt}}
\toprule
Id & Hypothesis or test & Sec. & Status & Data used & Outcome \\
\midrule
Q1-a & Accepted pairs persist in the second half & 5.1 & fixed in advance & first half / second half & persist, but not beyond what co-buy counts predict \\
Q1-b & Pairs with independent non-trading links co-buy more than matched random pairs & 5.1 & fixed in advance & first half / second half & supported; no added persistence at equal co-buy counts \\
Q1-c & Discovery groups beat raw co-buy clustering as a partition & 5.1 & fixed in advance & first half / second half & supported (giant component vs bounded groups) \\
Q1-d & Discovery groups beat naive clustering as a signal at equal numbers of flagged tokens & 5.1 & added in response to review & first half / second half & not supported: naive scores rank at least as well \\
Q1-e & Hub threshold and discovery parameters do not change the findings & 5.1, 9 & fixed in advance & slices of the window & supported \\
Q1-f & Each evidence channel adds information at equal co-buy count & 5.1 & added in response to review & first half / second half & not supported: acceptance adds nothing; single channels mixed and activity-dependent \\
Q2-a & Group activity is concentrated (Gini above the uniform null) & 5.2 & fixed in advance & full window & supported \\
Q2-b & Groups fall into distinct types & 5.2 & developed on this window & full window & four families; exploratory \\
Q3-a & Campaigns reuse a small pool of wallets more than chance & 5.3 & fixed in advance & full window & supported \\
Q4-a & Known members buy earlier than other early buyers (permutation null) & 6.1 & fixed in advance & full window & supported \\
Q4-b & Tokens with members show a pump-like price path (stratified controls) & 6.2 & fixed in advance & full window & supported \\
Q4-c & Members extract more cash than a matched placebo set of the same token & 6.3, 6.4 & fixed in advance & full window & supported on three valuation bases \\
Q4-d & Price-path contrast survives inverse-probability weighting on first-minute covariates & 6.4 & fixed in advance & full window & supported untrimmed; not supported within common support \\
Q4-e & Price-path contrast with covariates fixed at creation, by creator-clustered intervals & 6.4 & added in response to review & full window & supported (large E-value) \\
Q4-f & Main contrasts hold on each day of the window & 5.4, 10 & added in response to review & day by day & supported on each of the four days (appendix table of the daily replication) \\
Q4-g & Exclusion of incomplete launches biases the group-control contrasts & 3.3 & added in response to review & full window & no differential exclusion by construction; base-rate lift up to about 1.3 times larger \\
Q5-a & A follower at the earliest group signal earns a positive net return & 7.1 & fixed in advance & 70/30 split in time & not supported (all configurations negative) \\
Q5-b & The follower does better than random entry & 7.1 & fixed in advance & 70/30 split in time & not supported \\
Q5-c & Longer latency lowers the return (same trades) & 7.1 & added in response to review & 70/30 split in time & supported, about one point per second \\
Q5-d & A learned ranking of entry moments earns a positive net return & 7.3 & fixed in advance & 55/15/30 split, token-level purge & not supported \\
Q5-e & Apparent profits arise from omitting the own position or from a label leak & 7.4 & developed on this window & full window & demonstrated \\
\bottomrule
\end{tabular}
\end{table}

\begin{table}[htbp]
\centering\scriptsize
\setlength{\tabcolsep}{3pt}
\caption{Ledger of the hypotheses and tests of the paper, part B (question 6). Status: fixed in advance (the design of the validation and of the follower simulation, and the primary contrasts), developed on the same window (exploratory), or added in response to review. No multiplicity correction is applied; Outcome states what each test shows and Data used says whether the figure comes from data that played a part in selecting the rule.}\label{tab:ledgerB (question 6)}
\begin{tabular}{lp{70pt}lllp{70pt}}
\toprule
Id & Hypothesis or test & Sec. & Status & Data used & Outcome \\
\midrule
Q6-a & Merging groups that share wallets repairs fragmentation & 8.1 & developed on this window & second half (tuned) & supported; thresholds tuned on the evaluation half \\
Q6-b & Fast-wallet communities exist independently of the group builder & 8.2 & developed on this window & full window & 26 communities; only some pump \\
Q6-c & The 126-wallet team's reach-20 excess survives pre-entry weighting, matching and within-creator contrast & 8.3 & developed on this window & full window & supported at the 20-SOL level; not at 60 SOL or graduation \\
Q6-d & Three further communities show the same excess & 8.4 & developed on this window & full window & supported for three, absent for others \\
Q6-e & Teams' success persists in rank across halves & 8.5 & developed on this window & two halves & weak \\
Q6-f & Reputation, agreement and scouts raise the precision of the team signal & 8.6 & developed on this window & second half (tuned) & supported on the development half \\
Q6-g & The refined rules keep their precision on an untouched block & 8.10 & added in response to review & block C, frozen rules & partly supported (about one half to three quarters) \\
Q6-h & The signal precedes the event & 8.10 & added in response to review & block C & not for 26-48\% of signals \\
Q6-i & Team signals do not depend on one team & 8.10 & added in response to review & block C, leave-one-team-out & precision holds; coverage depends on the largest team \\
Q6-j & Group features add information beyond the first minute (prospective outcome) & 8.10 & added in response to review & block C & not supported \\
Q6-k & The creator's initial buy and snipers predict success & 8.7 & developed on this window & 70/30 split & supported; initial buy strongest \\
Q6-l & Teams conceal creator links, funding or cash-out & 8.8 & developed on this window & full window & creator not linked; funding mostly invisible; cash-out to low-degree addresses for one team \\
Q6-m & Graph neural networks predict who joins a team better than simple smoothing & 8.9 & developed on this window & random half of candidates & not supported; with whole teams held out the AUC falls from about 0.9 to about 0.72 \\
Q6-n & Wallet rotation hides the teams & 8.5, 8.8 & developed on this window & full window & no sign within the window; longer horizons untestable \\
\bottomrule
\end{tabular}
\end{table}

\end{document}